\documentclass{article}
\usepackage{graphicx} 
\usepackage{orcidlink,thumbpdf,lmodern}
\usepackage{authblk}
\usepackage{listings}
\usepackage{framed}
\usepackage{caption, subcaption}
\usepackage{makecell}
\usepackage{subcaption} 
\usepackage{float}
\usepackage{booktabs}
\usepackage{tikz}
\def\checkmark{\tikz\fill[scale=0.4](0,.35) -- (.25,0) -- (1,.7) -- (.25,.15) -- cycle;} 

\usepackage{xfrac}

\usepackage[round, sort]{natbib}
\title{\texttt{cosimmr}: an R package for fast fitting of Stable Isotope Mixing Models with covariates}
\author[1]{Emma Govan}%
\author[2]{Andrew L Jackson}%
\author[3]{Stuart Bearhop}%
\author[3]{Richard Inger}%
\author[4]{Brian C Stock}%
\author[5]{Brice X Semmens}%
\author[6]{Eric J Ward}%
\author[1]{Andrew C Parnell}%
\affil[1]{Hamilton Institute, Insight Centre for Data Analytics, Maynooth University}%
\affil[2]{Zoology Department, Trinity College Dublin}%
\affil[3]{Centre for Ecology and Conservation, University of Exeter}%
\affil[4]{Institute of Marine Research, Norway}%
\affil[5]{Scripps Institution of Oceanography, University of California}%
\affil[6]{Northwest Fisheries Science Center, NOAA Fisheries}%

\begin{document}

\maketitle

\section{Abstract}
\begin{enumerate}
\item The study of animal diets and the proportional contribution that different foods make to their diets is an important task in ecology. Stable Isotope Mixing Models (SIMMs) are an important tool for studying an animal's diet and understanding how the animal interacts with its environment.
\item We present \texttt{cosimmr}, a new R package designed to include covariates when estimating diet proportions in SIMMs, with simple functions to produce plots and summary statistics. The inclusion of covariates allows for users to perform a more in-depth analysis of their system and to gain new insights into the diets of the organisms being studied.
\item A common problem with the previous generation of SIMMs is that they are very slow to produce a posterior distribution of dietary estimates, especially for more complex model structures, such as when covariates are included. The widely-used Markov chain Monte Carlo (MCMC) algorithm used by many traditional SIMMs often requires a very large number of iterations to reach convergence. In contrast, \texttt{cosimmr} uses Fixed Form Variational Bayes (FFVB), which we demonstrate gives up to an order of magnitude speed improvement with no discernible loss of accuracy.
\item We provide a full mathematical description of the model, which includes corrections for trophic discrimination and concentration dependence, and evaluate its performance against the state of the art MixSIAR model. Whilst MCMC is guaranteed to converge to the posterior distribution in the long term, FFVB converges to an approximation of the posterior distribution, which may lead to sub-optimal performance. However we show that the package produces equivalent results in a fraction of the time for all the examples on which we test. The package is designed to be user-friendly and is based on the existing \texttt{simmr} framework. 
\end{enumerate}

Keywords: bayesian statistics; mixing models; stable isotopes; trophic ecology; variational bayes; 

\section{Introduction}\label{sec:intro}
Stable Isotope Mixing Models (SIMMs) are commonly used in ecology to study the proportional contribution that different foods make to an animal's diet \citep{phillips2012converting}. This information can be important as it allows scientists to look at diet, which resources are important for different species \citep{mcdonald2020ecology}, and consequently niche overlap and competition \citep{teixeira2021niche, aksu2023high}, as well as being useful in looking at trophic position and energy flow in an ecosystem \citep{manlick2022stable}. These models have been extensively used by ecologists over the past 20  years with recent papers revealing the foraging behaviour in Dugongs \citep{thibault2024c}, overlap in trophic niche between native and non-native species of carp \citep{aksu2023high}, and assessment of nursery areas used by the scalloped hammerhead shark \citep{paez2024assessment}. SIMMs have been shown to produce results comparable to direct observation \citep{swan2020evaluating}.
The approach relies on the fact that the stable isotopes of several elements, but most usefully those of hydrogen($\delta^{2}H$), carbon ($\delta^{13}C$), nitrogen ($\delta^{15}N$) and sulphur ($\delta^{34}S$), are incorporated into animal tissues from the diet in a predicable manner \citep{inger2008applications}. Thus, if the isotope ratios of potential dietary items are known then animal diets can be reconstructed from the stable isotope ratios from proteinaceous tissues using SIMMs. 
\texttt{cosimmr} is a new R package \citep{R_ref} developed to allow for the fast running of SIMMs, especially but not limited to those that include covariates. It has been designed to be easy to use for non-expert R users, with S3 classes used throughout. SIMMs are widely used and cited in other fields, such as geology \citep{munoz2019radiogenic} and pollution studies \citep{zaryab2022determining}, amongst many others. In other scientific areas, SIMMs and similar models can be referred to in the literature as `source apportionment models' \citep{hopke1991receptor}, `end member analysis' \citep{hooper1990modelling}, or `mass balance analysis' \citep{miller1972chemical}. Further discussion of these models can be found in \cite{govan2023simmr}.

The basic mathematical equation for a statistical SIMM is:
$$y = \sum_{k=1}^K p_k s_k + \epsilon.$$
Here $y$ is the mixture (consumer tissues) value (for example, the $\delta^{13}C$ or $\delta^{15}N$ values for the species we wish to study), $p_k$ are the proportions contributed by each source (dietary item)  $k$ (of $K$ total sources), $s_k$ is the source tracer value for source $k$, and $\epsilon$ is a residual term. The parameters $p_k$ are usually the main focus of scientific interest. These models are commonly expanded in diet analyses to allow for processes that cause the mixture and source tracer values to differ besides the source proportions, such as Trophic Discrimination Factors \citep[TDFs;][]{inger2008applications} and concentration dependence \citep{phillips2002incorporating}. Other expansions include process error on the dietary proportions \citep{moore2008incorporating} as well as hierarchical source fitting \citep{ward2010including}. The models are further made richer by incorporating random effects on the source values \citep{semmens2009quantifying}. Here, whilst we include many of these extensions, our focus is on the inclusion of covariate dependence on the proportions $p_k$. The restriction that these must sum to unity (i.e. a simplex) makes their estimation more complex, and specialist link functions are required to map their values on to covariates. We provide a more detailed explanation of the mathematical model behind \texttt{cosimmr} in Section \ref{sec:math}.

Modern SIMMs are fitted using the standard tools of Bayesian inference. Most commonly this involves using Markov chain Monte Carlo (MCMC) to obtain samples from the posterior. For complex models with covariates this can be extremely slow, with models requiring millions of posterior samples and taking several days to converge, if they converge at all. By contrast in \texttt{cosimmr} we use Fixed Form Variational Bayes (FFVB), specifically Gaussian Variational Bayes with Cholesky Decomposed Variance \citep{titsias2014doubly, tan2018gaussian}. FFVB is an optimisation-based algorithm which works by first defining the form of the posterior distribution (here a multivariate normal distribution), and then minimising the Kullback-Leibler divergence between this VB approximation and the true posterior distribution. More details on the assumed distributions used in \texttt{cosimmr} can be found in Section \ref{sec:FFVBalg}. The main advantage of using FFVB, and thus of \texttt{cosimmr}, is that it works by optimisation rather than sampling, and therefore can be much faster to produce a posterior distribution. However, convergence issues can still occur \citep{yao2018yes}. Our approach gives \texttt{cosimmr} a key advantage over other packages, which tend to use JAGS \citep{plummer2003jags} to implement the MCMC algorithm. Additionally, \texttt{cosimmr} runs using C++ code via Rcpp \citep{rcpp_ref} which allows for an additional speed boost.

SIMMs have been a popular method for studying animal diets, with thousands of citations across different papers, and many packages have been developed in order to make this easier. Some of the most popular packages are listed below. A summary is also provided in Table \ref{table:Other_pkgs}. A more detailed comparison of \texttt{cosimmr}, \texttt{simmr}, and MixSIAR is provided in Table \ref{table:comparison}.

\begin{itemize}
    \item Isosource \citep{Isosource_paper} was one of the earliest developed packages for running SIMMs. It worked by simulating possible values of each proportion to produce many potential combinations of proportions. Importantly, it lacked an explicit statistical basis.
    \item MixSIR \citep{moore2008incorporating} adopted a Bayesian framework and allowed for the inclusion of Trophic Discrimination Factors (TDFs). MixSIR utilised Importance Sampling, generating many samples of possible proportion combinations and calculating importance weights to produce the final posterior sample. MixSIR also allowed for the inclusion of prior information and allowed uncertainty to be incorporated into SIMMs.
    \item SIAR \citep{parnell2010source} was developed as an R package. It utilised Markov chain Monte Carlo (MCMC) sampling. SIAR also included a residual component $\epsilon$ in the model. SIAR is no longer updated and simmr was developed to replace it.
    \item simmr \citep{govan2023simmr} is an R package that follows a Bayesian framework, and provides the option of using either JAGS \citep[Just Another Gibbs Sampler;][]{plummer2003jags} or Fixed Form Variational Bayes \citep[FFVB;][]{tran2021practical} for running the models. simmr allows for the inclusion of concentration dependence and Trophic Discrimination Factors but does not allow for covariates on the dietary proportions.
    \item FRUITS \citep{fruits} allowed for the inclusion of concentration dependence and prior information in a Bayesian framework. FRUITS is encoded in  Visual Basic and runs via the FRUITS computer programme. This package simplifies the incorporation of prior information.
    \item IsotopeR \citep{hopkins2012estimating} adopted a Bayesian framework and allowed for inclusion of concentration dependence and TDFs, and also includes covariance. It uses MCMC for running the models.
    \item MixSIAR \citep{stock2018analyzing} is an R package that fits models in JAGS. MixSIAR allows for the inclusion of covariates as fixed, random, or continuous effects. It fits the source means hierarchically, either using raw data or sample statistics (means, variances, and sample sizes).
\end{itemize}

\begin{table}[ht]
\centering
\resizebox{\columnwidth}{!}{%
\begin{tabular}{|c|c|c|c|c|c|c|c|c|c|}
\hline
Software&Language&Algorithm&TDFs&{\makecell{Concentration\\Dependence}}&Covariates&{\makecell{Prior\\Info}}&Comments&{\makecell{Hierarchical/\\Source Fitting}}&Reference\\
\hline
Isosource &Visual Basic&Trial and Error&N&N&N&N&Frequentist&N&\cite{Isosource_paper}\\
simmr &R&MCMC and FFVB&Y&Y&N&Y&Ease-Of-Use Design&N&\cite{govan2023simmr}\\
FRUITS &Visual Basic&MCMC&Y&Y&N&Y&-&N&\cite{fruits}\\
IsotopeR &R&MCMC&Y&Y&N&Y&Hierarchical Model&Y&\cite{hopkins2012estimating}\\
MixSIAR &R&MCMC&Y&Y&Y&Y&Allows for Raw Data&Y&\cite{stock2018analyzing}\\
cosimmr &R&FFVB&Y&Y&Y&Y&Aims for Speed&N&This Paper\\
\hline
\end{tabular}%
}
\caption{Table showing summary of current SIMM software and the features they offer}
\label{table:Other_pkgs}
\end{table}

\begin{table}[ht]
\centering
\resizebox{\columnwidth}{!}{%
\begin{tabular}{|c|c|c|c|}
\hline
&simmr&MixSIAR&cosimmr\\\hline
TDFs&\checkmark&\checkmark&\checkmark\\
Concentration Dependence&\checkmark&\checkmark&\checkmark\\
Covariates&-&\checkmark&\checkmark\\
Prior Information&\checkmark&\checkmark&\checkmark\\
Hierarchical/Source Fitting&-&\checkmark&-\\
Prior Visualisation&\checkmark&-&\checkmark\\
MCMC&\checkmark&\checkmark&-\\
FFVB&\checkmark&-&\checkmark\\
\hline
\end{tabular}%
}
\caption{Table comparing simmr, MixSIAR and cosimmr features}
\label{table:comparison}
\end{table}

TDFs (Trophic Discrimination Factors) account for the fact that consumers may differentially lose `light' versions of isotopes with respect to `heavy' versions during the process of assimilation \citep{inger2008applications}. Thus, while TDFs are important in ecological applications, they have less relevance to geological or pollution-based applications as these same processes do not occur (although similar processes such as isotopic fractionation may need to be accounted for). TDFs can be calculated in the lab, or calculated mathematically, such as in \cite{greer2015simple}. Alternatively \texttt{SIDER} \citep{healy2018sider} is an R package that allows for estimates of TDFs based on phylogenetic relatedness of species. Estimates can also be obtained from the scientific literature.

Concentration dependence accounts for the fact that different food sources can contribute proportionally different amounts of each isotope \citep{phillips2002incorporating}. The standard two-isotope model assumes all food sources contribute both isotopes equally. However, there are often occasions where a food source can be rich in one isotope and poor in another, thus not contributing equally to both. Instead, concentration dependence assumes a source's contribution to each isotope is proportional to the mass of the food source times the elemental concentration of the isotope within the food source. Inclusion of concentration depedence facilitates conversion between consideration of either the total mass of food sources assimilated and the mass of specific elements within them.

The inclusion of covariates in the SIMM allows users to avoid pseudo-replication \citep{hurlbert1984pseudoreplication}, because if a covariate is important to the diet proportions, its exclusion violates the assumption that all mixtures are independently and identically distributed. Including covariates in SIMMs allows users to determine the potentially causal relationships between covariates and diet proportions. Although the model returns these coefficients, they are only available in a transformed space (via the link function) and not directly interpretable. We have designed \texttt{cosimmr} to produce interpretable output in `coefficient space' where users can determine the direction of the relationship and evaluate uncertainty, and also in `proportion space' ($p$-space) which allows users to see the effect of the covariate directly on the dietary proportions. These tools are defined via a \texttt{predict} function that to allow the user to predict dietary proportions based on combinations of covariates that may not necessarily be present in the original dataset. We follow the `tidyverse' \citep{tidyverseref} style guide, with the Snake case naming convention and S3 classes used throughout. Being able to evaluate the model at new values of the covariates allows for a more detailed picture to be seen. Uncertainty intervals, in the form of Bayesian credible intervals, are provided for all estimated quantities. Advanced users have access to the full posterior distributions as created by the FFVB algorithm. 

\begin{figure}[h!]
\centering
\includegraphics[width=0.75\textwidth]{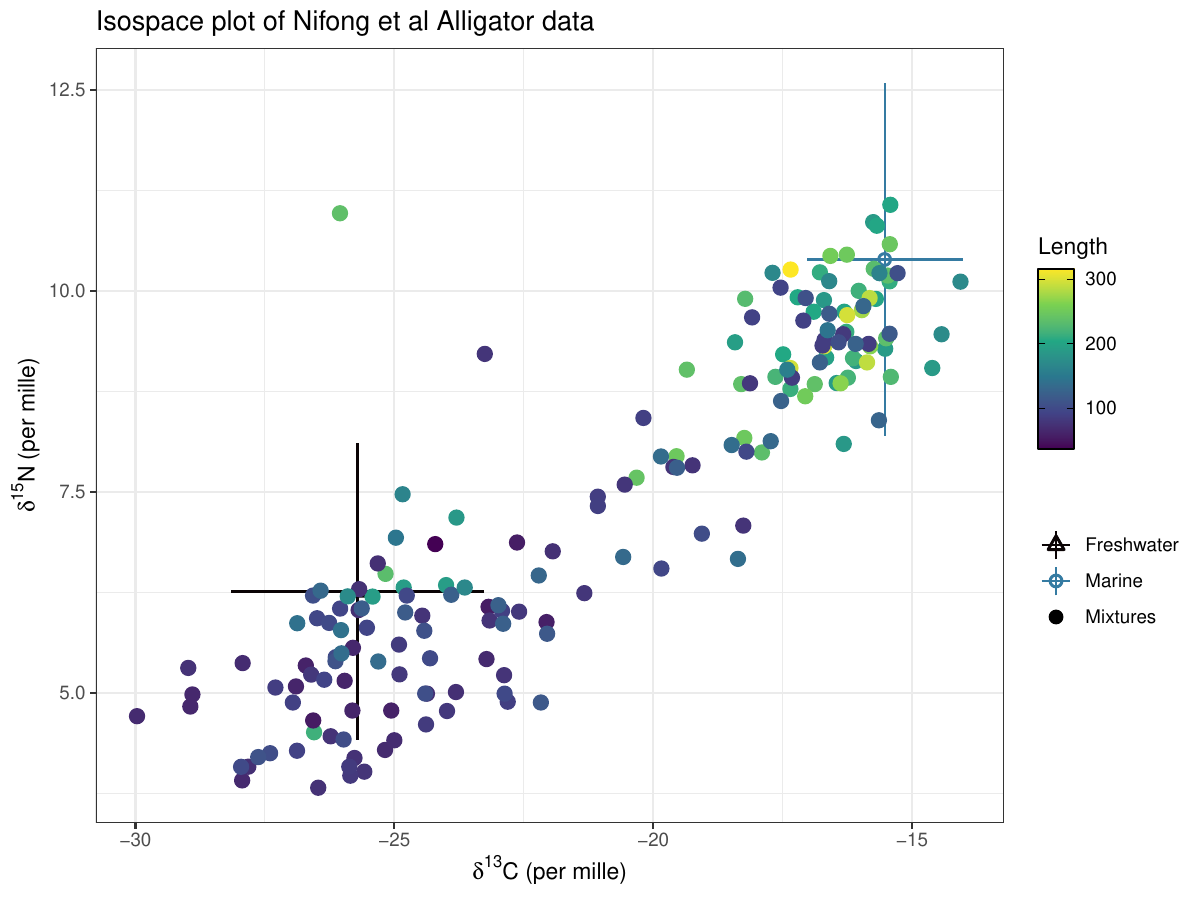}
\caption{\label{fig:alligator_iso} Iso-space plot for Alligator dataset (discussed in Section \ref{subsec:Ali}). Individuals are represented by circles coloured by covariate (length) and their isotope-ratio values are adjusted by TDFs. Two diet sources, Freshwater and Marine, are represented by a black triangle and blue circle, respectively, with 1 standard deviation also plotted. The axes on this plot are the Carbon-13 and Nitrogen-15 ratios of the isotope with respect to the `standard' measurement.}
\end{figure}

The data required by \texttt{cosimmr} can be illustrated by Figure \ref{fig:alligator_iso}. This example is discussed more fully in Section \ref{subsec:Ali}. This figure shows an `iso-space' plot generated by \texttt{cosimmr}. It shows each individual plotted in iso-space, with the axes representing $\delta^{13}C$ and $\delta^{15}N$. The diet sources (Marine and Freshwater in this case) are also plotted on the graph. It is important that all individuals lie within the mixing polygon created by the source means. If they do not, it indicates that the mixing system does not follow the model assumptions. Possible reasons for observations lying outside the mixing polygon include: issues with data collection; inaccurate TDFs; or missing food sources, amongst others. However, note that the mixing polygon vertices are sample means subject to sampling error, so there is some uncertainty in their exact position. Hierarchical Bayesian models allow for the source means to deviate from the source sample means by maximizing the likelihood of the source and mixture tracer data together \citep{ward2010including,hopkins2012estimating,stock2018analyzing}.

\section{Statistical approaches to stable isotope mixing models}\label{sec:math}

The fundamental SIMM we fit can be written as:

$$y_{ij} = \sum_{k=1}^K p_k(\mathbf{x}_i) q_{jk} (s_{ijk} + c_{ijk}) + \epsilon_{ij}$$

\noindent Where: \begin{itemize}
    \item $y_{ij}$ are the mixture/consumer tracer values of individual $i$ for tracer $j$,
    \item $p_k(\mathbf{x}_i)$ are the proportions of each source $k$ contributing to the mixture value at each covariate value $\mathbf{x}_i$ where $\mathbf{x}_i$ is an $L$-vector of covariate values for individual $i$,
    \item $q_{jk}$ represents the concentration dependence for tracer $j$ on source $k$,
    \item $s_{ijk}$ is the consumed source value by individual $i$ of the food source $k$ on tracer $j$,
    \item $c_{ijk}$ is the trophic discrimination factor of individual $i$ for source $k$ on tracer $j$
    \item $\epsilon_{ij}$ is the residual error for individual $i$ on tracer $j$
\end{itemize}
We index individuals as $i = 1, \ldots, N$, tracers as $j = 1, \ldots, J$, and sources as $k = 1, \ldots, K$. We assume there are $l = 1, \ldots, L$ covariates so that $\mathbf{x}_i = \{x_{i1}, \ldots, x_{iL}\}$. For notational brevity we write $p_k(\mathbf{x}_i)$ as $p_{ik}$. It is common to make the prior assumptions that $\epsilon_{ij} \sim N(0, \sigma_j^2)$, $s_{ijk} \sim N(\mu_{s, jk}, \sigma_{s, jk}^2)$, and $c_{ijk} \sim N(\mu_{c, jk}, \sigma_{c, jk}^2)$. Here $\mu_{s, jk}, \mu_{c, jk}, \sigma_{s, jk}, \sigma_{c, jk}$ may be assumed fixed as they are commonly available from other data sources, or learnt as part of a Bayesian hierarchical model. The residual standard deviation $\sigma_j$ is usually given a Uniform or (inverse) Gamma weakly informative prior. Other approaches have used multivariate normal distributions for these terms \citep[e.g.][]{hopkins2012estimating, parnell2013bayesian, stock2018analyzing} but these are not implemented in \texttt{cosimmr} as yet.
The model of \citet{stock2018analyzing} adds an additional multiplicative parameter to the first variance term to account for the assimilation of food items according to whether organisms are specialising in certain regions of the source probability distribution. Following their approach, we give the parameter $\epsilon$ a $U(0,20)$ prior when this additional process error is required. A table of prior values set can be seen at \ref{table:priors}.

\begin{table}[ht]
\centering
\resizebox{0.3\columnwidth}{!}{%
\begin{tabular}{|c|c|}
\hline
Term&Prior\\ \hline
$\epsilon$ & U(0,20)\\
$\beta_{kl}$ & N(0,1)\\
$\frac{1}{\sigma_j}$ & Ga(1,1)\\
\hline
\end{tabular}%
}
\caption{Table showing default priors used in \texttt{cosimmr}}
\label{table:priors}
\end{table}

The source and TDF random effects add an additional burden of $2KJ$ parameters into the model which can cause a significant computational slowdown. \cite{moore2008incorporating} proposed proposed marginalising across these parameters to produce a more complex, but more computationally tractable likelihood:

$$y_{ij} \sim N\left(\frac{\sum_{k=1}^Kp_{ik}q_{kj}\mu_{sc, kj}}{\sum_{k=1}^Kp_{ik}q_{kj}}, \frac{\sum_{k=1}^Kp_{ik}^2q_{kj}^2\sigma_{sc,kj}^2}{\sum_{k=1}^Kp_{ik}^2q_{kj}^2} +\sigma_j^{2}\right)$$

\noindent where $\mu_{sc, kj} = \mu_{s, kj} + \mu_{c, kj}$ and $\sigma_{sc,kj}^2 = \sigma_{s,kj}^2 + \sigma_{c,kj}^2$. 

The remaining prior distribution is that of the $p_{ik}$ terms which must retain the constraint that $\sum_{k} p_{ik} = 1$, but also allow for the terms to be dependent on the covariates $\mathbf{x}_i$. We use a Centralised Log-Ratio (CLR; \cite{clrdistref}) link function so that:
$$[p_{i1},...p_{iK}] = \left[ \frac{\exp(f_{i1})}{\sum_j {\exp(f_{ij})}}, \ldots, \frac{\exp(f_{iK})}{\sum_j {\exp(f_{ij})}} \right]$$
This prior has the advantage that the resulting terms $f_{ik}$ are unconstrained and made to depend directly on $\mathbf{x}_i$. We model this dependence linearly, but extensions that capture more nuanced dependence seem like a fruitful avenue for future research. We thus set: 
$f_{ik} = \mathbf{x}_i^T \mathbf{\beta}_k$. In other words, we can write the proportion for individual $i$ consuming food $k$ as $p_{ik} = CLR(\mathbf{x}_i^T \mathbf{\beta}_k)$.
where the parameters $\mathbf{\beta}_k$ model the dependence of the covariates across source $k$. We require a further prior distribution on $\mathbf{\beta}_k$ to ensure identifiability. By default we thus set $\beta_{kl} \sim N(0, 1)$. In certain circumstances where prior information is available on the $\beta_{kl}$ values we may use an informative prior of the form $\beta_{kl} \sim N(\mu_{\beta, kl}, \sigma_{\beta, kl}^2)$. The prior distribution for $\beta$ can be changed by the user in \texttt{cosimmr} via the \texttt{cosimmr\_ffvb} function.

\section{The Fixed Form Variational Bayes Algorithm}\label{sec:FFVBalg}
Fixed Form Variational Bayes (FFVB) is an optimisation-based algorithm that aims to approximate the posterior distribution of a Bayesian model in a pre-defined form \citep{salimans2013fixed}. It aims to finds the parameters of the `closest' probability distribution to that of the true posterior. Unlike traditional MCMC sampling methods, the greedy nature of the optimisation can usually find this approximate posterior far faster. In \texttt{cosimmr} we use a sub-type of FFVB known as Gaussian Variational Bayes with Cholesky decomposed covariance \citep{titsias2014doubly, tan2018gaussian}. To avoid becoming diverted in mathematical detail, we defer a full description of our approach to Appendix \ref{app:algorithm}. However here we provide an intuitive guide to how the fitting process works. 

Our model assumes that the joint posterior distribution of all the parameters is multivariate normal. The parameters for our model are $\mathbf{\beta}$, and $\mathbf{\sigma^2}$, representing the regression parameters across sources and covariates, and the residual variances across tracers. Recall that $p$ is a deterministic function of $\beta$ so not included in the algorithm. Since the variances are all restricted to be positive, we model these on the log scale. We thus write $\mathbf{\theta} = \{\mathbf{\beta}, \mathbf{\log(\sigma^2}) \} = \{ \beta_{11}, \ldots, \beta_{KL}, \log(\sigma^2_1), \ldots, \log(\sigma^2_J)\}$ as the set of parameters for which we want a posterior. 

For FFVB we need to define the form of the posterior distribution. We use:
$$\mathbf{\theta} \sim MVN(\mathbf{\mu}_\mathbf{\theta}, \mathbf{\Sigma}_\mathbf{\theta})$$
where $\mathbf{\mu}_\theta$ and $\mathbf{\Sigma}_\mathbf{\theta}$ are the mean and covariance matrix of the approximated posterior distribution. To avoid the positive semi-definite constraints on $\mathbf{\Sigma}_\theta$ we model the Cholesky decomposition of this matrix so that $\mathbf{\Sigma}_\mathbf{\theta} = \mathbf{LL}^T$. Together these terms are vectorised and written as: $\mathbf{\lambda} = \{\mbox{vec}(\mathbf{\mu}_\mathbf{\theta}), \mbox{vec}(\mathbf{L)} \}$. The goal of the algorithm is to provide a set of optimal values for $\mathbf{\lambda}$ which captures this posterior distribution. 
The main steps of the algorithm are as follows:
\begin{itemize}
\item Get starting values and use these to get an estimate of the difference between the posterior and variational posterior.
\item Calculate the gradient of this difference and use this to update $\mathbf{\lambda}$. The gradient of the posterior is calculated using automatic differentation and the gradient of the variational posterior is calculated manually.
\item Use new values of $\mathbf{\lambda}$ in place of starting values and repeat until stopping condition is met.
\end{itemize}

The algorithm requires several user-set hyper-parameters for fitting the model. The main hyper-parameters are: the patience $P$, which determines when the algorithm stops; and $S$ which provides the number of parameter samples used at each stage of the algorithm. For other algorithm parameters, we have used the default values from \cite{tran2021practical}. These parameters include: the fixed (\texttt{beta\_1} and \texttt{beta\_2}) and adaptive (\texttt{eps\_0}) learning rates; the size of the window to use when calculating stopping conditions (\texttt{t\_W}); and the threshold for exploring the learning space before a the learning rate is decreased (\texttt{tau}).The stopping conditions work by calculating a moving average of the lower bound (the difference between the log of the posterior and the log of the variational posterior). When the moving average does not improve after $P$ iterations then the algorithm stops. All hyper-parameters can be changed by the user if they wish when running the FFVB algorithm through \texttt{cosimmr}, though we have provided reasonable defaults which should work in most circumstances

The use of FFVB in covariate-dependent SIMMs is novel, and confers an advantage over MCMC due to the increase in speed.  This method is flexible and can be extended in the future, to include hierarchical source fitting, raw data, as well as random effects, features which are currently available in other SIMM software. We see up to an order of magnitude speed increase when comparing FFVB to MCMC, with comparable results produced.

\section{Running the \texttt{cosimmr} package}
The \texttt{cosimmr} package is available via CRAN. The first step to use it is to install and load the package:
\begin{verbatim}
R> install.packages("cosimmr")
R> library(cosimmr)
\end{verbatim}

The user must provide mixture data ($y$), source means ($\mu_{s,kj}$), and source standard deviations ($\sigma_{s,kj}$) to run \texttt{cosimmr}. TDFs, Concentration Dependence, and any covariates are not necessary to run the model but they may need to be included in order for the model to be ecologically valid.
The \texttt{cosimmr\_load} function can be used to read the data into R. This ensures the data is loaded in the correct format.
For illustration purposes we will use an artificially generated dataset, though see later sections for real case studies:
\begin{verbatim}
R> y = matrix(c(5, 5.1, 4.7, 3.6, 3.2, 0, -1, -2, -3, -7, 
+               3.1, 5.6, 3.6, 4.7, 1.3, 1, -4, -3, -7, -9), 
+               ncol = 2)
R> colnames(y) = c("iso1", "iso2")
R> mu_s = matrix(c(-10, 0, 10, -10, 10, 0), ncol = 2, nrow = 3)
R> sigma_s = matrix(c(1, 1, 1, 1, 1, 1), ncol = 2, nrow = 3)
R> s_names = c("A", "B", "C")
R> x = c(1.6, 1.7, 2.1, 2.5, 1.1, 3.7, 4.5, 6.8, 7.1, 7.7)
\end{verbatim}

This dataset contains measurements of two isotopic ratios, `iso1' and `iso2', as well as three food sources named `A', `B', and `C'. There is one continuous covariate, named `x'. These data can then be loaded into \texttt{cosimmr} using \texttt{cosimmr\_load} to create an object of class \texttt{cosimmr\_in}:
\begin{verbatim}
R> cosimmr_in_1 = cosimmr_load(
+                              formula = y ~ x,
+                              source_names = s_names,
+                              source_means = mu_s,
+                              source_sds = sigma_s
)
\end{verbatim}

As discussed in the introduction, it is recommended that these data are plotted on an `iso-space' plot before modelling. The iso-space plot shows tracer values of the mixtures as well as sources, with each axis representing a tracer. These tracers are often isotope ratios. It is important that the mixture data lies within the polygon formed when the sources are joined with straight lines (this polygon is referred to as the mixing polygon). If the mixtures do not lie within this polygon it indicates that there is an issue - potential reasons are that TDFs are inappropriate or a food source has been omitted. The polygon vertices are subject to sampling error and this could be another potential source of error. The iso-space plot can be generated by \texttt{cosimmr} by running the following code:
\begin{verbatim}
R> plot(cosimmr_in_1, col_by_cov = TRUE, cov_name = "x")
\end{verbatim}

\begin{figure}[h!]
\centering
\includegraphics[width=0.75\textwidth]{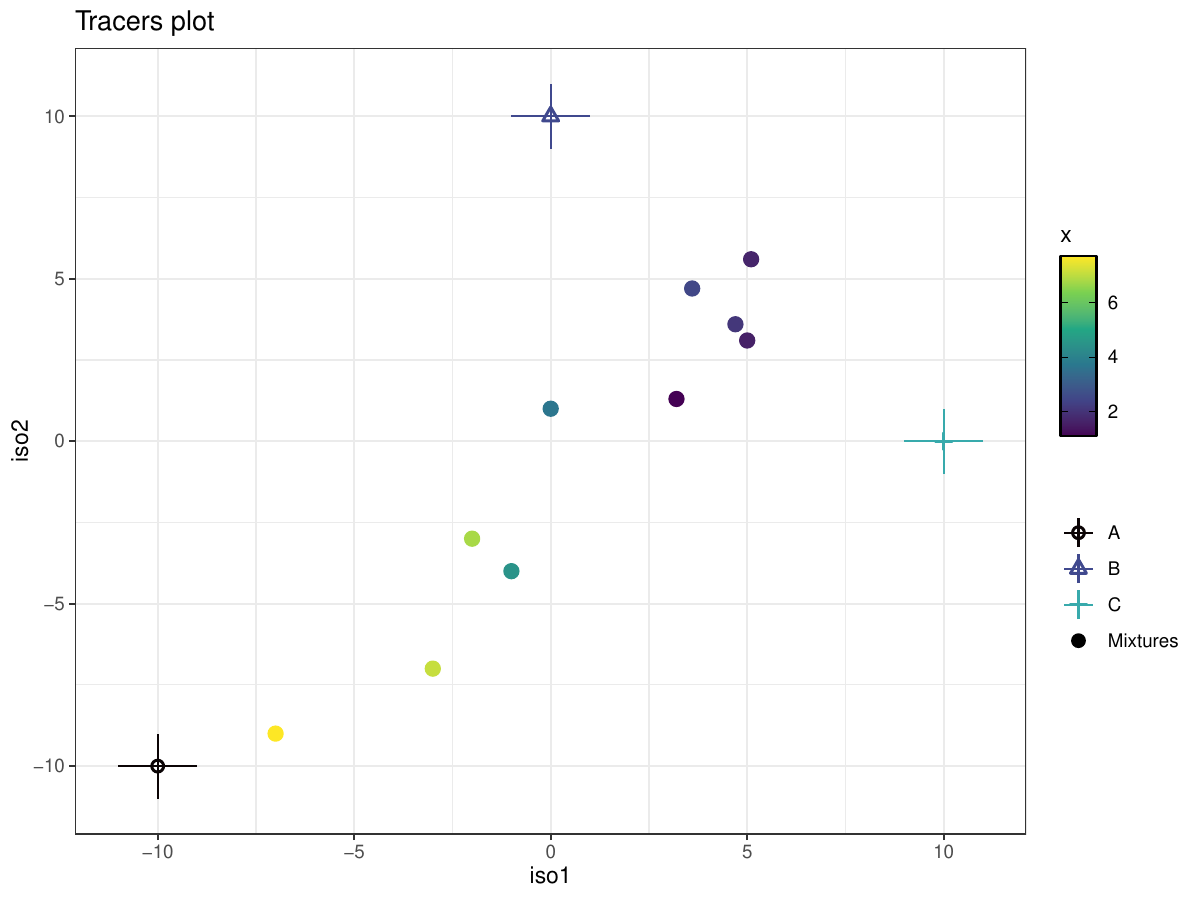}
\caption{\label{fig:simple_simm_example} An iso-space plot generated for artificial dataset showing iso1 on the $x$-axis and iso2 on the $y$-axis. The food sources A, B, and C are shown. Individuals are shown by circles coloured by covariate x.}
\end{figure}

The resulting plot can be seen in Figure \ref{fig:simple_simm_example}. It shows the mixtures lie within the mixing polygon. These data can then be run through the \texttt{cosimmr\_ffvb} function to produce an output of class \texttt{cosimmr\_out}:

\begin{verbatim}
R> cosimmr_out_1 = cosimmr_ffvb(cosimmr_in_1)
\end{verbatim}

\texttt{cosimmr} has built-in functions to produce summaries of the model run. Both graphical and textual summaries can be produced, as shown below.
\begin{verbatim}
R> summary(cosimmr_out_1, type = "statistics")

Summary for Observation 1

         mean    sd
P(A)    0.093 0.026
P(B)    0.424 0.052
P(C)    0.482 0.045
sd_iso1 1.449 1.133
sd_iso2 1.790 1.437

\end{verbatim}

In this example, we have specified that we wish to produce `statistics'. As we have not specified an observation then the function defaults to returning statistics for observation 1. Any/multiple individuals can be selected and `statistics', `quantiles' or `correlations' can be produced for each individual. The `statistics' summary produces a table of the means and standard deviations for the estimates of the proportion of each food eaten by the individual. An estimate of the marginal residual error of the isotope ratios is produced. In this summary we can see that individual 1's diet is mostly composed of foods B and C, with A contributing very little to their diet. This matches the observations we can make from the iso-space plot, where individual 1 lies at (5, 3.1), equidistant from food sources B and C and quite far away from food source A. The `quantiles' summary produces the 2.5\%, 25\%, 50\%, 75\% and 97.5\% quantiles for the same values as provided by the `statistics' summary. The `correlations' option produces the correlation values between each source and the residual error of the isotope ratios.

\begin{figure}[h!]
\centering
\includegraphics[width=0.75\textwidth]{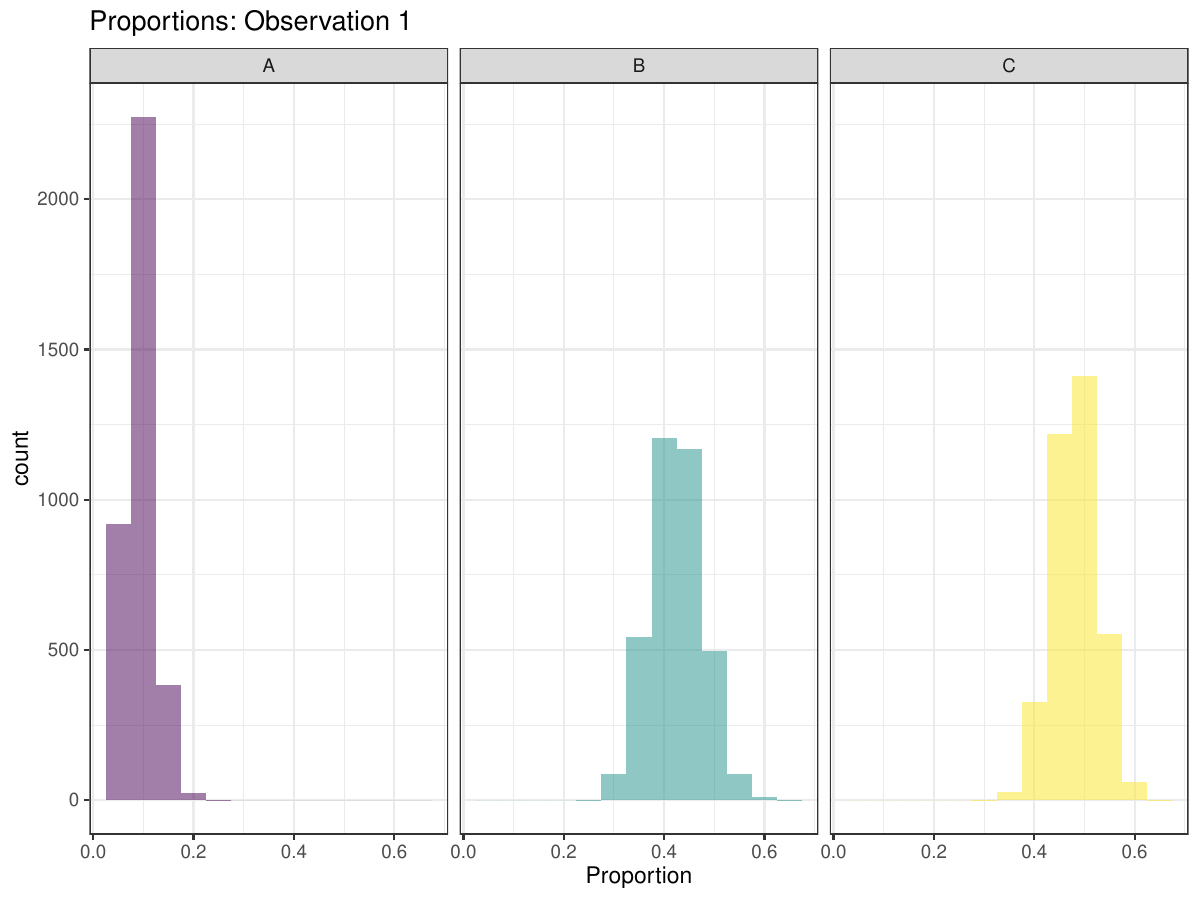}
\caption{\label{fig:simple_simm_prop_plot} \texttt{cosimmr} proportion plot showing consumption of different food
sources for observation 1 for the simple example}
\end{figure}

\begin{figure}[h!]
\centering
\includegraphics[width=0.75\textwidth]{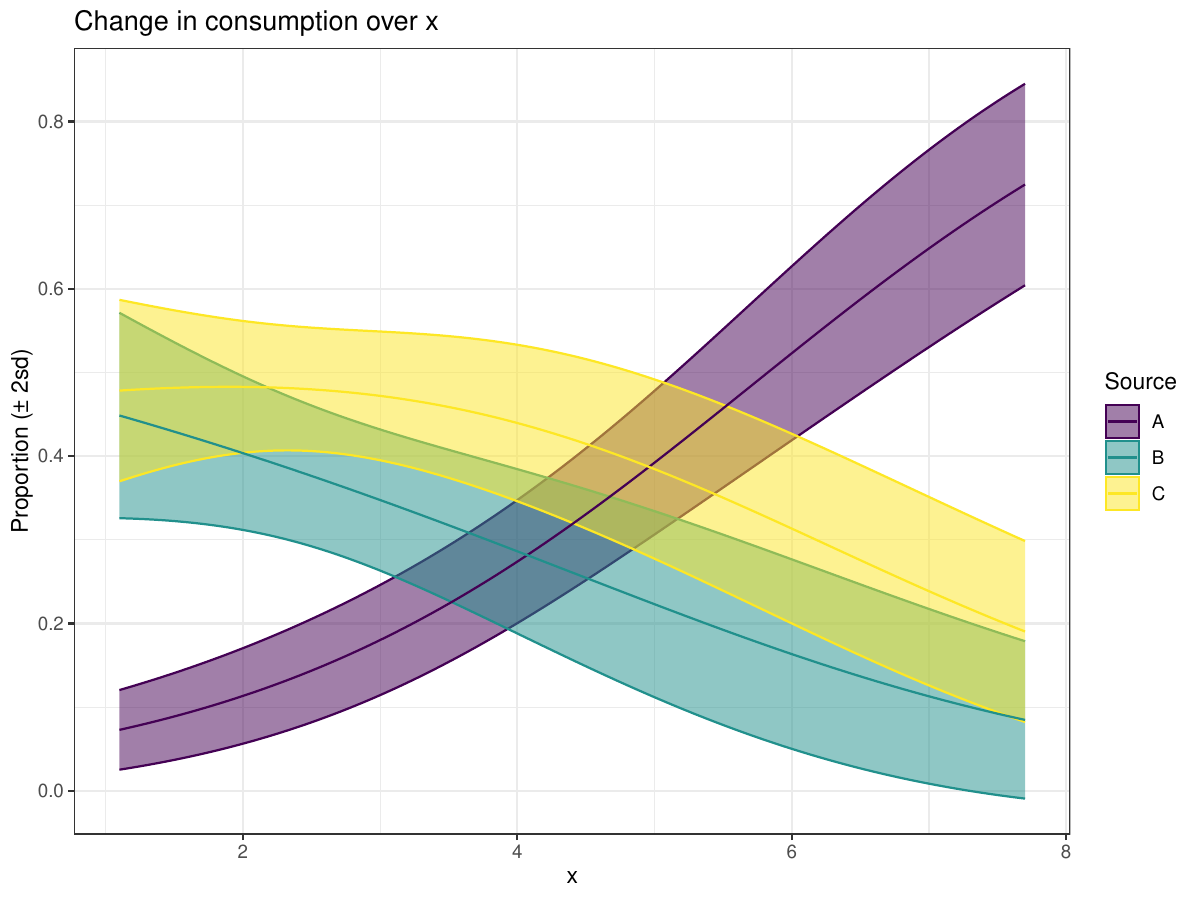}
\caption{\label{fig:line_simple} Covariates plot showing the change in the consumption of three food sources A, B, and C, as the covariate `x' increases for the simple example. Shaded interval shows mean $\pm$ 2 standard deviations.}
\end{figure}

Graphical plots can be produced in \texttt{cosimmr}. For example, below we create a proportion plot for observation 1. We can see the plot in Figure \ref{fig:simple_simm_prop_plot}. This shows the range of the proportion estimates for each food source for individual 1. We also create a `covariates plot' which shows the change in the proportion of foods consumed as the covariate changes. This plot can be seen in Figure \ref{fig:line_simple}.

\begin{verbatim}
R> plot(cosimmr_out_1, type = c("prop_histogram", "covariates_plot"), 
+                      obs = 1, cov_name = "x")
\end{verbatim}

Another important function within \texttt{cosimmr} is the ability to predict values based on covariate values, using the \texttt{predict} function, as illustrated below:
\begin{verbatim}
R> x_pred = data.frame(x = c(3, 5))
R> pred_out = predict(cosimmr_out_1, x_pred)
R> summary(pred_out, type = "statistics", obs = c(1,2))

Summary for Observation 1

      mean    sd
P(A) 0.181 0.033
P(B) 0.345 0.041
P(C) 0.474 0.039

Summary for Observation 2

      mean    sd
P(A) 0.393 0.043
P(B) 0.221 0.054
P(C) 0.386 0.054

\end{verbatim}

Here we can select values of the covariate for which we have no observations with those values, add these to the predict function along with out \texttt{cosimmr\_out} object, and return a new object which can then be used in the \texttt{summary} and \texttt{plot} functions as with a \texttt{cosimmr\_out} object. In this example we return dietary predictions for observations with covariate values 3 and 5, and can see what their estimated diet would look like if there were individuals with these predicted values.

\begin{figure}[h!]
\centering
\includegraphics[width=0.75\textwidth]{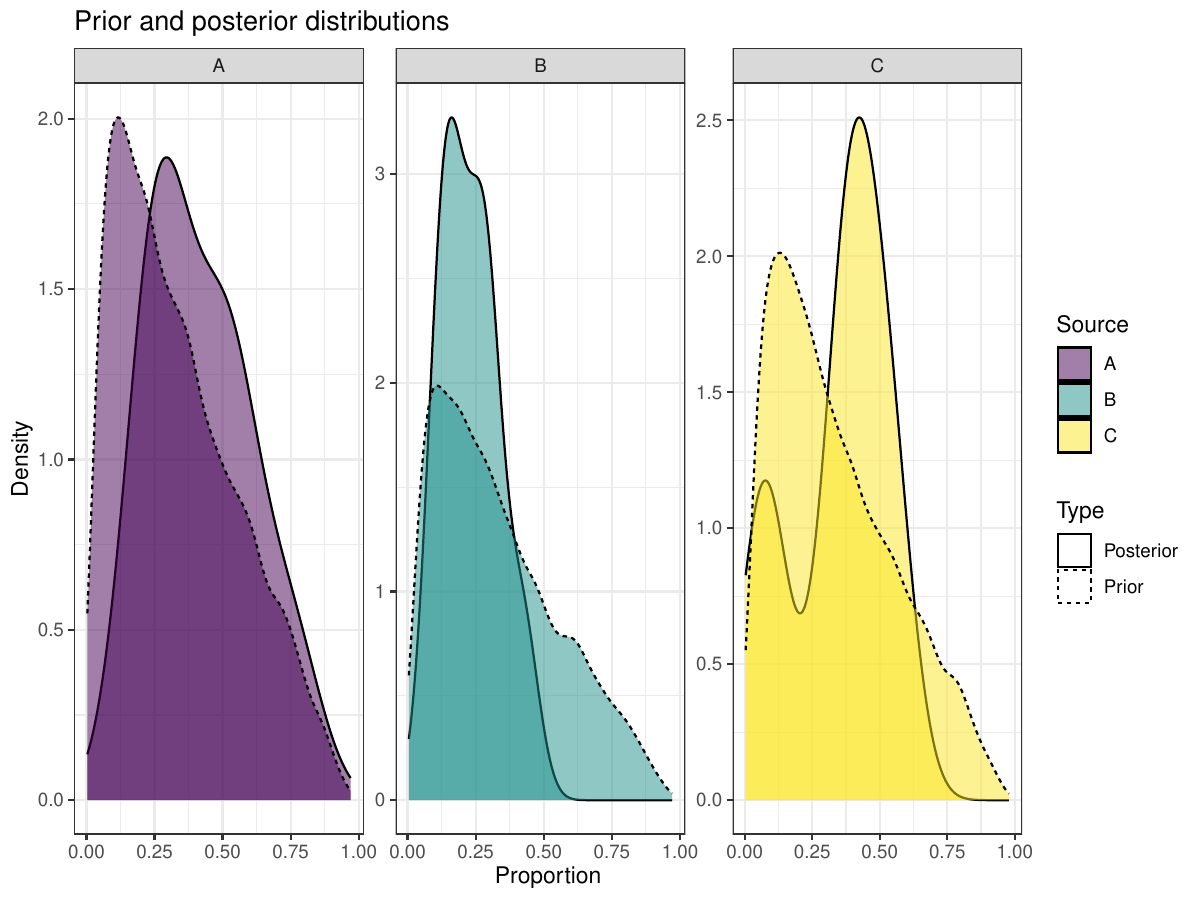}
\caption{\label{fig:simple_prior_viz} Three density plots showing the prior and posterior for each of the 3 food sources A, B, and C, in the simple example. Posterior estimates are shown for individual 1.}
\end{figure}

We can use the \texttt{prior\_viz} function to visualise how the posterior has changed from the prior. This plot overlays the posterior distribution and the prior distribution to see how it has changed or if the posterior has not changed much from the prior. This plot can be seen in Figure \ref{fig:simple_prior_viz}. We can see that our posterior estimates have changed from the prior estimate. One figure is shown here but it is recommended that users create plots for multiple individuals when running a model.

For convenience a summary of the main \texttt{cosimmr} functions is presented in Table \ref{table:functions}.

\begin{table}[h]
\centering
\begin{tabular}{|l|l|}
\hline
\texttt{cosimmr\_load} & Load in data in correct order/format\\ \hline
\texttt{cosimmr\_ffvb} & Run SIMM using Fixed Form Variational Bayes algorithm \\ \hline
\texttt{summary} & Produce summary of proportion values. Options include\\ 
& statistics, quantiles, and correlations \\ \hline
\texttt{plot} & Create plots, options include histogram or boxplot of beta \\
&values, iso-space plot, histogram or density plot of estimated \\
&proportions, plot of covariates vs proportions \\ \hline
\texttt{predict} & Predict proportion values for covariates not present in\\ 
&original dataset \\ \hline
\texttt{posterior\_predictive} & Create posterior predictive distribution of observations\\
&and plots for each observation \\ \hline
\texttt{prior\_viz} & Create plots showing prior values set vs posterior obtained\\
\hline
\end{tabular}
\caption{Main functions available in \texttt{cosimmr}}
\label{table:functions}
\end{table}

\section{Simulation Checks}\label{sec:sim}

In this section we use simulated data to verify that \texttt{cosimmr} returns valid estimates when a run is performed. We simulate data from the model using a variety of different data set sizes, varying $N$, $J$, $K$ and $L$, and changing the main parameters in the model. We evaluate the performance of the model by looking at how often the posterior distribution obtained by \texttt{cosimmr} matches the true values. The code for performing the runs in this section can be found at \url{https://github.com/emmagovan/cosimmr_paper/}. 

The values selected for several different simulations are presented in Table \ref{table:1}. We run low ($N = 50$, $J = 2$, $K = 3$, $L = 2$), medium ($N = 200$, $J = 3$, $K = 4$, $L = 5$) and high ($N = 500$, $J = 4$, $K = 5$, $L = 10$) versions to capture a range of scenarios, where N = no. of individuals, J = no. of tracers, K = no. of food sources, and L = no. of covariates. For each of these we simulate data using the default prior distribution of $\beta_{kl} \sim N(0, 1)$ and with $\sigma \sim Ga(1,1)$. We set $\mu_s \sim U(-10, 10)$, and $\sigma_s \sim U(0, 2)$. TDFs and Concentration dependence are ignored for this example.
\begin{table}[h]
\centering
\begin{tabular}{|c|c|c|c|}
\hline
     & Low & Medium & High \\ \hline
N &50 & 200 & 500 \\
J& 2 & 3 & 4 \\
K &3 & 4 & 5\\
L& 2 & 5 & 10\\
\hline
\end{tabular}
\caption{Values of parameters for different runs of our simulation checks}
\label{table:1}
\end{table}

After running each model, we produce posterior uncertainty intervals at the 50\% level and calculate the proportion of posterior samples inside these values. As an example, the posterior predictive plot for tracer 1 in the Medium run is shown in Figure \ref{fig:post_pred_sim1_med}. Posterior predictive plots for other simulations and tracers are found in the Appendix \ref{app:extra_plots} (Figures \ref{fig:post_pred_sim1_low}, \ref{fig:med_post_pred_app} and \ref{fig:high_post_pred_app}).

\begin{figure}[h!]
\centering
    \includegraphics[width=0.75\textwidth]{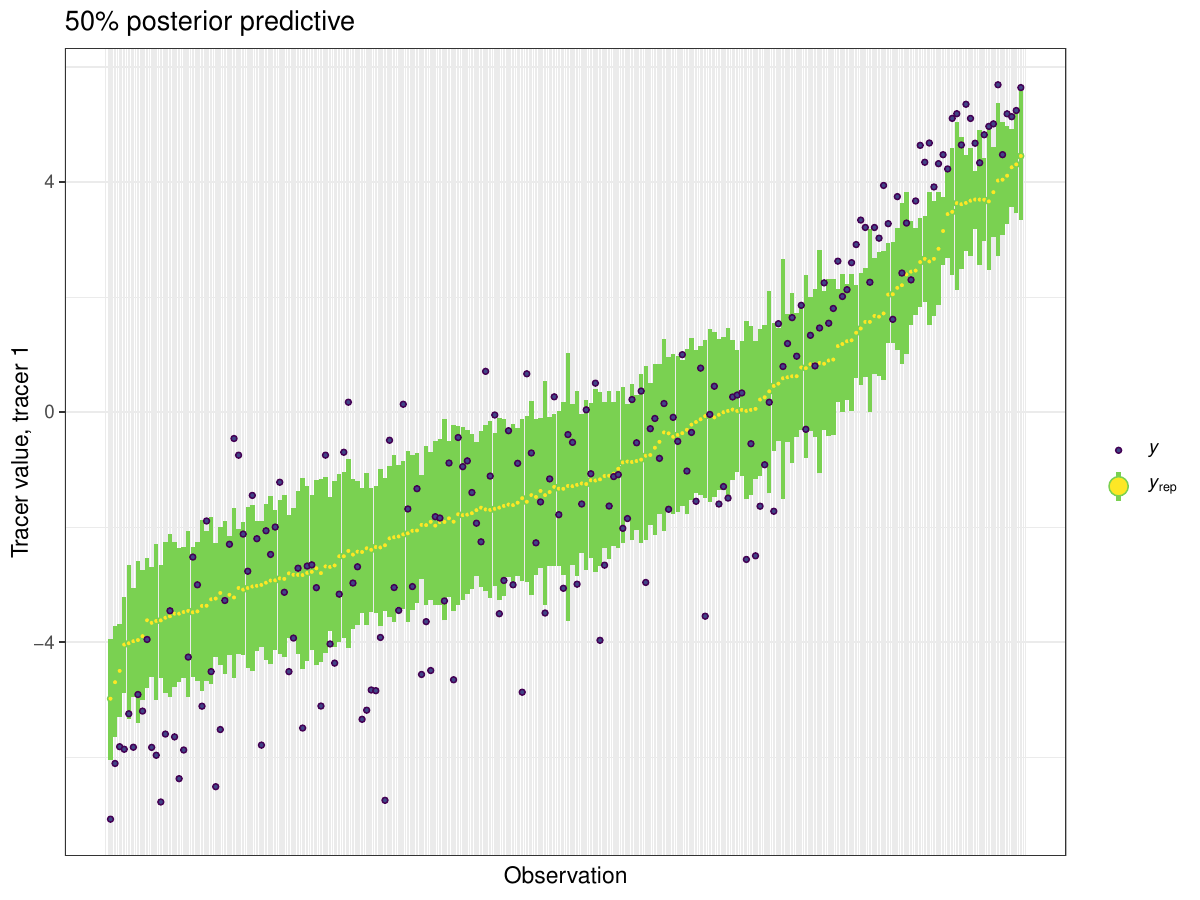}
 \caption{\label{fig:post_pred_sim1_med} Plot showing posterior uncertainty intervals at the 50\% level for the `Medium' simulated model run for tracer 1. The proportion of posterior values inside these values was 51\%.}
\end{figure}

The results show that \texttt{cosimmr} produces accurate estimates for the posterior values, even with more complex data and increased numbers of tracers and covariates.

\begin{figure}[h!]
\centering
\includegraphics[width=0.75\textwidth]{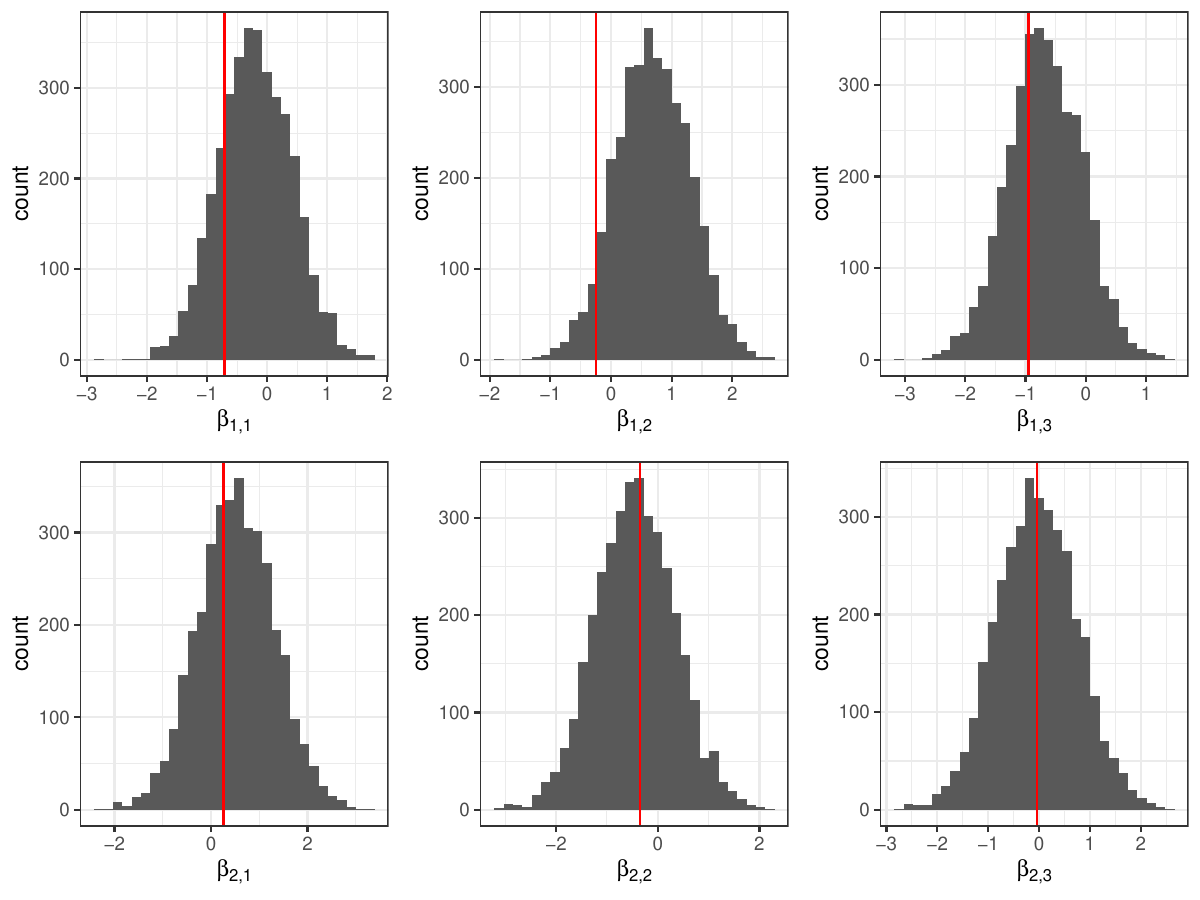}
\caption{\label{fig:beta low} Histograms showing posterior samples for beta values generated via \texttt{cosimmr} for the `Low' example, and red line showing `true' value of $\beta$ used to generate the mixture data.}
\end{figure}

 We show an example posterior distribution of $\mathbf{\beta}$ in Figure \ref{fig:beta low} for the `Low' run.  From Figure \ref{fig:beta low} we can see that \texttt{cosimmr} is performing well and producing $\mathbf{\beta}$ values similar to the original values used to generate data for this example, as illustrated by the red lines visible on the plots. This holds true for the Medium and High examples. Plots are omitted to avoid repetition.

\section{Case Studies}\label{sec:casestudies}

We now perform a direct comparison of \texttt{cosimmr} and MixSIAR to evaluate both the accuracy of the FFVB posterior and check the computational gains. We provide three case studies: the first using the Geese data of \cite{inger2006temporal}, for which we include a single categorical covariate (Group number). The second uses the Isopod data of \cite{galloway2014quantitative} which contains 8 tracers and a single covariate (Site). The third uses the Alligator data of \cite{nifong2015size}, for which we provide a detailed model comparison across 8 different potential covariate panels. In each case we compare the posterior distributions of the parameters, the posterior predictive performance, and the computational speed differences. All of the data for our model fits is available in the \texttt{cosimmr} package, available at \url{https://github.com/emmagovan/cosimmr} and on CRAN, and the code for running the models is available at \url{https://github.com/emmagovan/cosimmr_paper}. 

\subsection{Geese data \citep{inger2006temporal}}\label{subsec:Geese}
\begin{figure}[h!]
\centering
\includegraphics[width=0.75\textwidth]{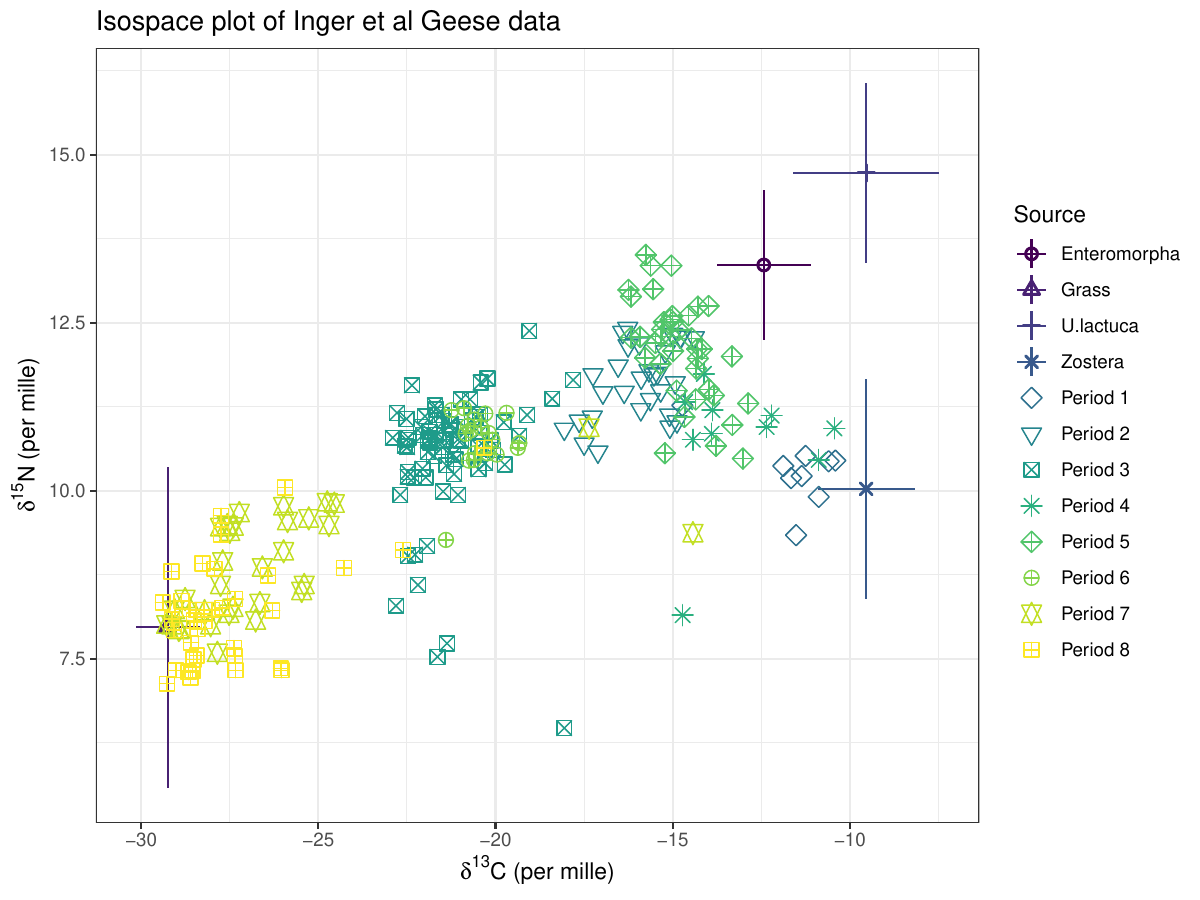}
\caption{\label{fig:geese_iso} Iso-space plot for geese dataset, coloured by covariate to highlight the differences between groups.}
\end{figure}

Our first example looks at the Brent Geese (\textit{Branta bernicla hrota}) dataset originally from \cite{inger2006temporal}. The covariate in this example is `Group' which is discrete. There are eight different groups which represent different time points at which individuals were sampled. $\delta^{13}C$ and $\delta^{15}N$ are the two isotopes used in this study. The iso-space plot for this data is seen in Figure \ref{fig:geese_iso}. TDFs and concentration dependence are accounted for in this model.

By looking at the Group covariate and how the proportion of each food in the diet of the geese differs between groups, we can see how their diet changes over time. This example highlights the usefulness and importance of covariates in SIMMs. The time of year influences the diet of these geese. They are known to consume \textit{Zostera spp.} between October and December at Strangford Lough (where these data were collected), and as time passes the geese remaining on the lough consume more \textit{Enteromorpha spp.} and \textit{Ulva lactuca} \citep{mathers1997quality}. To exclude time as a covariate in this example would violate the assumption that the data are IID, as their diets are influenced by the time of year, and consequently, by what food is easily available to the geese.

For this example, the Geese data was run through both \texttt{cosimmr} and MixSIAR. For MixSIAR a `long' run was needed for convergence. The first thing to note from these model runs is that \texttt{cosimmr} produces these results in a much quicker timeframe than MixSIAR, as we can see in Table \ref{table:geese_time}. \texttt{cosimmr} is over three times faster than MixSIAR for this example. The `Group' covariate is discrete and therefore in \texttt{cosimmr} it is treated as eight covariates when transformed into numeric covariates. Therefore this example is slower in \texttt{cosimmr} than other examples with only one numeric covariate. The code for using a categorical covariate in \texttt{cosimmr} is demonstrated below:
\begin{verbatim}
R> data(geese_data)
R> cosimmr_geese_in = cosimmr_load(
    formula = geese_data$mixtures ~ as.factor(geese_data$groups), 
    source_names = geese_data$source_names, 
    source_means = geese_data$source_means, 
    source_sds = geese_data$source_sds, 
    correction_means = geese_data$correction_means, 
    correction_sds = geese_data$correction_sds, 
    concentration_means = geese_data$concentration_means)
\end{verbatim}

\begin{table}[h]
\centering
\resizebox{\columnwidth}{!}{%
\begin{tabular}{|c|c|c|c|c|c|c|c|c|}
\hline
      &min &      lq &    mean &  median &      uq&     max &neval\\ \hline
cosimmr&29.5&30.4&38.0&33.7&45.0&61.2&   10 \\
 MixSIAR &113.6&121.5&130.1&124.5&139.3&155.3&   10\\
 \hline
\end{tabular}%
}
\caption{Table showing computation time (minutes) of \texttt{cosimmr} and MixSIAR (`long' run needed for convergence) for Geese example, showing the minimum (min), lower quartile (lq), mean (mean), median (median), upper quartile (uq) and number of evaluations (neval).}
\label{table:geese_time}
\end{table}

\begin{figure}[h!]
\centering
\begin{subfigure}[b]{0.45\textwidth}
    \includegraphics[width=\textwidth]{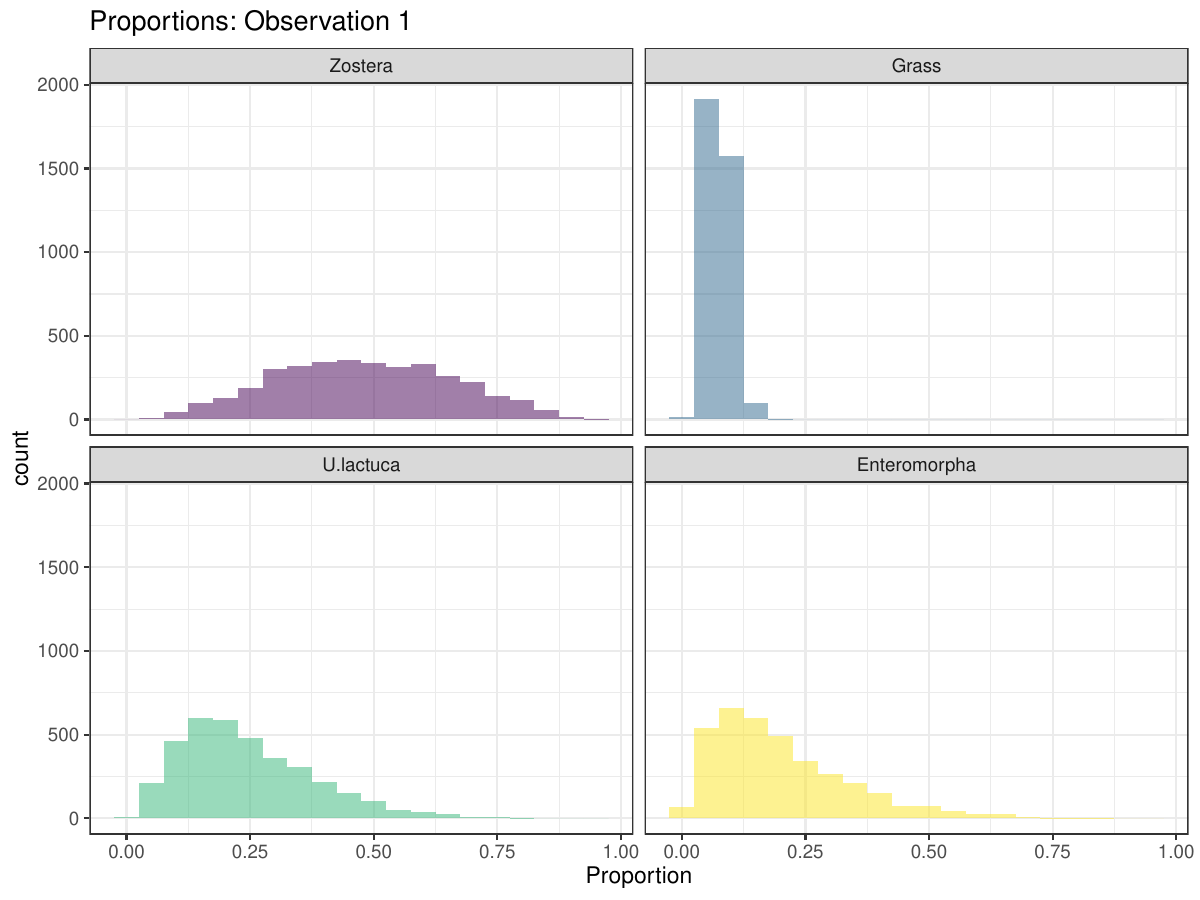}
    \caption{\texttt{cosimmr}}
    \label{fig:geese_cosimmr_small}
  \end{subfigure}
  \begin{subfigure}[b]{0.45\textwidth}
    \includegraphics[width=\textwidth]{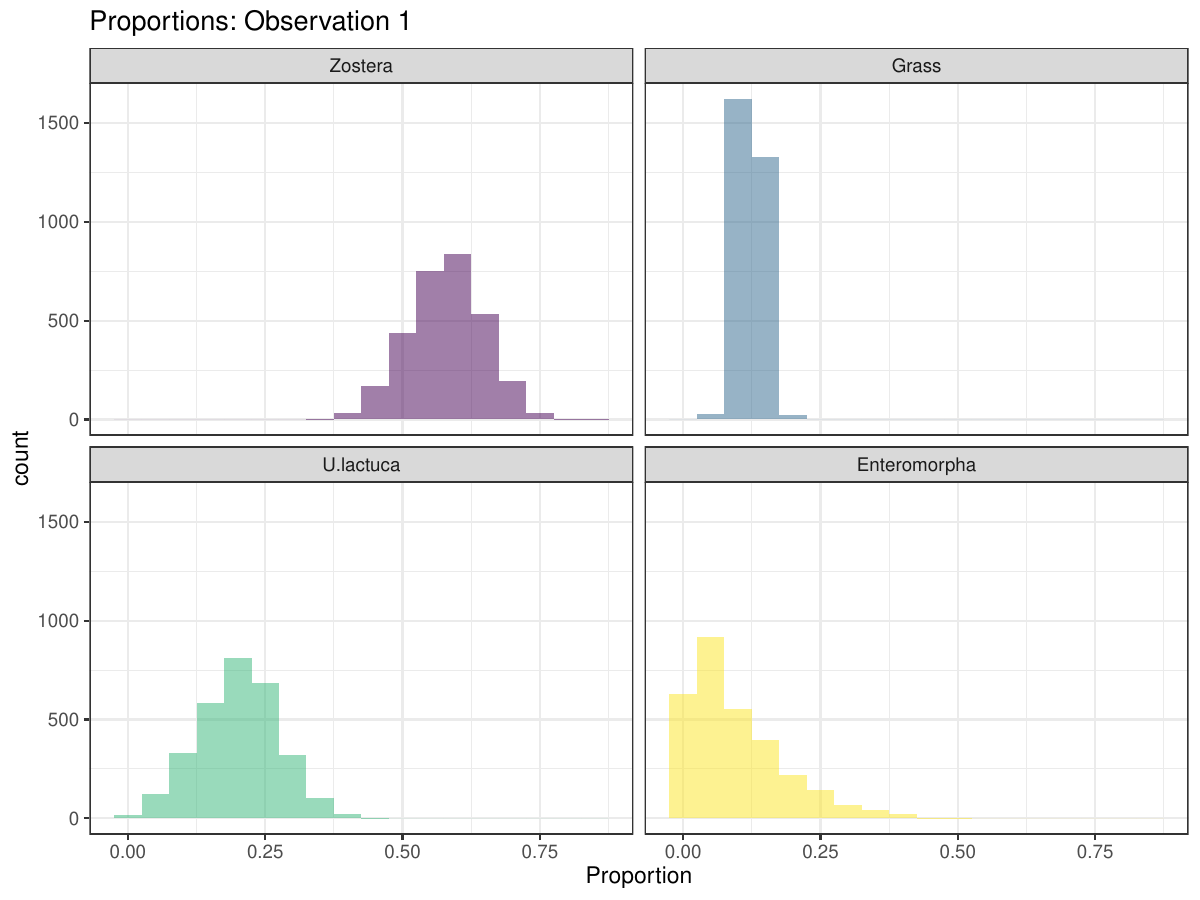}
    \caption{MixSIAR}
    \label{fig:geese_mixsiar}
  \end{subfigure}
\caption{\label{fig:geese_cosimmr} Proportion plot showing consumption of different food
sources for observation 1 for Geese example for \texttt{cosimmr} and MixSIAR}
\end{figure}

As well as comparing computational time, it is important that \texttt{cosimmr} produces results that are comparable to other SIMM software. From looking at Figure \ref{fig:geese_cosimmr} we can see that \texttt{cosimmr} and MixSIAR produce comparable results in terms of proportional estimates. These figures show the estimated dietary proportions for group 1. We can see that both \texttt{cosimmr} and MixSIAR produce similar estimates for the percentage that each food makes up in the diet of the first group. This result holds for the other seven groups in this dataset. Differences in results may be due to a slight difference in error structure between \texttt{cosimmr} and MixSIAR.

\begin{figure}[h!]
\centering
    \includegraphics[width=0.75\textwidth]{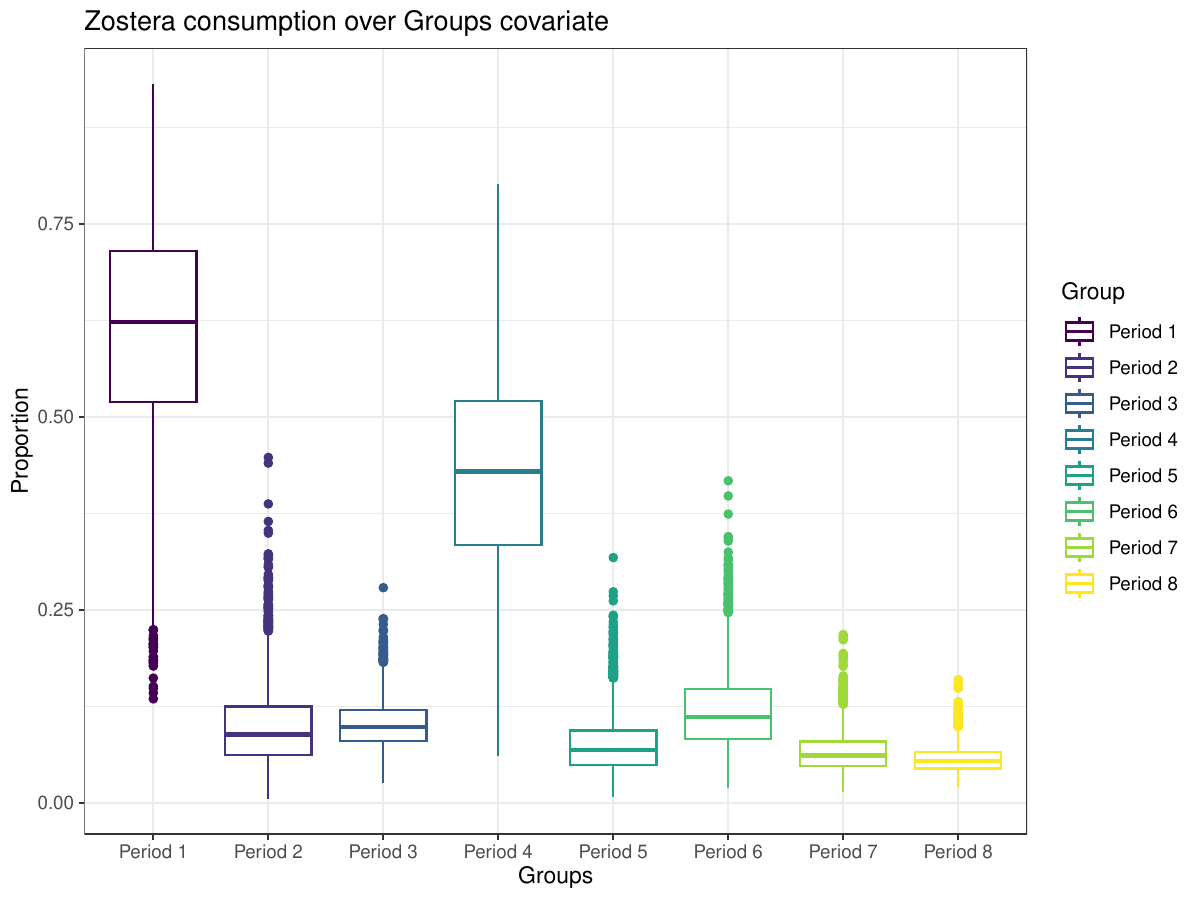}
\caption{\label{fig:geese_cov_change_boxplot} Boxplots showing change in consumption of \textit{Zostera} for Geese in different periods of the year.}
\end{figure}

\begin{figure}[h!]
\centering
    \includegraphics[width=0.75\textwidth]{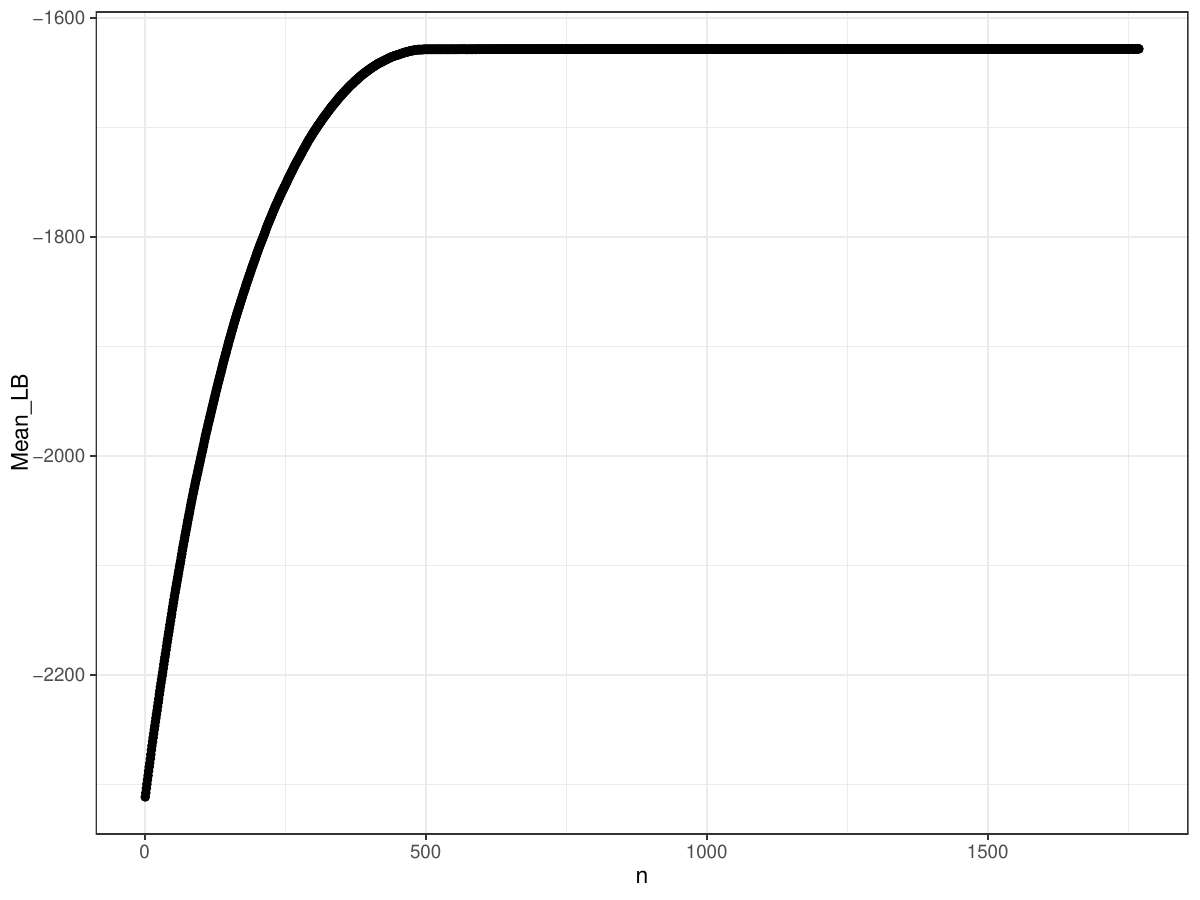}
\caption{\label{fig:convergence} Lineplot showing convergence of mean lower bound values for geese example, produced using \texttt{convergence\_check} function in \texttt{cosimmr}.}
\end{figure}

We can generate more complex plots using the \texttt{plot} function in \texttt{cosimmr}, to see how the consumption of a specific food changes between groups. In Figure \ref{fig:geese_cov_change_boxplot} we see the difference in consumption of \textit{Zostera spp.} across different groups. This highlights the usefulness of including covariates, as this detail would otherwise be lost. A convergence check can be performed using the function \texttt{convergence\_check}. This function returns the mean lower bound values produced and shows that this value converges. The result of this can be seen in Figure \ref{fig:convergence}.

\begin{figure}[h!]
\centering
\includegraphics[width=0.45\textwidth]{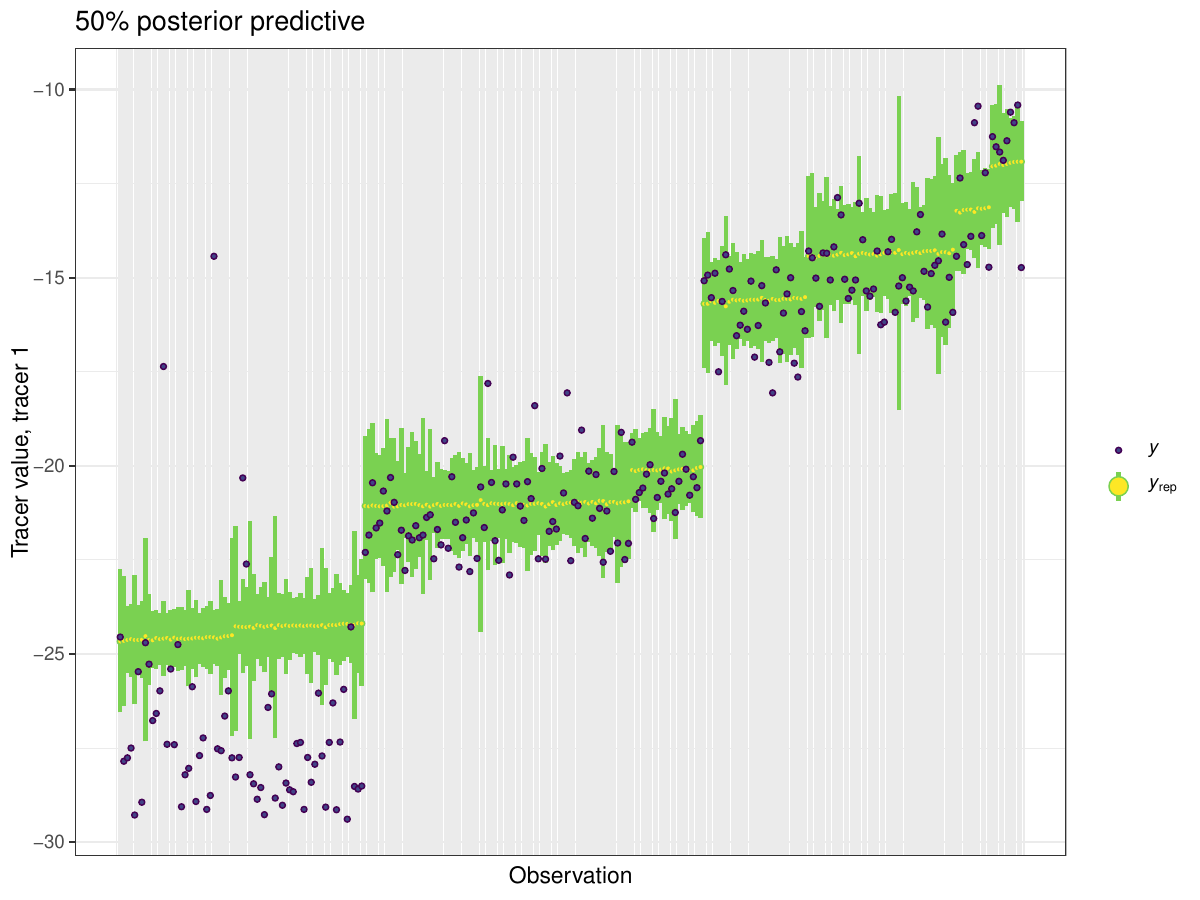}
\includegraphics[width=0.45\textwidth]{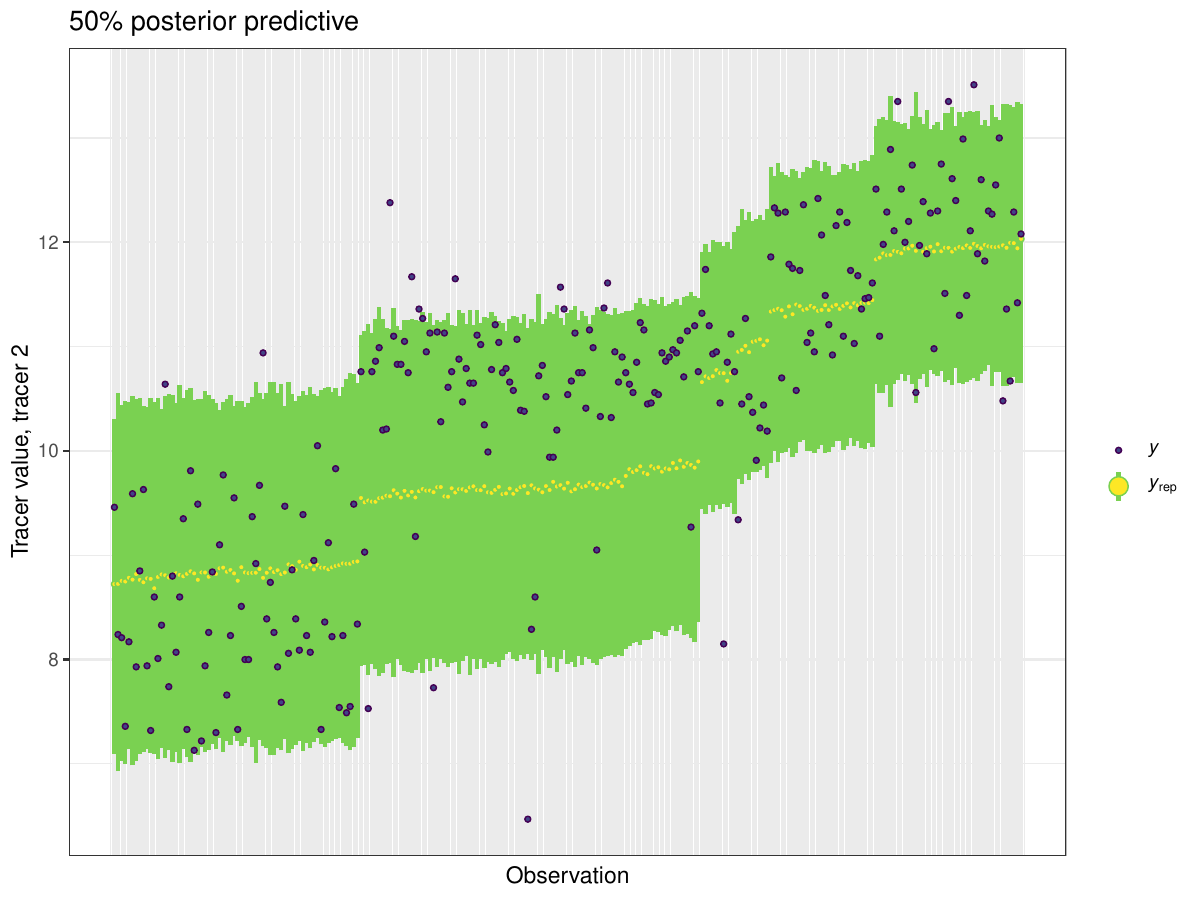}
\caption{\label{fig:geese_post_pred} Plot showing posterior uncertainty intervals at the 50\% level for the Geese data. The proportion of posterior values inside these values was 77\%. Groups 7 and 8 in this example contain outliers which may be a reason for the proportion being higher than we would expect.}
\end{figure}

The \texttt{posterior\_predictive} function can be used to produce a plot showing the posterior uncertainty intervals for this dataset. The produced plot is seen in Figure \ref{fig:geese_post_pred}.  We can see that 77\% of the data lies within the 50\% uncertainty intervals, showing that \texttt{cosimmr} is fitting the data adequately, although this value is above what we would expect. The posterior predictive values are a useful check of model fit and these are not available in other packages, so easy comparison is not available, but can be accessed using the \texttt{posterior\_predictive} function in \texttt{cosimmr}. Groups 7 and 8 contain the outliers seen in Figure \ref{fig:geese_post_pred} and this may be a potential reason for these results. For 75\% uncertainty intervals 87\% of observations are inside these intervals and 93\% are inside for 95\% confidence intervals. From this example we can see the importance and usefulness of including covariates, as it allows for us to look at the diet of the geese over time to see how the proportions of different foods in their diets change as the season progresses. It also highlights observations that may require further scrutiny.

\subsection{Isopod data \citep{galloway2014quantitative}}\label{subsec:Iso}
\begin{figure}[h!]
\centering
\includegraphics[width=0.75\textwidth]{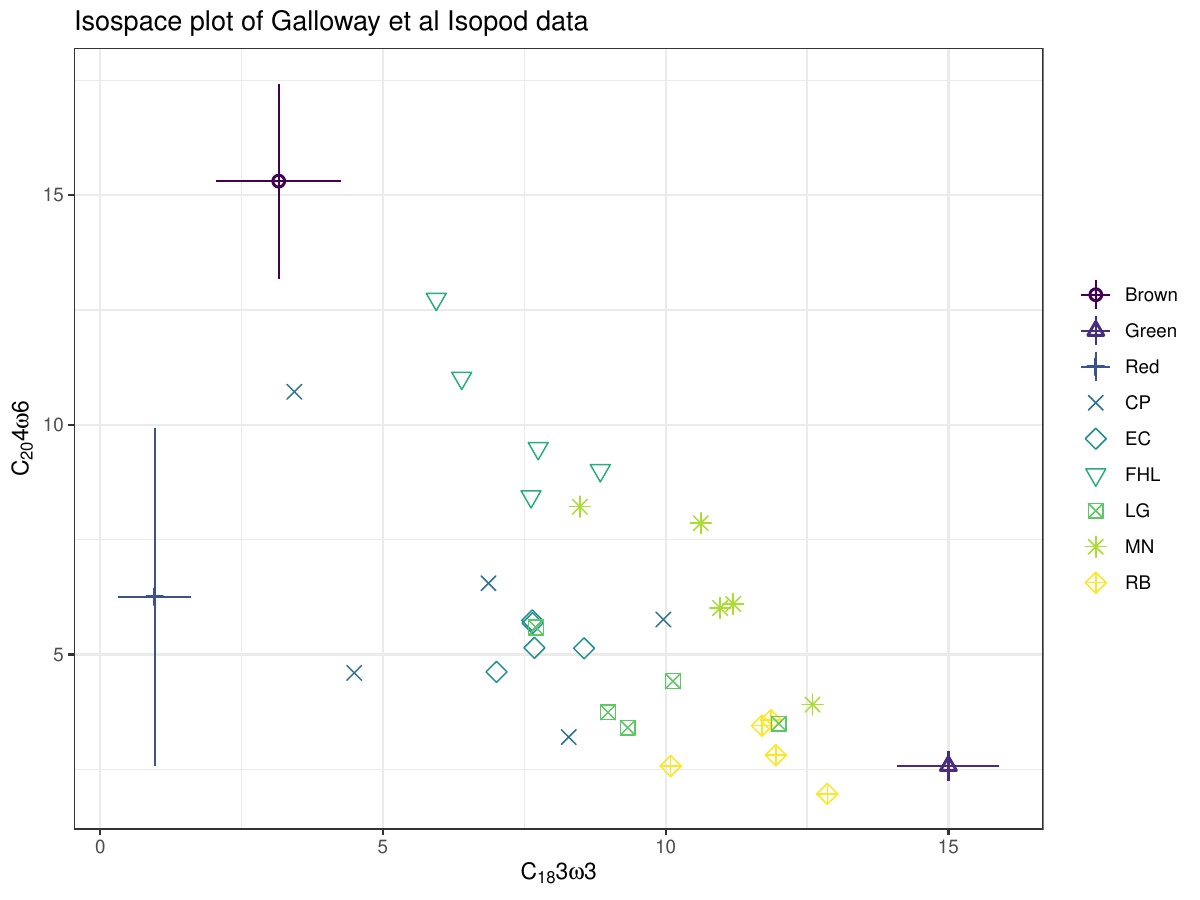}
\caption{\label{fig:iso_iso} Iso-space plot for isopod dataset showing 6 different Sites across 2 of 8 possible tracers}
\end{figure}

The second case study is the isopod dataset (\textit{Pentidotea wosnesenskii}) from \cite{galloway2014quantitative}. Six sites were used, which varied in algal cover, and this is included as the sole covariate. Three food sources are included in this example: Green (phylum Chlorophyta), Brown (phylum Ochrophyta), and Red (phylum Rhodophyta) algae. There are eight tracers - fatty acids instead of stable isotopes, as fatty acid signatures are shown to differ significantly between algal phyla \citep{galloway2012fatty}. An iso-space plot for these data is seen in Figure \ref{fig:iso_iso}, which is 2-dimensional and can therefore only show two of the eight tracers. \texttt{cosimmr} allows for users to specify which tracers they would like to plot in the iso-space plot. More than two tracers can make it difficult to check visually that all individuals lie within the multidimensional mixing polytope so caution is needed to ensure accurate TDFs are included and all relevant food sources are included. The posterior predictive plots can be particularly helpful when using $>$2 tracers to discover problem observations (or tracers themselves) because these are available per tracer as opposed to per pair of tracers in the iso-plot. 

\begin{figure}[h!]
\centering
\begin{subfigure}[b]{0.45\textwidth}
    \includegraphics[width=\textwidth]{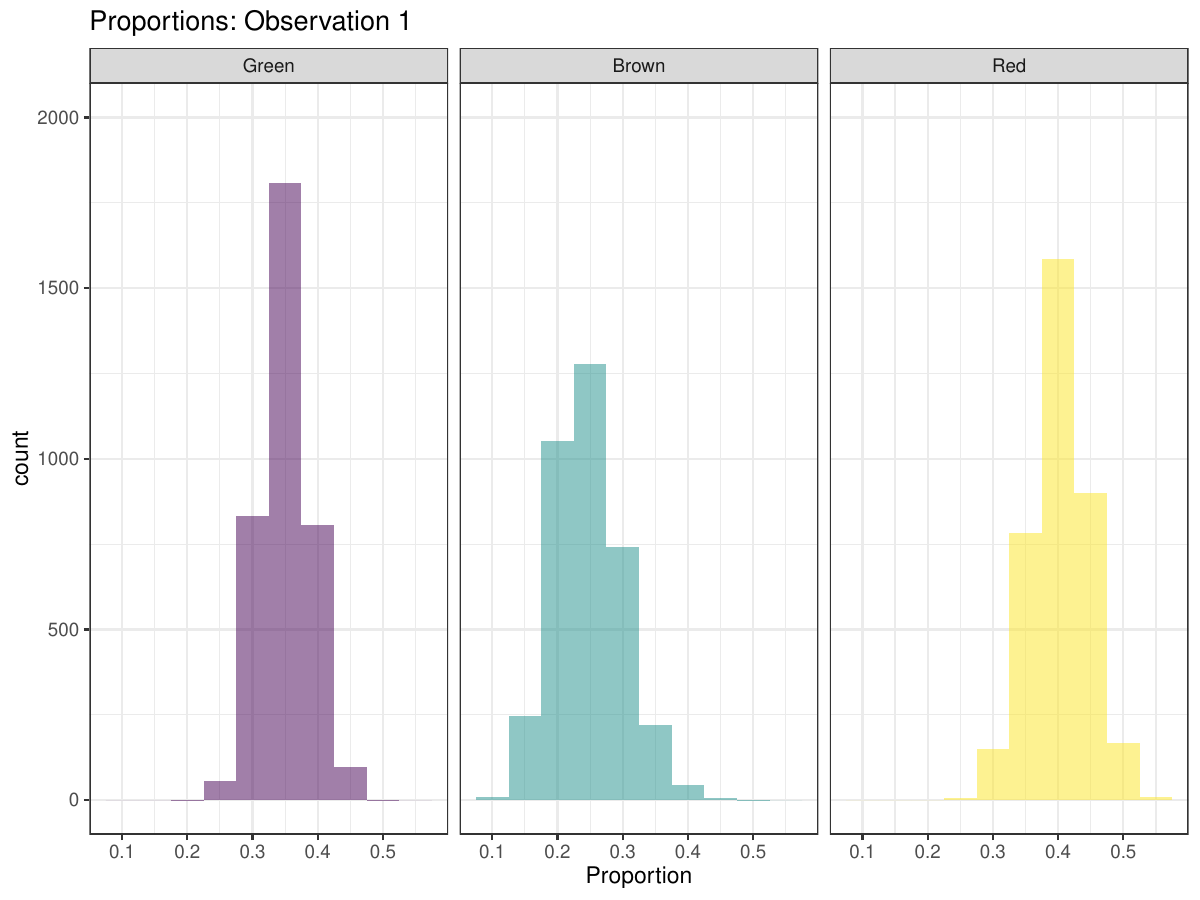}
    \caption{\texttt{cosimmr}}
    \label{fig:iso_cosimmr_small}
  \end{subfigure}
  \begin{subfigure}[b]{0.45\textwidth}
    \includegraphics[width=\textwidth]{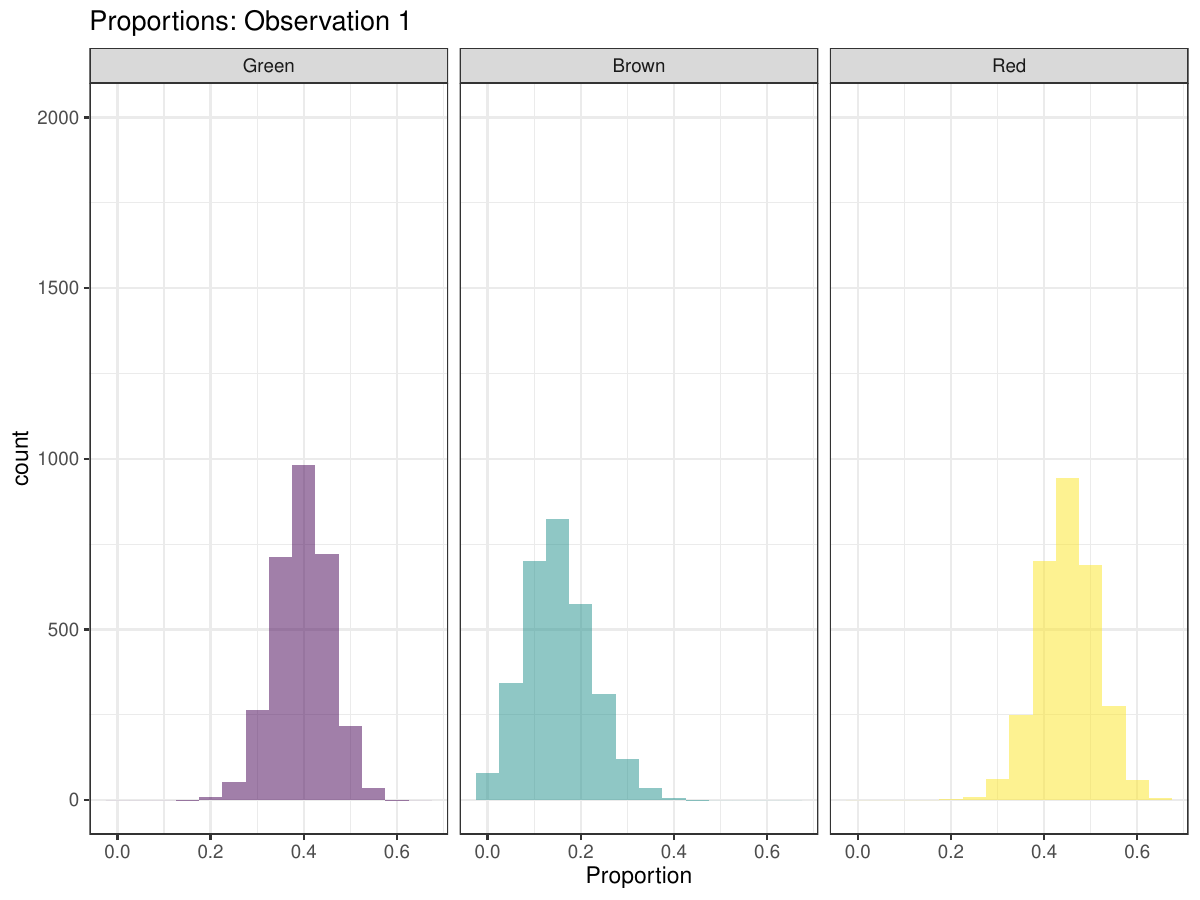}
    \caption{MixSIAR}
    \label{fig:iso_mixsiar}
  \end{subfigure}
\caption{\label{fig:iso_cosimmr} Proportion plot showing consumption of different food
sources for observation 1 for Isopod example for \texttt{cosimmr} and MixSIAR}
\end{figure}

As in the previous example, comparing the proportion estimates for \texttt{cosimmr} and MixSIAR (Figure \ref{fig:iso_cosimmr}) across different covariates levels, we can see that both are returning similar estimates, with \texttt{cosimmr} returning those estimates in a much shorter time (Table \ref{table:isopod_time}). A `normal' run in MixSIAR was enough to ensure convergence with this model. Numerical results are presented for both \texttt{cosimmr} and MixSIAR in Table \ref{table:isopod_obs1_est}. Slight differences in results may be due to the fact MixSIAR treats Site as a random effect vs \texttt{cosimmr} treating it as a fixed effect. 

\begin{table}[h]
\centering
\resizebox{\columnwidth}{!}{%
\begin{tabular}{|c|c|c|c|c|}
\hline
Programme&Source&25\%&50\%&75\%\\ \hline
\texttt{cosimmr} &Green &0.322 & 0.345 & 0.370\\
MixSIAR &Green &0.362 & 0.401 & 0.437\\
\texttt{cosimmr} & Brown &0.214 & 0.248 & 0.287\\
MixSIAR & Brown & 0.095 & 0.144 & 0.196 \\
\texttt{cosimmr} & Red & 0.372 & 0.401 & 0.431 \\
MixSIAR & Red & 0.411 & 0.454 & 0.495 \\
\hline
\end{tabular}%
}
\caption{Table showing estimates of food consumed for Observation 1 for both \texttt{cosimmr} and MixSIAR for Isopod example.}
\label{table:isopod_obs1_est}
\end{table}

\begin{table}[h]
\centering
\resizebox{\columnwidth}{!}{%
\begin{tabular}{|c|c|c|c|c|c|c|c|c|}
\hline
&min&lower quartile&mean&median&upper quartile&max& no. evaluations\\ \hline
cosimmr &418&  455&  587&  582&  748&  791&    10 \\
MixSIAR &1182& 1246& 1268& 1283& 1292& 1298&    10\\
\hline
\end{tabular}%
}
\caption{Table showing computation time (seconds) of \texttt{cosimmr} and MixSIAR for Isopod example.}
\label{table:isopod_time}
\end{table}

\begin{figure}[h!]
\centering
\includegraphics[width=0.75\textwidth]{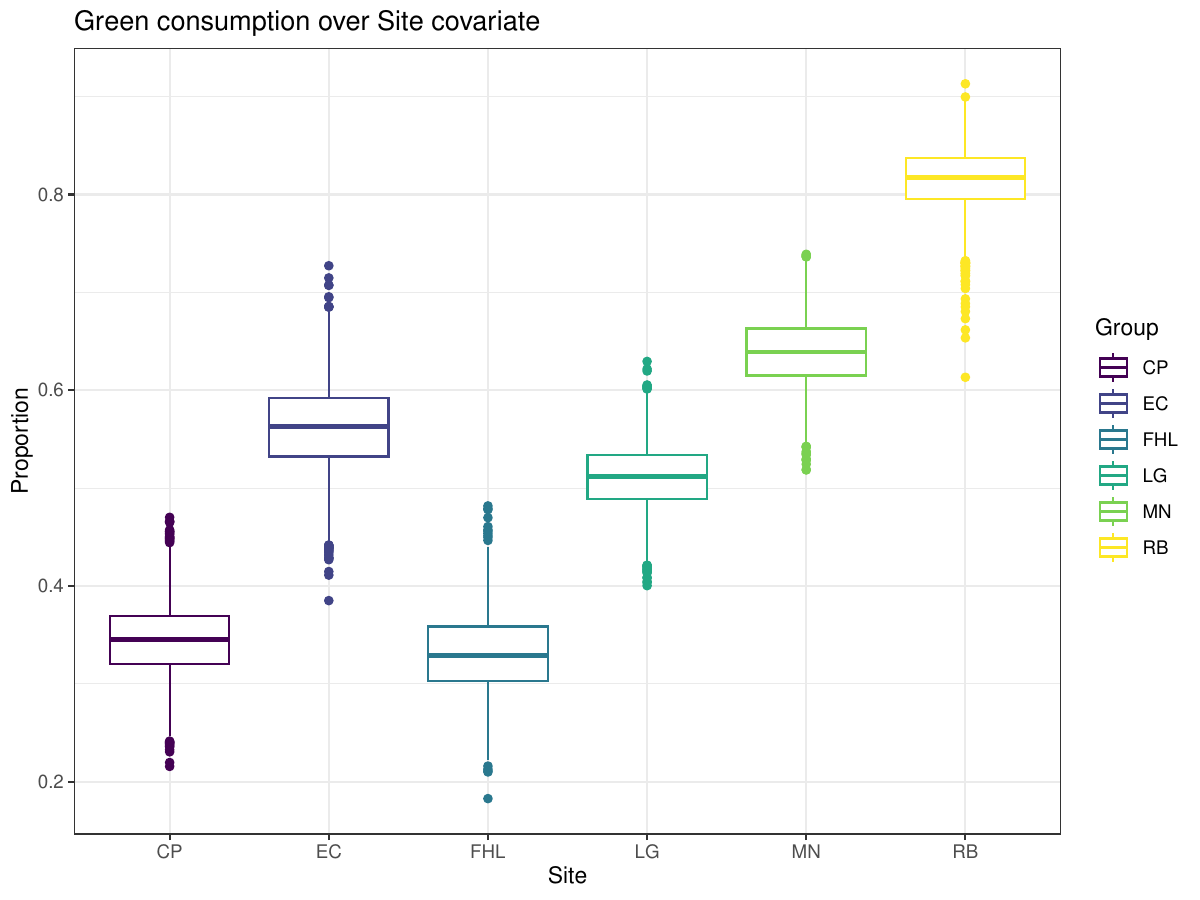}

\caption{\label{fig:iso_boxplot} Boxplots showing change in Algae consumption across sites for Green Algae for Isopod example}
\end{figure}

The importance of the covariate in this example is seen in Figure \ref{fig:iso_boxplot}. This plot shows the difference in average consumption of Green algae across different sites. This allows us to see the importance of the included covariate and how it affects the dietary proportions of individuals at that site.

\begin{figure}[h!]
\centering
\includegraphics[width=0.75\textwidth]{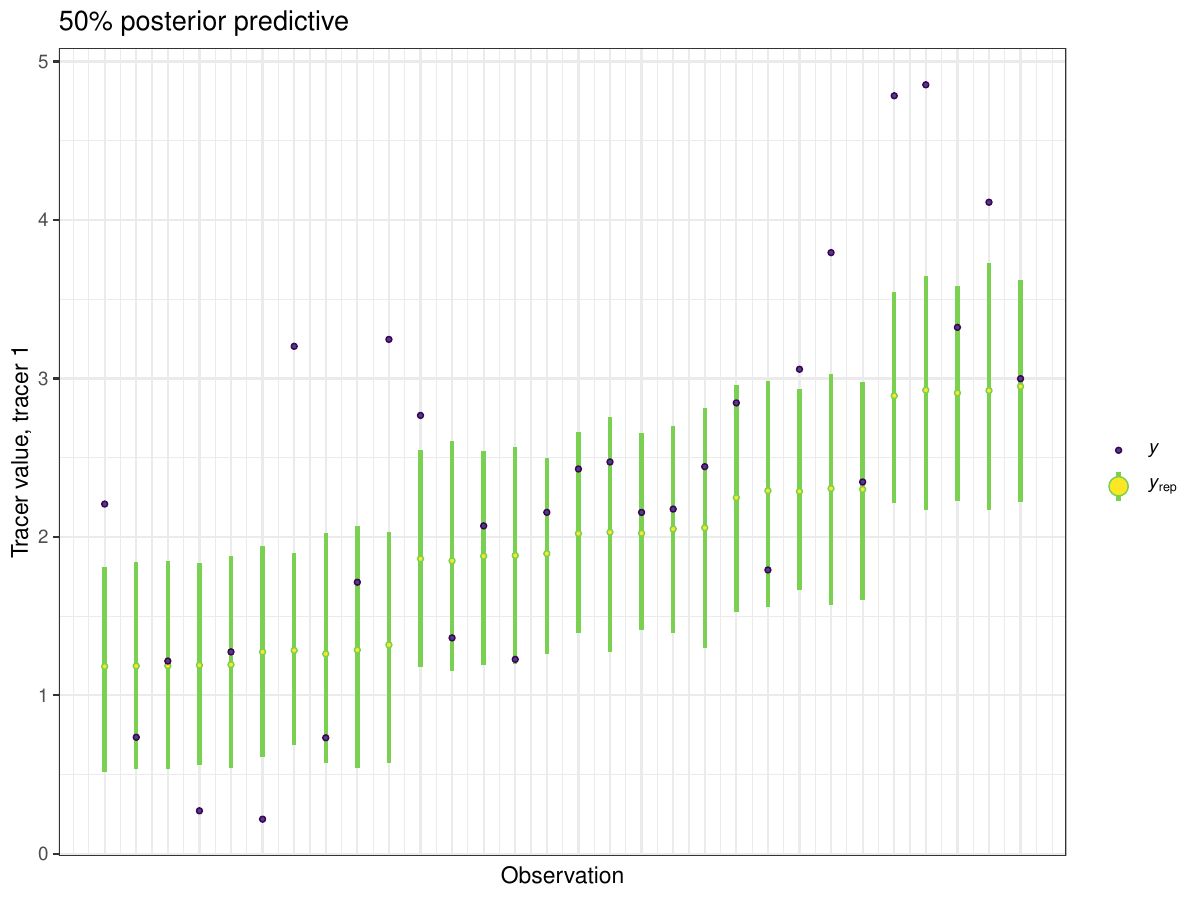}
\caption{\label{fig:isopod_post_pred} Plot showing posterior uncertainty intervals at the 50\% level for the Isopod data for tracer 1. The proportion of posterior values inside these values was 59\%.}
\end{figure}

The posterior predictive plot can be produced using the \texttt{posterior\_predictive} function. The resulting plot for tracer 1 can be seen in Figure \ref{fig:isopod_post_pred}. 59\% of values are inside the 50\% interval for this overall run. The posterior predictive for the other tracers can be viewed in the Appendix (Figure \ref{fig:iso_appendix_post_pred}).

This example highlights the computational efficiency of \texttt{cosimmr} over MixSIAR and other SIMM software. \texttt{cosimmr} is producing similar results to other SIMMs for this example, but is much quicker thanks to the use of FFVB.

\subsection{Alligator data \citep{nifong2015size}}\label{subsec:Ali}
The final example utilises alligator (\textit{Alligator mississippiensis}) data from \cite{nifong2015size}. In this example we run 8 alternative models with both \texttt{cosimmr} and MixSIAR, where each model utilises a different combination of covariates, and determine the best model fit, with the aim being that both algorithms present the same model as the best fit.
The eight models are described in Table \ref{table:3}.
\begin{table}[h]
\centering
\begin{tabular}{|c|l|}
\hline
Model & Covariate(s) \\ \hline
1 & NULL \\
2 & Habitat (Fresh, Intermediate, Marine)\\
3 & Sex (Male, Female)\\
4 & Sclass (Small Juvenile, Large Juvenile, Sub-Adult, Adult)\\
5 & Length (continuous effect)\\
6 & Sex + Sclass\\
7 & Sex + Length \\
8 & Sex * Sclass\\
\hline
\end{tabular}
\caption{Table showing different model options for Alligator example}
\label{table:3}
\end{table}

The iso-space data for this example can be seen in Figure \ref{fig:alligator_iso}. There are only two food sources in this set, Marine and Terrestrial. All food sources in this example were grouped into one of these two categories. The iso-space plot is coloured by covariate `Length' (this covariate is used in Model 5 and Model 7).

All eight models were fitted in both \texttt{cosimmr} and MixSIAR. 
MixSIAR uses `LOO' (leave-one-out cross validation) for choosing the best fitting model \citep{loo_package}. We used the same package and method on results from \texttt{cosimmr}  to choose the best fitting model. For both, model 5 (Length) is selected as the best model. The output of this model comparison can be seen in Table \ref{table:4}. The error structure of \texttt{cosimmr} and MixSIAR is slightly different. MixSIAR also utilises hierarchical source fitting which is not implemented in \texttt{cosimmr}. This may explain the slight differences in results obtained.
\begin{table}[h]
\centering
\resizebox{\columnwidth}{!}{%
\begin{tabular}{ccccc}
\hline
& \multicolumn{2}{c}{\texttt{cosimmr}} & \multicolumn{2}{c}{MixSIAR}\\ 
Model    &looic& se\_looic& looic& se\_looic \\ 
\midrule
Model 1 &   1990.9 & 31.2& 1834.6 & 16.7 \\
Model 2 &   1943.1 & 61.0& 1747.9 & 28.8 \\
Model 3 &   2072.8 & 45.6&1831.3 & 17.6 \\
Model 4 &   1833.3 & 60.0&1687.5 & 31.8\\
\textbf{Model 5} &   \textbf{1754.1} & \textbf{41.8}&\textbf{1678.3} & 3\textbf{1.3}\\
Model 6 &   1844.7 & 57.5&1689.2 & 31.5\\
Model 7 &   1822.5 & 54.6&1681.2 & 31.4\\
Model 8 &   1770.8 & 37.8&1690.4 & 29.8\\
\hline
\end{tabular}%
}
\caption{Table showing LOO output for \texttt{cosimmr} and MixSIAR alligator models, where looic is the LOO information criterion ($-2 \times elpd\_loo$) and se\_looic is the standard error of looic}
\label{table:4}
\end{table}

\begin{figure}[h!]
\centering
\begin{subfigure}[b]{0.45\textwidth}
    \includegraphics[width=\textwidth]{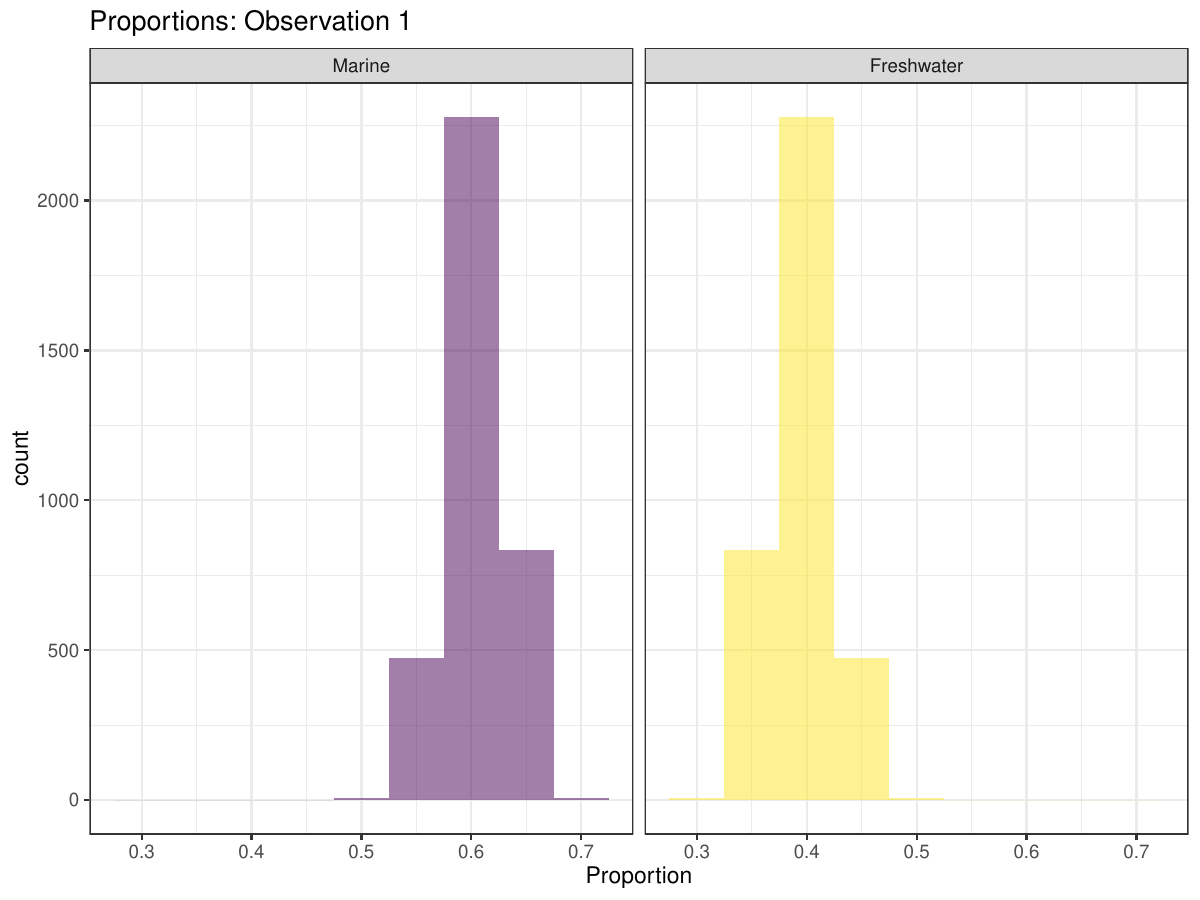}
    \caption{\texttt{cosimmr}}
    \label{fig:alli_cosimmr}
  \end{subfigure}
  \begin{subfigure}[b]{0.45\textwidth}
    \includegraphics[width=\textwidth]{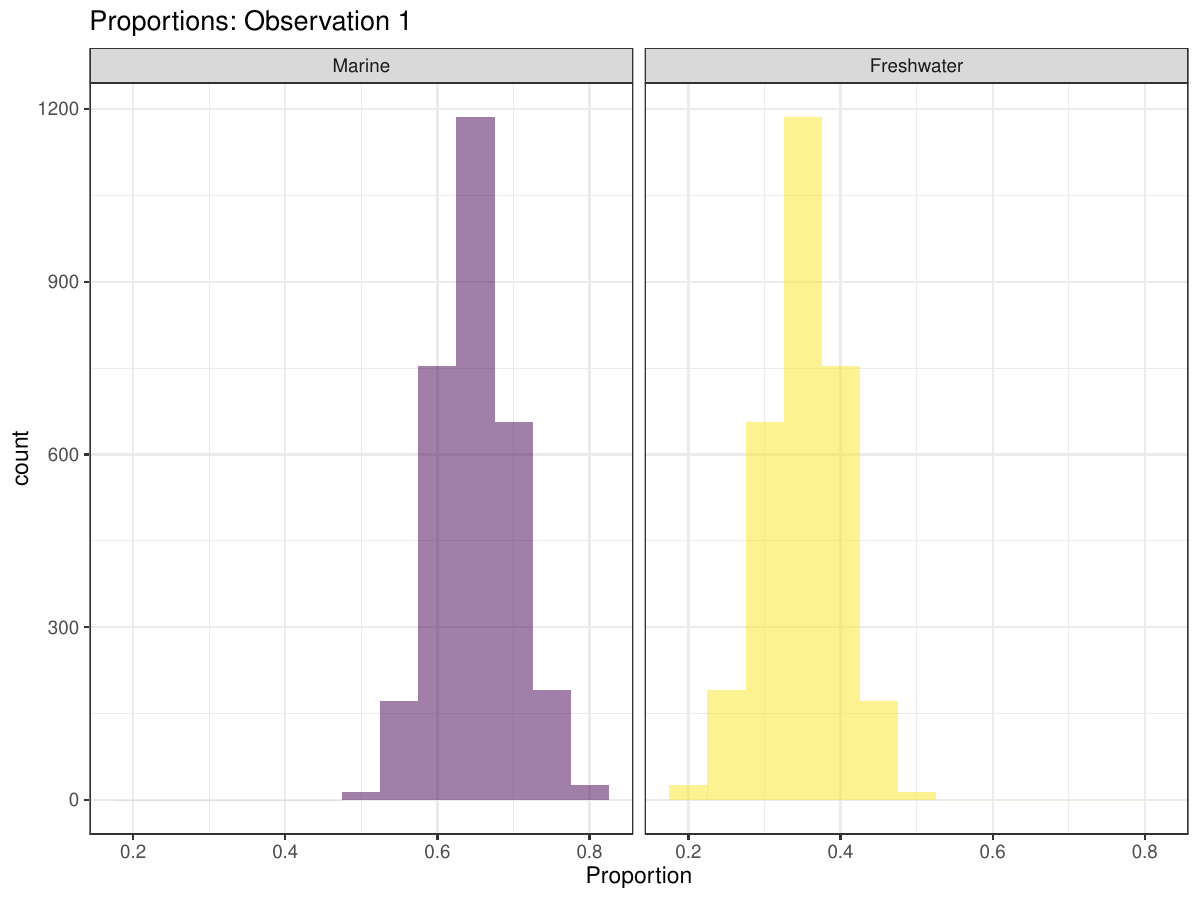}
    \caption{MixSIAR}
    \label{fig:alli_mixsiar}
  \end{subfigure}
\caption{\label{fig:alligator_co_mix} Proportion plot showing consumption of different food
sources for observation 1 for Alligator example for \texttt{cosimmr} and MixSIAR}
\end{figure}

We can compare the output of both \texttt{cosimmr} and MixSIAR and see that they are returning comparable results. In Figure \ref{fig:alligator_co_mix}  we plot the estimates for Model 5 for observation 1, an individual alligator of length 186 cm. Comparing the time for both runs shows that \texttt{cosimmr} is approximately 10 times faster (See Table \ref{table:alligator_timing}). The `short' version of MixSIAR is a long enough run for convergence for this example.

\begin{table}[h]
\centering
\resizebox{\columnwidth}{!}{%
\begin{tabular}{|c|c|c|c|c|c|c|c|}
\hline
    &       min&       lower quartile&      mean&    median&        upper quartile&       max\\ \hline
cosimmr&  111&  121&  126&  127&  132&  140&   10 \\  
MixSIAR& 1330& 1349& 1383& 1379& 1410& 1449&   10  \\
\hline
\end{tabular}%
}
\caption{Table showing computation time (seconds) of \texttt{cosimmr} and MixSIAR for Alligator example for Model 5 run.}
\label{table:alligator_timing}
\end{table}

\begin{figure}[h!]
\centering
\includegraphics[width=0.75\textwidth]{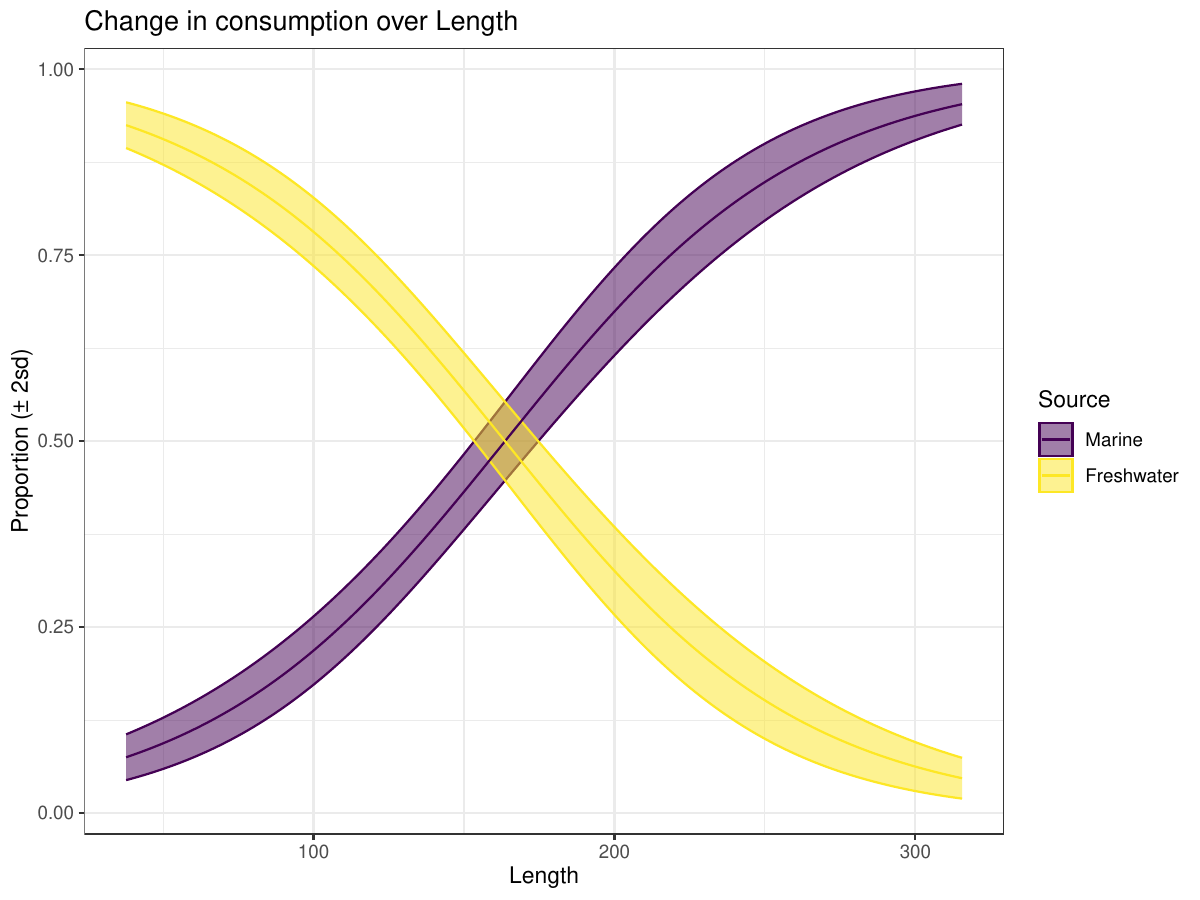}
\caption{\label{fig:aligator_line_1} Plot showing the change in Freshwater and Marine consumption vs change in Length for Alligator example, where proportion is plus or minus two standard deviations.}
\end{figure}

Figure \ref{fig:aligator_line_1} shows the predicted consumption of each food source varying with Length. The variable Length was provided to the \texttt{predict} function as a regular grid. This figure highlights why covariates can be a useful tool in SIMMs, as without covariates we get an average diet across all individuals. By including Length as a covariate, we get a much deeper insight into the animals diet. We can see that as an individual increases in length, it increases its consumption of Marine sources and consequently its consumption of Freshwater sources drops. Marine consumption increases from below 10\% to above 90\%. These findings agree with stomach content analysis performed by \cite{nifong2015size}.

\begin{figure}[h!]
\centering
\includegraphics[width=0.45\textwidth]{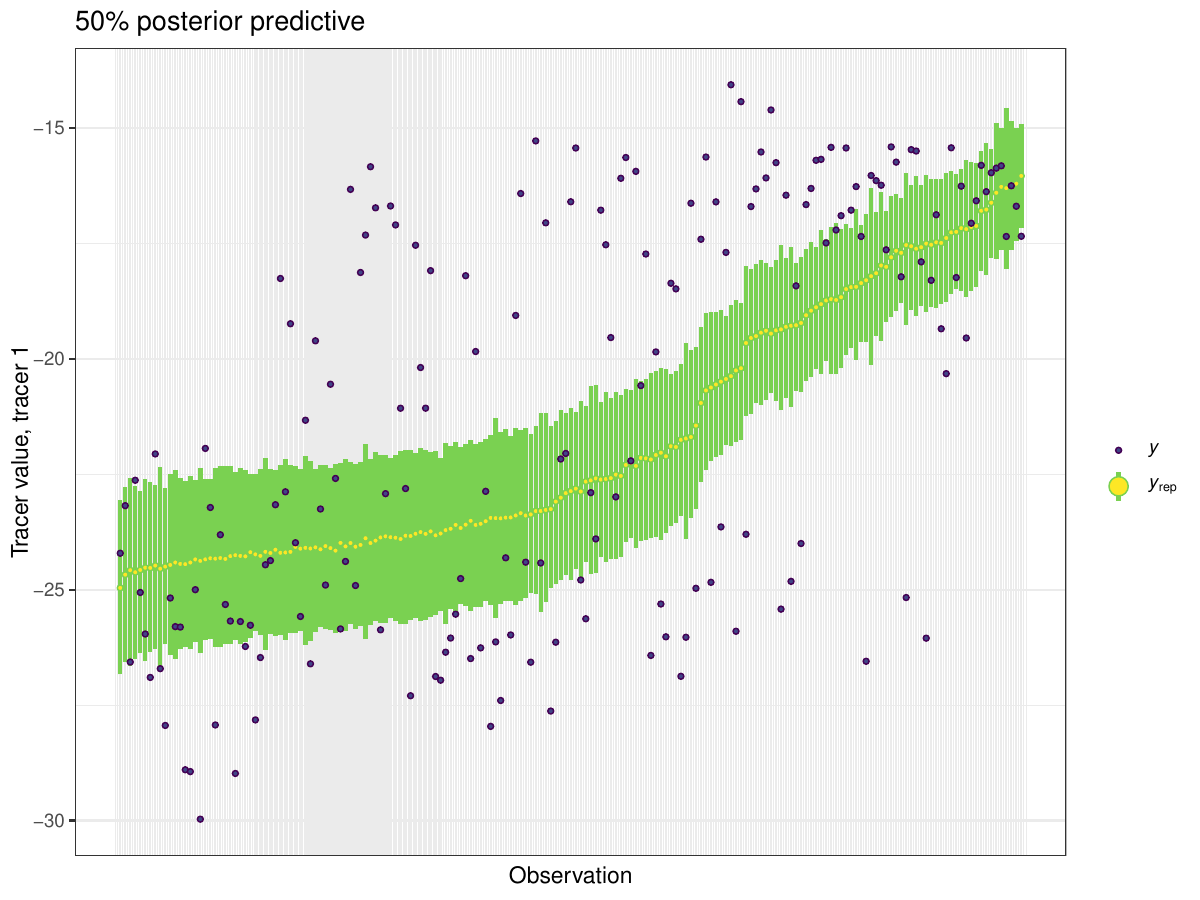}
\includegraphics[width=0.45\textwidth]{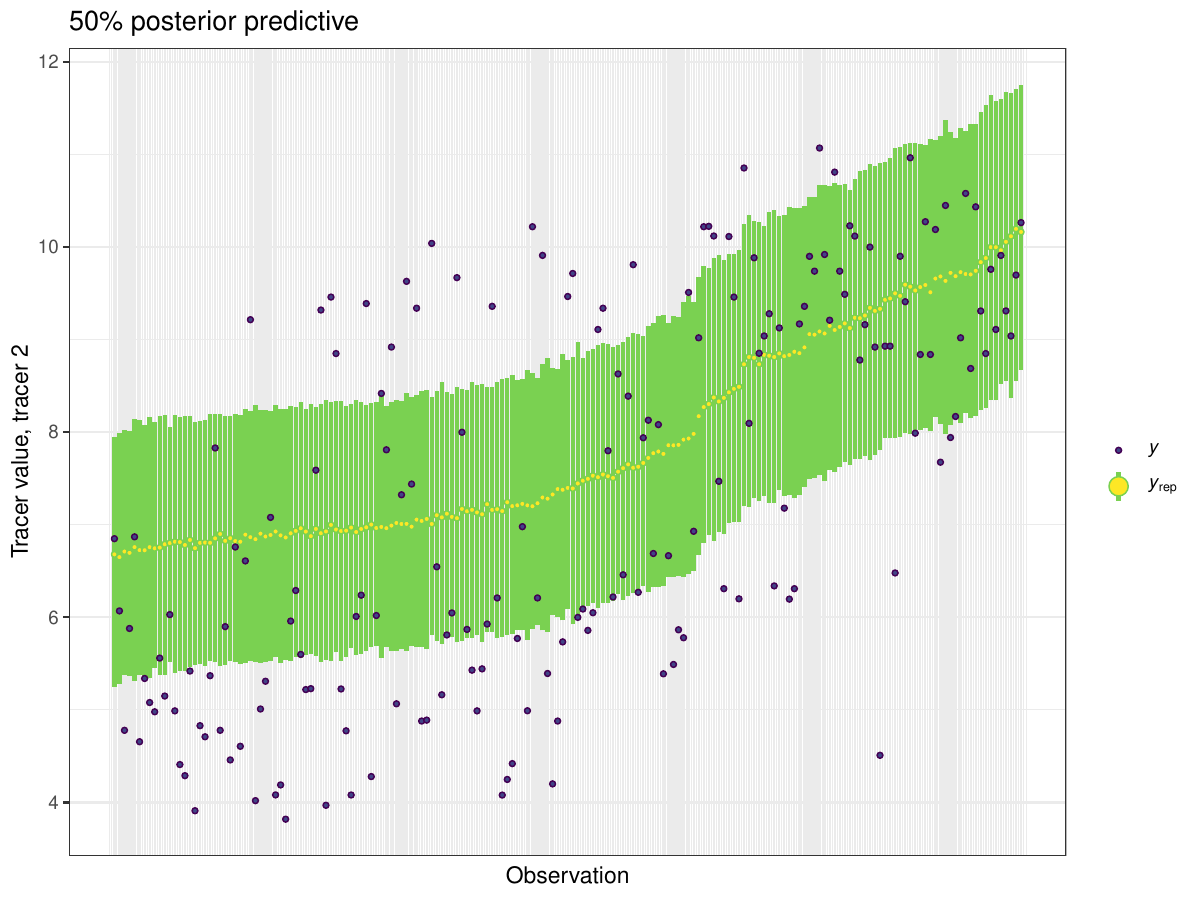}
\caption{\label{fig:aligator_post_pred}Plot showing posterior uncertainty intervals at the 50\% level for the Alligator example for tracer 1 and 2. The proportion of posterior values inside these values was 42\%.}
\end{figure}

For this example we generate a posterior predictive plot, seen in Figure \ref{fig:aligator_post_pred}. Here for a 50\% interval 42\% of individuals lie inside. For a 75\% interval 87\% of individuals lie inside and this climbs to 93\% for the 95\% interval. This indicates a good level of fit for this model. Like previous examples, the posterior predictive plot highlights outliers and points that lie far outside the 50\% interval. This is an indicator that these observations may require further scrutiny.

This example highlights the computational efficiency of \texttt{cosimmr}. The ten-fold increase in performance speed of \texttt{cosimmr} over MixSIAR highlights the high value of this package for those fitting multiple models. We see that the package produces comparable results to MixSIAR, both in terms of proportional output as well as when multiple models are compared against each other. The addition of the \texttt{posterior\_predictive} function provides strong guidance (not available in other packages) as to how well the model is fitting, as well as highlighting observations for further inspection.

\section{Conclusions}
Fixed Form Variational Bayes is a novel technique within the field of Stable Isotope Mixing Models. It is an optimization-based algorithm which contrasts with the sampling-based approaches traditionally used in SIMMs, such as MCMC. FFVB works by estimating a variational posterior which approximates the true posterior. Through the examples presented in this paper we have demonstrated that it performs as well as MCMC in terms of results produced, while also offering a significant speed improvement of up to one order of magnitude. 

The use of FFVB over MCMC allows users to run more complex models in a shorter time. Alternatively it may allow users to compare more models across differing covariates with a view to finding one that matches the data best. We believe that this speed advantage (without loss of accuracy) is an important development for SIMMs. It is important to remember that the FFVB method only ever produces an approximation of the posterior so model checking through, e.g. posterior predictive distributions, is especially important.

We have introduced the package \texttt{cosimmr} to implement these new methods. The inclusion of covariates allows users to avoid violating the assumption of IID data. The package contains functions that allow users to make predictions for combinations of covariates not found or recorded during data collection, allowing for a deeper understanding of the system being studied. We have shown that \texttt{cosimmr} is demonstrably faster than previous packages (due to FFVB), while returning comparable results. We have designed it to be user-friendly for non-expert users, with built-in summary, plotting and predict functions. It is important that users are aware plots are generally created for an individual observation, with the option to specify the individual(s) for which to create plots built into the package functions. Other functions help users to see which of their covariates are having an impact on the diet of the animal being studied, and how well the model fits the data. 

Future work on \texttt{cosimmr} could include allowing for random effects, hierarchical modelling, or source fitting - these are all options that are currently available in MixSIAR but could also be implemented using FFVB in order to speed up model fitting. The challenge here is ensuring the optimization still converges satisfactorily despite the additional parameters. We plan to implement these options in a future evolution of the package.

\section*{Computational details}
The results in this paper were obtained using R~4.4.1. R is available from the Comprehensive R Archive Network (CRAN) at \url{https://CRAN.R-project.org/}. 
C++ compiler gcc version 13.1.0

\section*{Acknowledgments}

This publication has emanated from research conducted with the financial support of Science Foundation Ireland under Grant number SFI/12/RC/2289\_P2\_Parnell to ACP. For the purpose of Open Access, the authors have applied a CC BY public copyright licence to any Author Accepted Manuscript version arising from this submission.

\section*{Author Contributions}
Emma Govan and Andrew C. Parnell conceived the ideas, designed methodology and led the writing of the manuscript; All authors contributed critically to the drafts and gave final approval for publication.

\bibliography{biblio}

\begin{thebibliography}{44}
\providecommand{\natexlab}[1]{#1}
\providecommand{\url}[1]{\texttt{#1}}
\expandafter\ifx\csname urlstyle\endcsname\relax
  \providecommand{\doi}[1]{doi: #1}\else
  \providecommand{\doi}{doi: \begingroup \urlstyle{rm}\Url}\fi

\bibitem[Aitchison(1986)]{clrdistref}
J.~Aitchison.
\newblock \emph{The Statistical Analysis of Compositional Data}.
\newblock Chapman and Hall, 1986.
\newblock ISBN 13:978-94-010-8324-9.

\bibitem[Aksu et~al.(2023)Aksu, Emiro{\u{g}}lu, Balzani, Britton, K{\"o}se, Kurtul, Ba{\c{s}}kurt, Mol, {\c{C}}{\i}nar, Haubrock, et~al.]{aksu2023high}
S.~Aksu, {\"O}.~Emiro{\u{g}}lu, P.~Balzani, J.~R. Britton, E.~K{\"o}se, I.~Kurtul, S.~Ba{\c{s}}kurt, O.~Mol, E.~{\c{C}}{\i}nar, P.~J. Haubrock, et~al.
\newblock High trophic similarity between non-native common carp and gibel carp in {T}urkish freshwaters: {I}mplications for management.
\newblock \emph{Aquaculture and Fisheries}, 2023.
\newblock \doi{10.1016/j.aaf.2023.08.003}.

\bibitem[Eddelbuettel and Fran\c{c}ois(2011)]{rcpp_ref}
D.~Eddelbuettel and R.~Fran\c{c}ois.
\newblock {Rcpp}: Seamless {R} and {C++} integration.
\newblock \emph{Journal of Statistical Software}, 40\penalty0 (8):\penalty0 1--18, 2011.
\newblock \doi{10.18637/jss.v040.i08}.

\bibitem[Fernandes et~al.(2014)Fernandes, Millard, Brabec, Nadeau, and Grootes]{fruits}
R.~Fernandes, A.~R. Millard, M.~Brabec, M.-J. Nadeau, and P.~Grootes.
\newblock Food reconstruction using isotopic transferred signals ({FRUITS}): A {B}ayesian model for diet reconstruction.
\newblock \emph{PloS one}, 9\penalty0 (2):\penalty0 e87436, 2014.
\newblock \doi{10.1371/journal.pone.0087436}.

\bibitem[Galloway et~al.(2014)Galloway, Eisenlord, Dethier, Holtgrieve, and Brett]{galloway2014quantitative}
A.~Galloway, M.~Eisenlord, M.~Dethier, G.~Holtgrieve, and M.~Brett.
\newblock Quantitative estimates of isopod resource utilization using a {B}ayesian fatty acid mixing model.
\newblock \emph{Marine Ecology Progress Series}, 507:\penalty0 219--232, 2014.
\newblock \doi{10.3354/meps10860}.

\bibitem[Galloway et~al.(2012)Galloway, Britton-Simmons, Duggins, Gabrielson, and Brett]{galloway2012fatty}
A.~W. Galloway, K.~H. Britton-Simmons, D.~O. Duggins, P.~W. Gabrielson, and M.~T. Brett.
\newblock Fatty acid signatures differentiate marine macrophytes at ordinal and family ranks.
\newblock \emph{Journal of Phycology}, 48\penalty0 (4):\penalty0 956--965, 2012.
\newblock \doi{10.1111/j.1529-8817.2012.01173.x}.

\bibitem[Govan et~al.(2023)Govan, Jackson, Inger, Bearhop, and Parnell]{govan2023simmr}
E.~Govan, A.~L. Jackson, R.~Inger, S.~Bearhop, and A.~C. Parnell.
\newblock simmr: A package for fitting stable isotope mixing models in {R}.
\newblock \emph{arXiv preprint arXiv:2306.07817}, 2023.

\bibitem[Greer et~al.(2015)Greer, Horton, and Nelson]{greer2015simple}
A.~L. Greer, T.~W. Horton, and X.~J. Nelson.
\newblock Simple ways to calculate stable isotope discrimination factors and convert between tissue types.
\newblock \emph{Methods in Ecology and Evolution}, 6\penalty0 (11):\penalty0 1341--1348, 2015.
\newblock \doi{10.1111/2041-210X.12421}.

\bibitem[Healy et~al.(2018)Healy, Guillerme, Kelly, Inger, Bearhop, and Jackson]{healy2018sider}
K.~Healy, T.~Guillerme, S.~B. Kelly, R.~Inger, S.~Bearhop, and A.~L. Jackson.
\newblock {SIDER}: an {R} package for predicting trophic discrimination factors of consumers based on their ecology and phylogenetic relatedness.
\newblock \emph{Ecography}, 41\penalty0 (8):\penalty0 1393--1400, 2018.
\newblock \doi{10.1111/ecog.03371}.

\bibitem[Hooper et~al.(1990)Hooper, Christophersen, and Peters]{hooper1990modelling}
R.~P. Hooper, N.~Christophersen, and N.~E. Peters.
\newblock Modelling streamwater chemistry as a mixture of soilwater end-members—an application to the {P}anola mountain catchment, {G}eorgia, {USA}.
\newblock \emph{Journal of Hydrology}, 116\penalty0 (1-4):\penalty0 321--343, 1990.
\newblock \doi{10.1016/0022-1694(90)90131-G}.

\bibitem[Hopke(1991)]{hopke1991receptor}
P.~K. Hopke.
\newblock \emph{Receptor modeling for air quality management}.
\newblock Elsevier, 1991.

\bibitem[Hopkins~III and Ferguson(2012)]{hopkins2012estimating}
J.~B. Hopkins~III and J.~M. Ferguson.
\newblock Estimating the diets of animals using stable isotopes and a comprehensive {B}ayesian mixing model.
\newblock \emph{PLoS one}, 7\penalty0 (1):\penalty0 e28478, 2012.
\newblock \doi{10.1371/journal.pone.0028478}.

\bibitem[Hurlbert(1984)]{hurlbert1984pseudoreplication}
S.~H. Hurlbert.
\newblock Pseudoreplication and the design of ecological field experiments.
\newblock \emph{Ecological monographs}, 54\penalty0 (2):\penalty0 187--211, 1984.
\newblock \doi{10.2307/1942661}.

\bibitem[Inger and Bearhop(2008)]{inger2008applications}
R.~Inger and S.~Bearhop.
\newblock Applications of stable isotope analyses to avian ecology.
\newblock \emph{Ibis}, 150\penalty0 (3):\penalty0 447--461, 2008.
\newblock \doi{10.1111/j.1474-919X.2008.00839.x}.

\bibitem[Inger et~al.(2006)Inger, Ruxton, Newton, Colhoun, Robinson, Jackson, and Bearhop]{inger2006temporal}
R.~Inger, G.~D. Ruxton, J.~Newton, K.~Colhoun, J.~A. Robinson, A.~L. Jackson, and S.~Bearhop.
\newblock Temporal and intrapopulation variation in prey choice of wintering geese determined by stable isotope analysis.
\newblock \emph{Journal of Animal Ecology}, 75\penalty0 (5):\penalty0 1190--1200, 2006.
\newblock \doi{10.111 1/j.1365-2656.2006.01142.x}.

\bibitem[Manlick and Newsome(2022)]{manlick2022stable}
P.~J. Manlick and S.~D. Newsome.
\newblock Stable isotope fingerprinting traces essential amino acid assimilation and multichannel feeding in a vertebrate consumer.
\newblock \emph{Methods in Ecology and Evolution}, 13\penalty0 (8):\penalty0 1819--1830, 2022.
\newblock \doi{10.1111/2041-210X.13903}.

\bibitem[Mathers and Montgomery(1997)]{mathers1997quality}
R.~Mathers and W.~Montgomery.
\newblock Quality of food consumed by over wintering pale-bellied brent geese {B}ranta bernicla hrota and wigeon {A}nas penelope.
\newblock In \emph{Biology and Environment: Proceedings of the Royal Irish Academy}, pages 81--89. JSTOR, 1997.

\bibitem[McDonald et~al.(2020)McDonald, Wilson-Aggarwal, Swan, Goodwin, Moundai, Sankara, Biswas, and Zingeser]{mcdonald2020ecology}
R.~A. McDonald, J.~K. Wilson-Aggarwal, G.~J. Swan, C.~E. Goodwin, T.~Moundai, D.~Sankara, G.~Biswas, and J.~A. Zingeser.
\newblock Ecology of domestic dogs {C}anis familiaris as an emerging reservoir of guinea worm {D}racunculus medinensis infection.
\newblock \emph{PLoS neglected tropical diseases}, 14\penalty0 (4):\penalty0 e0008170, 2020.
\newblock \doi{10.1371/journal.pntd.0008170}.

\bibitem[Miller et~al.(1972)Miller, Friedlander, and Hidy]{miller1972chemical}
M.~Miller, S.~Friedlander, and G.~Hidy.
\newblock A chemical element balance for the pasadena aerosol.
\newblock \emph{Journal of Colloid and Interface Science}, 39\penalty0 (1):\penalty0 165--176, 1972.
\newblock \doi{10.1016/0021-9797(72)90152-X}.

\bibitem[Moore and Semmens(2008)]{moore2008incorporating}
J.~W. Moore and B.~X. Semmens.
\newblock Incorporating uncertainty and prior information into stable isotope mixing models.
\newblock \emph{Ecology letters}, 11\penalty0 (5):\penalty0 470--480, 2008.
\newblock \doi{10.1111/j.1461-0248.2008.01163.x}.

\bibitem[Munoz et~al.(2019)Munoz, Giosan, Blusztajn, Rankin, and Stinchcomb]{munoz2019radiogenic}
S.~E. Munoz, L.~Giosan, J.~Blusztajn, C.~Rankin, and G.~E. Stinchcomb.
\newblock Radiogenic fingerprinting reveals anthropogenic and buffering controls on sediment dynamics of the {M}ississippi river system.
\newblock \emph{Geology}, 47\penalty0 (3):\penalty0 271--274, 2019.
\newblock \doi{10.1130/G45194.1}.

\bibitem[Nifong et~al.(2015)Nifong, Layman, and Silliman]{nifong2015size}
J.~C. Nifong, C.~A. Layman, and B.~R. Silliman.
\newblock Size, sex and individual-level behaviour drive intrapopulation variation in cross-ecosystem foraging of a top-predator.
\newblock \emph{Journal of Animal Ecology}, 84\penalty0 (1):\penalty0 35--48, 2015.
\newblock \doi{10.1111/1365-2656.12306}.

\bibitem[Paez-Rosas et~al.(2024)Paez-Rosas, Suarez-Moncada, Arnes-Urgelles, Espinoza, Robles, and Salinas-De-Leon]{paez2024assessment}
D.~Paez-Rosas, J.~Suarez-Moncada, C.~Arnes-Urgelles, E.~Espinoza, Y.~Robles, and P.~Salinas-De-Leon.
\newblock Assessment of nursery areas for the scalloped hammerhead shark (sphyrna lewini) across the eastern tropical {P}acific using a stable isotopes approach.
\newblock \emph{Frontiers in Marine Science}, 10, 2024.
\newblock \doi{10.3389/fmars.2023.1288770}.

\bibitem[Parnell et~al.(2010)Parnell, Inger, Bearhop, and Jackson]{parnell2010source}
A.~C. Parnell, R.~Inger, S.~Bearhop, and A.~L. Jackson.
\newblock Source partitioning using stable isotopes: {C}oping with too much variation.
\newblock \emph{PloS one}, 5\penalty0 (3):\penalty0 e9672, 2010.
\newblock \doi{10.1371/journal.pone.0009672}.

\bibitem[Parnell et~al.(2013)Parnell, Phillips, Bearhop, Semmens, Ward, Moore, Jackson, Grey, Kelly, and Inger]{parnell2013bayesian}
A.~C. Parnell, D.~L. Phillips, S.~Bearhop, B.~X. Semmens, E.~J. Ward, J.~W. Moore, A.~L. Jackson, J.~Grey, D.~J. Kelly, and R.~Inger.
\newblock Bayesian stable isotope mixing models.
\newblock \emph{Environmetrics}, 24\penalty0 (6):\penalty0 387--399, 2013.
\newblock \doi{10.1002/env.2221}.

\bibitem[Phillips(2012)]{phillips2012converting}
D.~L. Phillips.
\newblock Converting isotope values to diet composition: {t}he use of mixing models.
\newblock \emph{Journal of Mammalogy}, 93\penalty0 (2):\penalty0 342--352, 2012.
\newblock \doi{10.1644/11-MAMM-S-158.1}.

\bibitem[Phillips and Gregg(2003)]{Isosource_paper}
D.~L. Phillips and J.~W. Gregg.
\newblock Source partitioning using stable isotopes: {C}oping with too many sources.
\newblock \emph{Oecologia}, 136:\penalty0 261--269, 2003.
\newblock \doi{10.1007/s00442-003-1218-3}.

\bibitem[Phillips and Koch(2002)]{phillips2002incorporating}
D.~L. Phillips and P.~L. Koch.
\newblock Incorporating concentration dependence in stable isotope mixing models.
\newblock \emph{Oecologia}, 130:\penalty0 114--125, 2002.
\newblock \doi{10.1007/s004420100786}.

\bibitem[Plummer(2003)]{plummer2003jags}
M.~Plummer.
\newblock {JAGS}: {A} program for analysis of {B}ayesian graphical models using {G}ibbs {S}ampling.
\newblock In \emph{Proceedings of the 3rd international workshop on distributed statistical computing}, volume 124, pages 1--10. Vienna, Austria., 2003.

\bibitem[{R Core Team}(2021)]{R_ref}
{R Core Team}.
\newblock \emph{R: {A} Language and Environment for Statistical Computing}.
\newblock R Foundation for Statistical Computing, Vienna, Austria, 2021.
\newblock URL \url{https://www.R-project.org/}.

\bibitem[Salimans and Knowles(2013)]{salimans2013fixed}
T.~Salimans and D.~A. Knowles.
\newblock {Fixed-Form Variational Posterior Approximation through Stochastic Linear Regression}.
\newblock \emph{Bayesian Analysis}, 8\penalty0 (4):\penalty0 837 -- 882, 2013.
\newblock \doi{10.1214/13-BA858}.
\newblock URL \url{https://doi.org/10.1214/13-BA858}.

\bibitem[Semmens et~al.(2009)Semmens, Ward, Moore, and Darimont]{semmens2009quantifying}
B.~X. Semmens, E.~J. Ward, J.~W. Moore, and C.~T. Darimont.
\newblock Quantifying inter-and intra-population niche variability using hierarchical {B}ayesian stable isotope mixing models.
\newblock \emph{PloS one}, 4\penalty0 (7):\penalty0 e6187, 2009.
\newblock \doi{10.1371/journal.pone.0006187}.

\bibitem[Stock et~al.(2018)Stock, Jackson, Ward, Parnell, Phillips, and Semmens]{stock2018analyzing}
B.~C. Stock, A.~L. Jackson, E.~J. Ward, A.~C. Parnell, D.~L. Phillips, and B.~X. Semmens.
\newblock Analyzing mixing systems using a new generation of {B}ayesian tracer mixing models.
\newblock \emph{PeerJ}, 6:\penalty0 e5096, 2018.
\newblock \doi{10.7717/peerj.5096}.

\bibitem[Swan et~al.(2020)Swan, Bearhop, Redpath, Silk, Goodwin, Inger, and McDonald]{swan2020evaluating}
G.~J. Swan, S.~Bearhop, S.~M. Redpath, M.~J. Silk, C.~E. Goodwin, R.~Inger, and R.~A. McDonald.
\newblock Evaluating bayesian stable isotope mixing models of wild animal diet and the effects of trophic discrimination factors and informative priors.
\newblock \emph{Methods in Ecology and Evolution}, 11\penalty0 (1):\penalty0 139--149, 2020.
\newblock \doi{10.1111/2041-210X.13311}.

\bibitem[Tan and Nott(2018)]{tan2018gaussian}
L.~S. Tan and D.~J. Nott.
\newblock Gaussian variational approximation with sparse precision matrices.
\newblock \emph{Statistics and Computing}, 28:\penalty0 259--275, 2018.
\newblock \doi{10.1007/s11222-017-9729-7}.

\bibitem[Teixeira et~al.(2021)Teixeira, Botta, Daura-Jorge, Pereira, Newsome, and Sim{\~o}es-Lopes]{teixeira2021niche}
C.~R. Teixeira, S.~Botta, F.~G. Daura-Jorge, L.~B. Pereira, S.~D. Newsome, and P.~C. Sim{\~o}es-Lopes.
\newblock Niche overlap and diet composition of three sympatric coastal dolphin species in the southwest {A}tlantic ocean.
\newblock \emph{Marine Mammal Science}, 37\penalty0 (1):\penalty0 111--126, 2021.
\newblock \doi{10.1111/mms.12726}.

\bibitem[Thibault et~al.(2024)Thibault, Letourneur, Cleguer, Bonneville, Briand, Derville, Bustamante, and Garrigue]{thibault2024c}
M.~Thibault, Y.~Letourneur, C.~Cleguer, C.~Bonneville, M.~J. Briand, S.~Derville, P.~Bustamante, and C.~Garrigue.
\newblock C and {N} stable isotopes enlighten the trophic behaviour of the dugong (dugong dugon).
\newblock \emph{Scientific Reports}, 14\penalty0 (1):\penalty0 896, 2024.
\newblock \doi{10.1038/s41598-023-50578-3}.

\bibitem[Titsias and L{\'a}zaro-Gredilla(2014)]{titsias2014doubly}
M.~Titsias and M.~L{\'a}zaro-Gredilla.
\newblock Doubly stochastic variational bayes for non-conjugate inference.
\newblock In \emph{International conference on machine learning}, pages 1971--1979. PMLR, 2014.

\bibitem[Tran et~al.(2021)Tran, Nguyen, and Dao]{tran2021practical}
M.-N. Tran, T.-N. Nguyen, and V.-H. Dao.
\newblock A practical tutorial on variational bayes.
\newblock \emph{arXiv preprint arXiv:2103.01327}, 2021.
\newblock \doi{10.48550/arXiv.2103.01327}.

\bibitem[Vehtari et~al.(2024)Vehtari, Gabry, Magnusson, Yao, Bürkner, Paananen, and Gelman]{loo_package}
A.~Vehtari, J.~Gabry, M.~Magnusson, Y.~Yao, P.-C. Bürkner, T.~Paananen, and A.~Gelman.
\newblock loo: Efficient leave-one-out cross-validation and waic for {B}ayesian models, 2024.
\newblock URL \url{https://mc-stan.org/loo/}.
\newblock R package version 2.7.0.

\bibitem[Ward et~al.(2010)Ward, Semmens, and Schindler]{ward2010including}
E.~J. Ward, B.~X. Semmens, and D.~E. Schindler.
\newblock Including source uncertainty and prior information in the analysis of stable isotope mixing models.
\newblock \emph{Environmental science \& technology}, 44\penalty0 (12):\penalty0 4645--4650, 2010.
\newblock \doi{10.1021/es100053v}.

\bibitem[Wickham et~al.(2019)Wickham, Averick, Bryan, Chang, McGowan, François, Grolemund, Hayes, Henry, Hester, Kuhn, Pedersen, Miller, Bache, Müller, Ooms, Robinson, Seidel, Spinu, Takahashi, Vaughan, Wilke, Woo, and Yutani]{tidyverseref}
H.~Wickham, M.~Averick, J.~Bryan, W.~Chang, L.~D. McGowan, R.~François, G.~Grolemund, A.~Hayes, L.~Henry, J.~Hester, M.~Kuhn, T.~L. Pedersen, E.~Miller, S.~M. Bache, K.~Müller, J.~Ooms, D.~Robinson, D.~P. Seidel, V.~Spinu, K.~Takahashi, D.~Vaughan, C.~Wilke, K.~Woo, and H.~Yutani.
\newblock Welcome to the {tidyverse}.
\newblock \emph{Journal of Open Source Software}, 4\penalty0 (43):\penalty0 1686, 2019.
\newblock \doi{10.21105/joss.01686}.

\bibitem[Yao et~al.(2018)Yao, Vehtari, Simpson, and Gelman]{yao2018yes}
Y.~Yao, A.~Vehtari, D.~Simpson, and A.~Gelman.
\newblock Yes, but did it work?: Evaluating variational inference.
\newblock In \emph{International Conference on Machine Learning}, pages 5581--5590. PMLR, 2018.

\bibitem[Zaryab et~al.(2022)Zaryab, Nassery, Knoeller, Alijani, and Minet]{zaryab2022determining}
A.~Zaryab, H.~R. Nassery, K.~Knoeller, F.~Alijani, and E.~Minet.
\newblock Determining nitrate pollution sources in the {K}abul plain aquifer ({A}fghanistan) using stable isotopes and {B}ayesian stable isotope mixing model.
\newblock \emph{Science of the Total Environment}, 823:\penalty0 153749, 2022.
\newblock \doi{10.1016/j.scitotenv.2022.153749}.

\end{thebibliography}
\newpage
\begin{appendix}
\section{Gaussian Variational Bayes with Cholesky Decomposed Covariance}
\label{app:algorithm}

 We use the Gaussian Variational Bayes with Cholesky decomposed covariance algorithm of \cite{tran2021practical}. If we define the joint set of parameters as $\mathbf{\theta} = (\mathbf{\beta}, \log(\mathbf{\sigma}^2))$ then we write our factorised variational posterior as:
\begin{eqnarray*}
q_\mathbf{\lambda}(\mathbf{\theta}) = q(\mathbf{\beta}, \log(\mathbf{\sigma})))
\end{eqnarray*}
where $\mathbf{\lambda} = (\mu_\mathbf{\beta}, \mu_\mathbf{\sigma}, vech(L))^T$ is the set of hyper-parameters associated with the variational posteriors:

\begin{eqnarray*}
q(\mathbf{\theta}) \equiv MVN(\mathbf{\mu}, \mathbf{\Sigma}) \\
\mathbf{\mu} \equiv (\mathbf{\mu}_\mathbf{\beta}, \mathbf{\mu}_\mathbf{\sigma}) \\
\mathbf{\Sigma} \equiv (\mathbf{\Sigma}_\mathbf{\beta}, \mathbf{\Sigma}_\mathbf{\sigma})
\end{eqnarray*}

To avoid the positive semi-definite constraints on $\mathbf{\Sigma}_\mathbf{\theta}$ we model the Cholesky decomposition of this matrix so that $\mathbf{\Sigma}_\mathbf{\theta} = \mathbf{LL}^T$.

To start the algorithm, initial values are required for $\mathbf{\lambda^{(0)}}$ (we use parenthetical super-scripts to denote iterations), the sample size $S$, the adaptive learning weights ($\beta_1, \beta_2$), the fixed learning rate $\epsilon_0$, the threshold $\alpha$, the rolling window size $t_W$, the maximum patience $P$.

Define $h$ to be the log of the joint distribution up to the constant of proportionality:
\begin{eqnarray*}
h(\mathbf{\theta}) = \log \left( p(y|\mathbf{\theta}) p(\mathbf{\theta}) \right)
\end{eqnarray*}

\noindent and $h_\lambda$ to be the log of the ratio between the joint and the VB posterior:
\begin{eqnarray*}
h_\mathbf{\lambda}(\mathbf{\theta}) = \log \left( \frac{ p(\mathbf{y}|\mathbf{\theta}) p(\mathbf{\theta)} }{ q_\mathbf{\lambda}(\mathbf{\theta)} } \right) = h(\mathbf{\theta}) - \log q_\mathbf{\lambda}(\mathbf{\theta}) 
\end{eqnarray*}

\noindent The initialisation stage proceeds with:
\begin{enumerate}
\item Generate samples from $\mathbf{\kappa_s} \sim N_d(0, I)$ for $s=1,...S$

\item Compute the unbiased estimate of the lower bound gradient:
\begin{eqnarray*}
\widehat{\nabla}_\lambda{LB}(\lambda^{(0)}) = \left(\widehat{\nabla_{\mu}{LB}}(\mathbf{\lambda}^{(0)})^T, \widehat{\nabla_{\mathbf{vech(L)}}{LB}}(\mathbf{\lambda}^{(0)})^T\right)^T \\
\widehat{\nabla}_\mu{LB}(\mathbf{\lambda}^{(0)}) =  \frac{1}{S}\sum_{s=1}^S\nabla_\mathbf{\theta} h_\mathbf{\lambda}(\mathbf{\theta}_s)\\
\widehat{\mathbf{\nabla}}_{\mathbf{vech(L)}}{LB}(\mathbf{\lambda}^{(0)}) = \frac{1}{S}\sum_{s=1}^Svech\left(\nabla_\theta h_\mathbf{\lambda}(\mathbf{\theta}_s)\mathbf{\kappa}_s^T\right)\\
\end{eqnarray*}

Create estimates of $\mathbf{\theta}_s$
\begin{eqnarray*}
    \mathbf{\theta}_s = \mathbf{\mu}^{(0)} + \mathbf{L}^{(0)}\mathbf{\kappa_s}
\end{eqnarray*}

\item Set

\begin{eqnarray*}
\bar{g}_0 := \nabla_\mathbf{\lambda}{LB}(\mathbf{\lambda}^{(0)})\\
\bar{\nu}_0 := \bar{g}_0^2\\
\bar{g} = g_0\\
\bar{\nu} = \nu_0\\
\end{eqnarray*}

\item Set $t=1$, patience = 0, and `stop = FALSE'.\\
\end{enumerate}
Now the algorithm runs with:
\begin{enumerate}
\item Generate $\mathbf{\kappa}_s \sim q_{\mathbf{\lambda}^{(t)}(\mathbf{\theta)}}$ for $s=1,...S$. Recalculate $\mathbf{\mu^{(t)}}$ and $\mathbf{L^{(t)}}$ from $\mathbf{\lambda^{(t)}}$

\item Compute the unbiased estimate of the lower bound gradient:
\begin{eqnarray*}
g_t := \widehat{\nabla}_\mathbf{\lambda{LB}}(\mathbf{\lambda}^{(t)}) = \left(\widehat{\nabla}_{\mathbf{\mu}}{LB}(\mathbf{\lambda}^{(t)})^T, \widehat{\nabla}_\mathbf{{vech(L)}}{LB}(\mathbf{\lambda}^{(t)})^T\right)^T
\end{eqnarray*}
where
\begin{eqnarray*}
    \widehat{\nabla}_\mathbf{\mu}{LB}(\mathbf{\lambda}^{(t)}) =  \frac{1}{S}\sum_{s=1}^S\nabla_\mathbf{\theta} h_\mathbf{\lambda}(\mathbf{\theta}_s)\\
\widehat{\nabla}_\mathbf{{vech(L)}}{LB}(\mathbf{\lambda}^{(t)}) = \frac{1}{S}\sum_{s=1}^Svech\left(\nabla_\mathbf{\theta} h_\mathbf{\lambda}(\mathbf{\theta}_s)\mathbf{\kappa}_s^T\right)\\
\end{eqnarray*}
with $\mathbf{\theta}_s = \mathbf{\mu}^{(t)} + \mathbf{L}^{(t)}\mathbf{\kappa_s}$

\item Compute: 

\begin{eqnarray*}
v_t = g_t^2 \\
\bar{g} = \beta_1\bar{g} + (1-\beta_1)g_t \\
\bar{v} = \beta_2\bar{v} + (1-\beta_2)v_t \\
\end{eqnarray*}

\item Update the learning rate:
\begin{eqnarray*}
l_t = min(\epsilon_0, \epsilon_0\frac{\alpha}{t})
\end{eqnarray*}
and the variational hyper-parameters:
\begin{eqnarray*}
\mathbf{\lambda}^{(t+1)} = \mathbf{\lambda}^{(t)} + l_t\frac{\bar{g}}{\sqrt{\bar{v}}}
\end{eqnarray*}

\item Compute the lower bound estimate:
\begin{eqnarray*}
\widehat{LB}(\mathbf{\lambda}^{(t)}) := \frac{1}{S} \sum_{s=1}^S h_{\mathbf{\lambda}^{(t)}}(\mathbf{\theta}_s)
\end{eqnarray*}

\item If $t \ge t_W$ compute the moving average LB
\begin{eqnarray*}
\overline{LB}_{t-t_W+1} := \frac{1}{t_W} \sum_{k=1}^{t_W} \widehat{LB}(\mathbf{\lambda}^{(t-k + 1)})
\end{eqnarray*}
If $\overline{LB}_{t-t_W+1} \ge \max(\overline{LB})$ patience = 0, else patience = patience +1

\item If patience $\ge$ P, `stop = TRUE`

\item Set $t:=t+1$
\end{enumerate}

\section{Further plots}
\label{app:extra_plots}

\begin{figure}[h!]
\begin{subfigure}[b]{0.45\textwidth}
    \includegraphics[width=\textwidth]{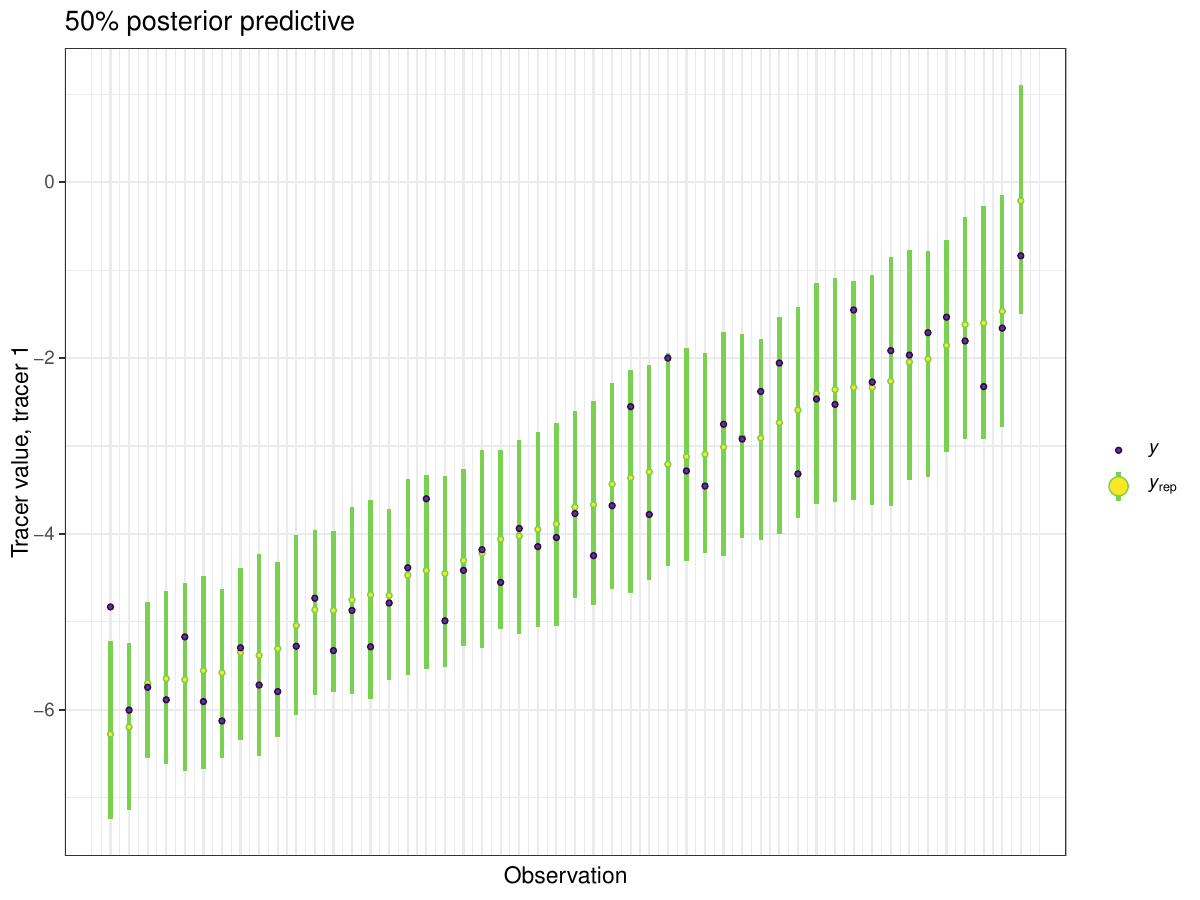}
    \caption{Tracer 2}
    \label{fig:t1low}
  \end{subfigure}
 \hfill
 \begin{subfigure}[b]{0.45\textwidth}
    \includegraphics[width=\textwidth]{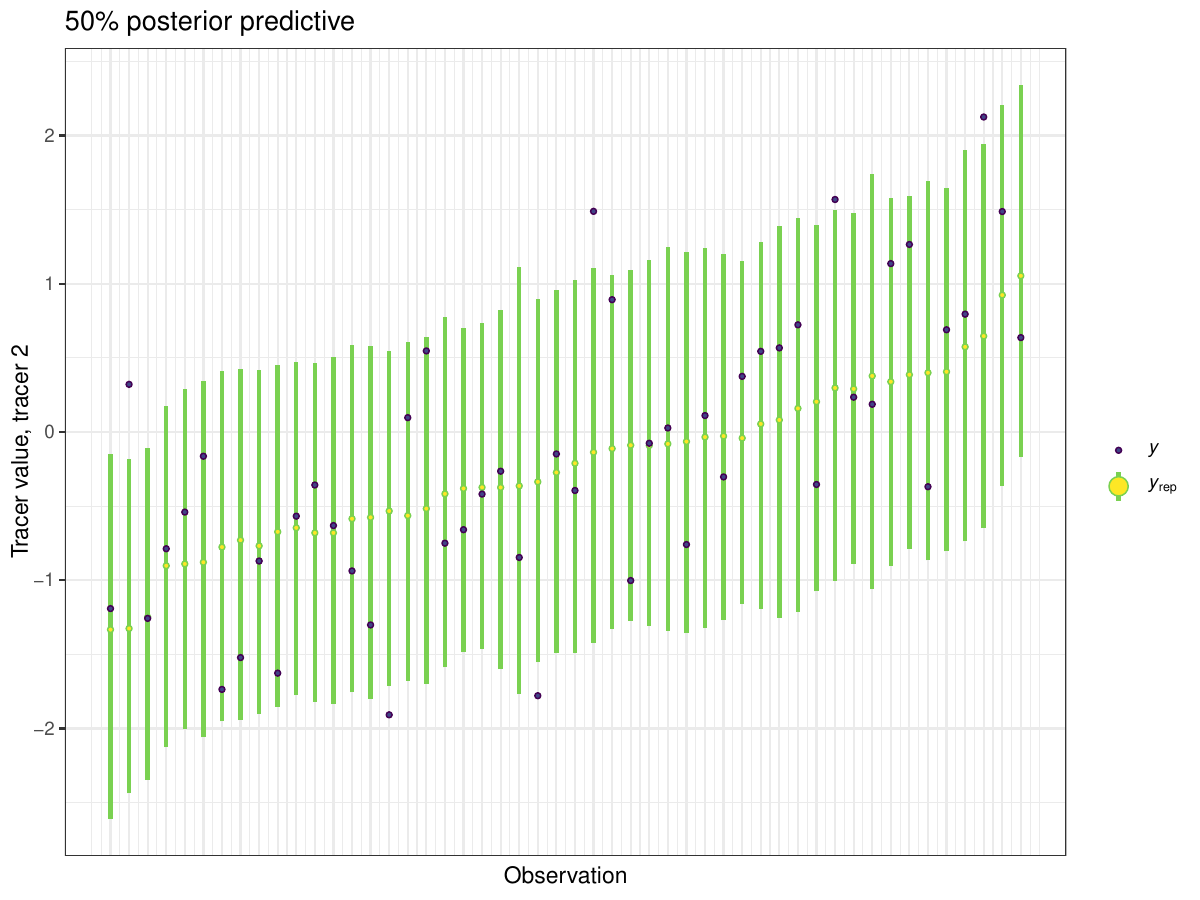}
    \caption{Tracer 3}
    \label{fig:t2_low}
  \end{subfigure}
    \hfill
\caption{\label{fig:post_pred_sim1_low} Plot showing posterior uncertainty intervals at the 50\% level for the `Low' simulated model run for tracer 1 and tracer 2. The proportion of posterior values inside these values was 93\%.}
\end{figure}

\begin{figure}[h!]
 \begin{subfigure}[b]{0.45\textwidth}
    \includegraphics[width=\textwidth]{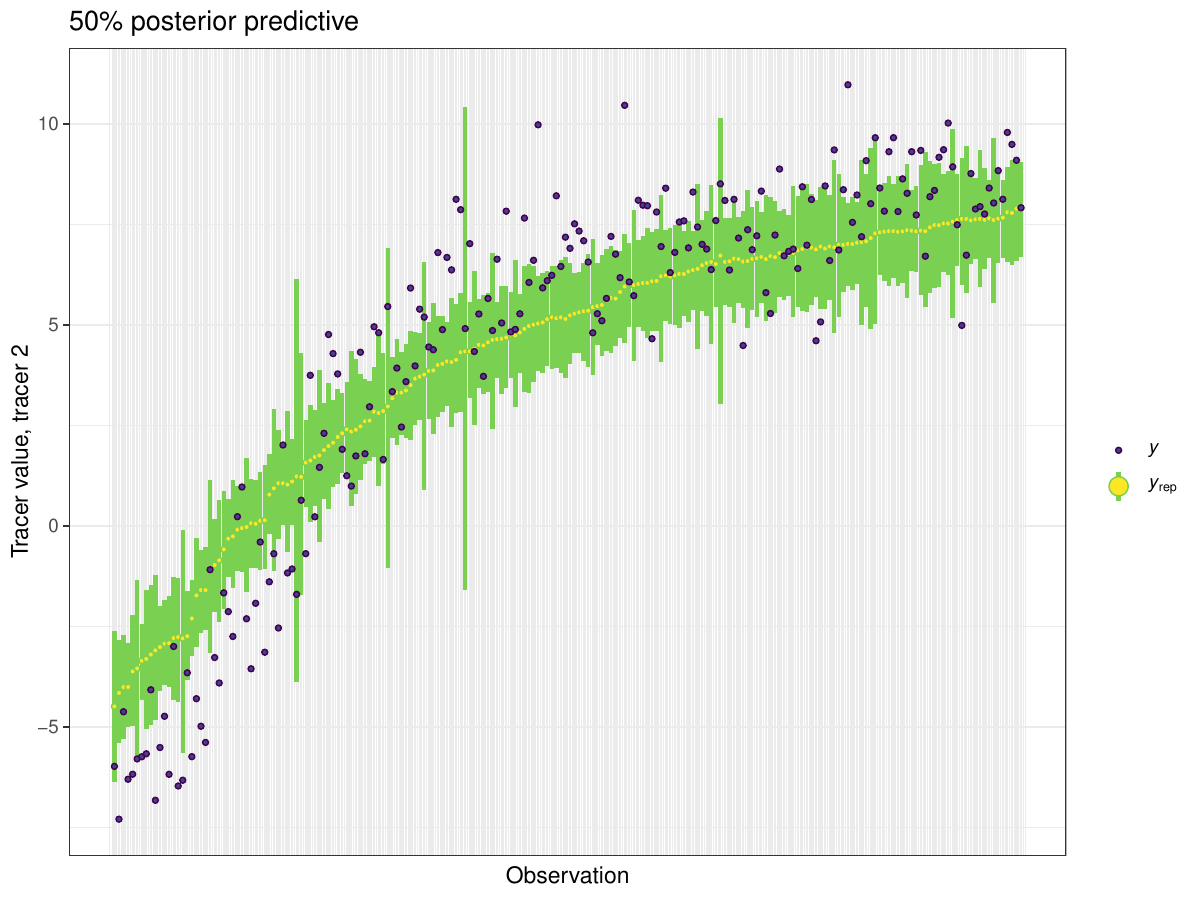}
    \caption{Tracer 2}
    \label{fig:t2_med}
  \end{subfigure}
 \hfill
 \begin{subfigure}[b]{0.45\textwidth}
    \includegraphics[width=\textwidth]{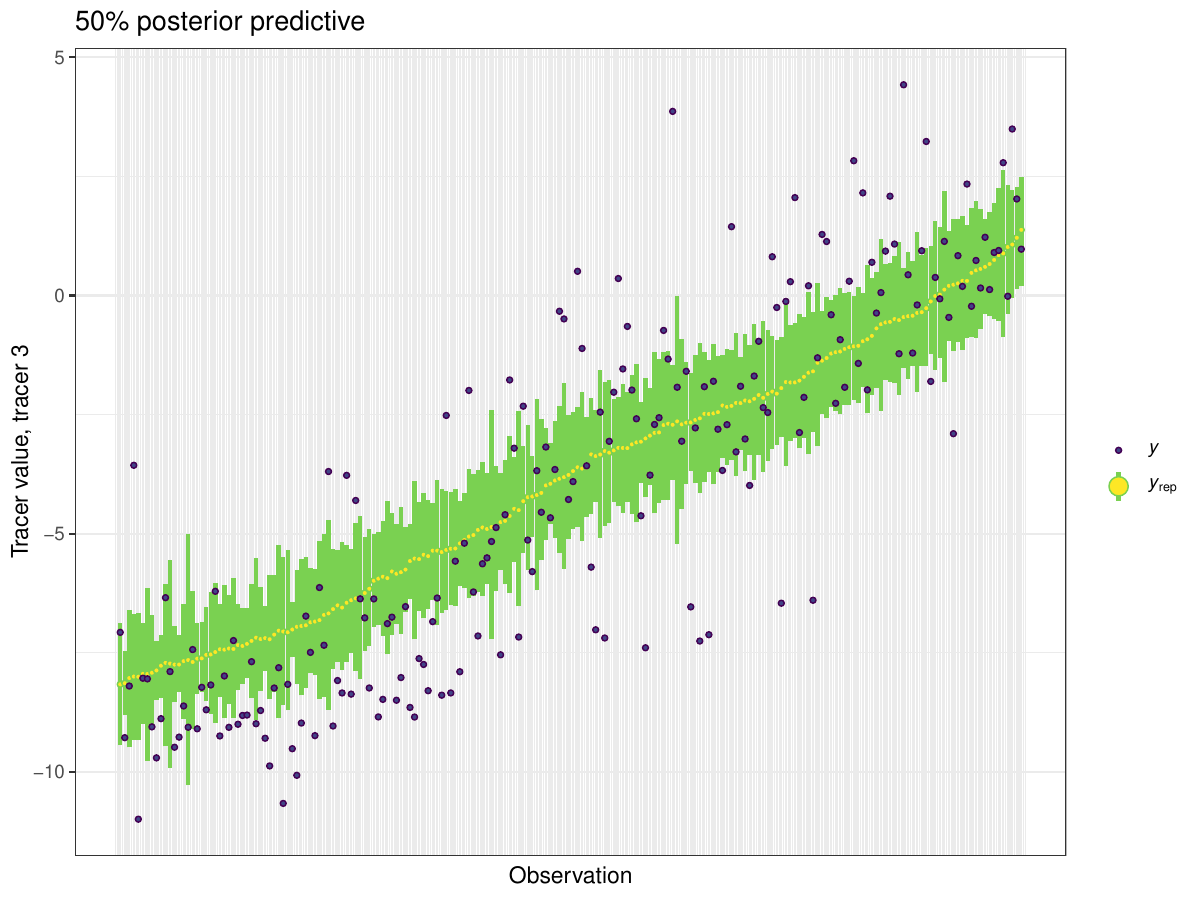}
    \caption{Tracer 3}
    \label{fig:t3_med}
  \end{subfigure}
    \hfill
  \caption{\label{fig:med_post_pred_app} Plot showing posterior uncertainty intervals at the 50\% level for the `Medium' simulated model run for tracers 2 and 3. The proportion of posterior values inside these values was 51\%.}
\end{figure}

\begin{figure}[h!]
 \begin{subfigure}[b]{0.45\textwidth}
   \includegraphics[width=\textwidth]{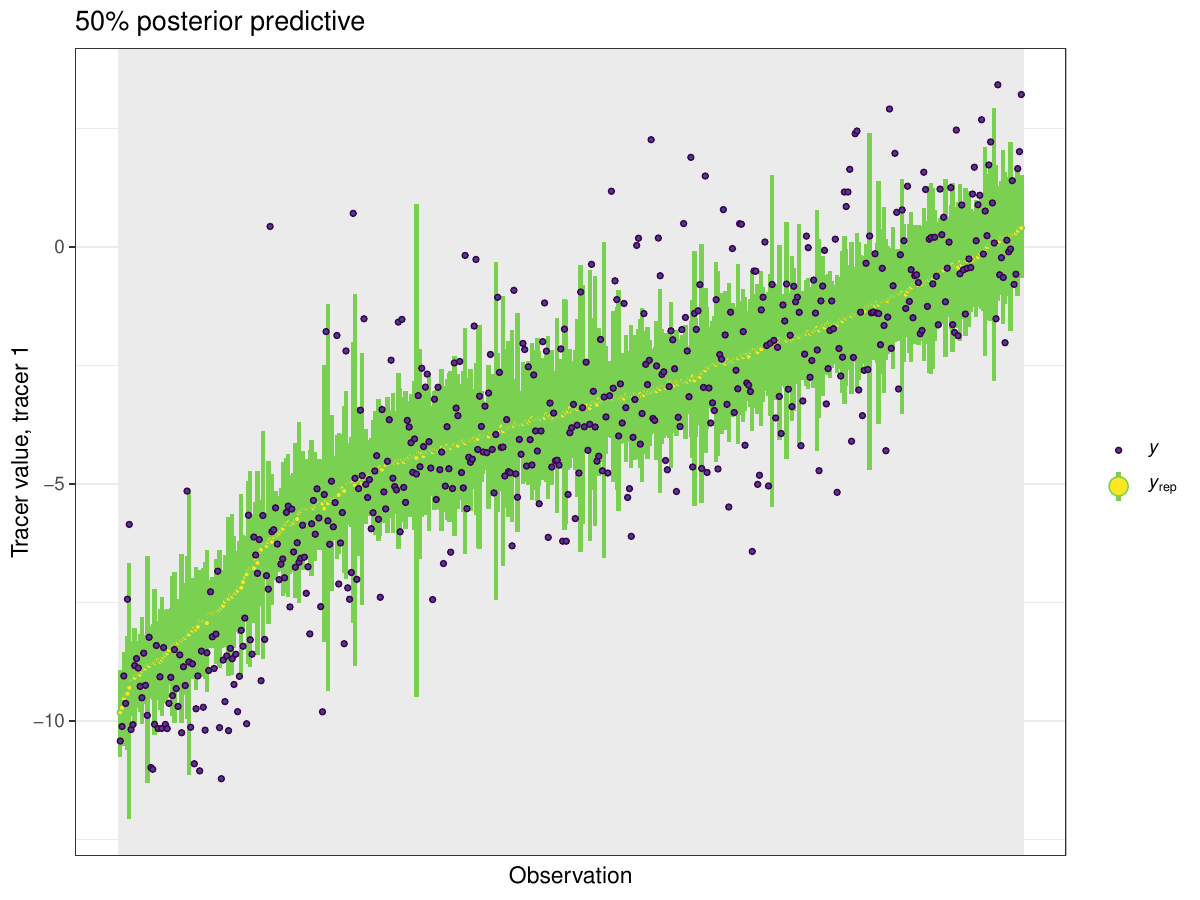}
       \caption{Tracer 1}
    \label{fig:t1_high}
  \end{subfigure}
   \hfill
 \begin{subfigure}[b]{0.45\textwidth}
    \includegraphics[width=\textwidth]{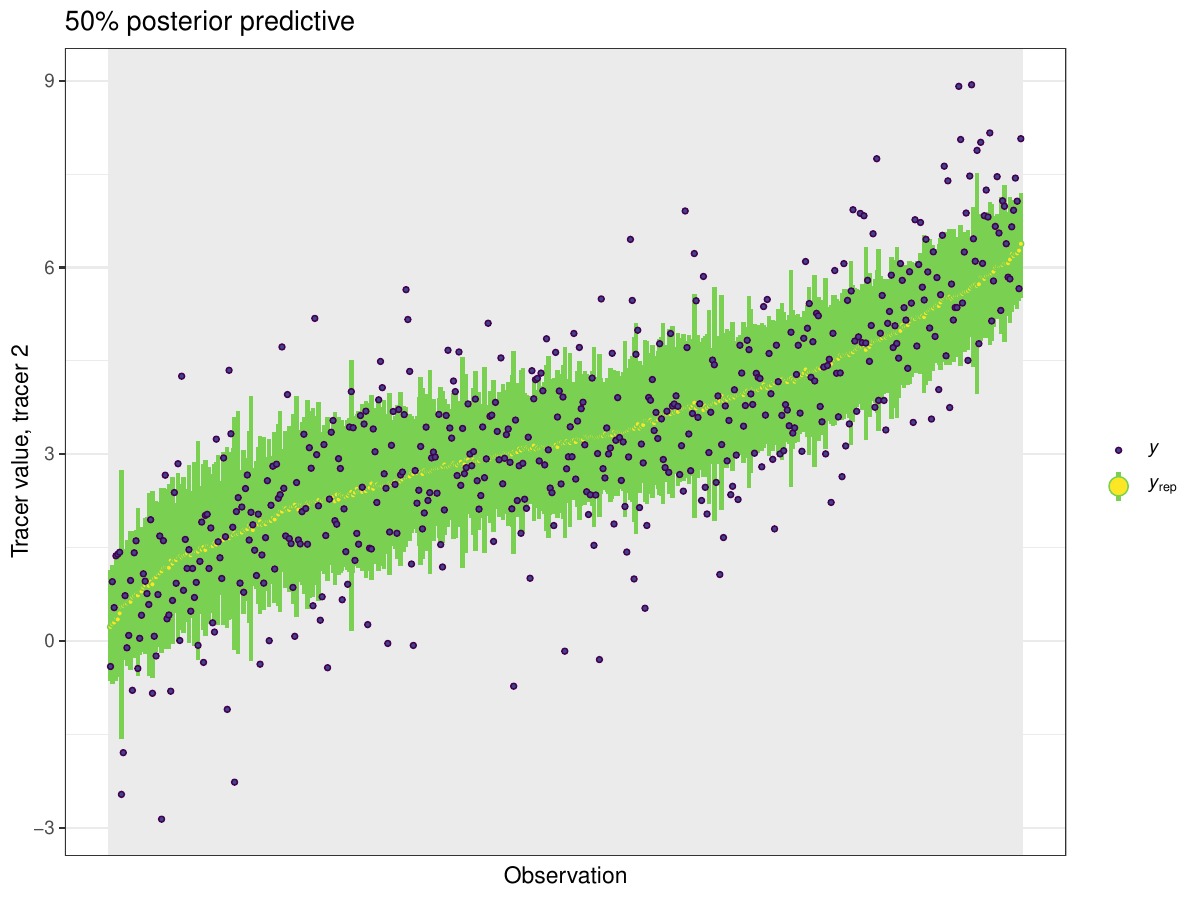}
        \caption{Tracer 2}
    \label{fig:t2_high}
  \end{subfigure}
   \hfill
 \begin{subfigure}[b]{0.45\textwidth}
    \includegraphics[width=\textwidth]{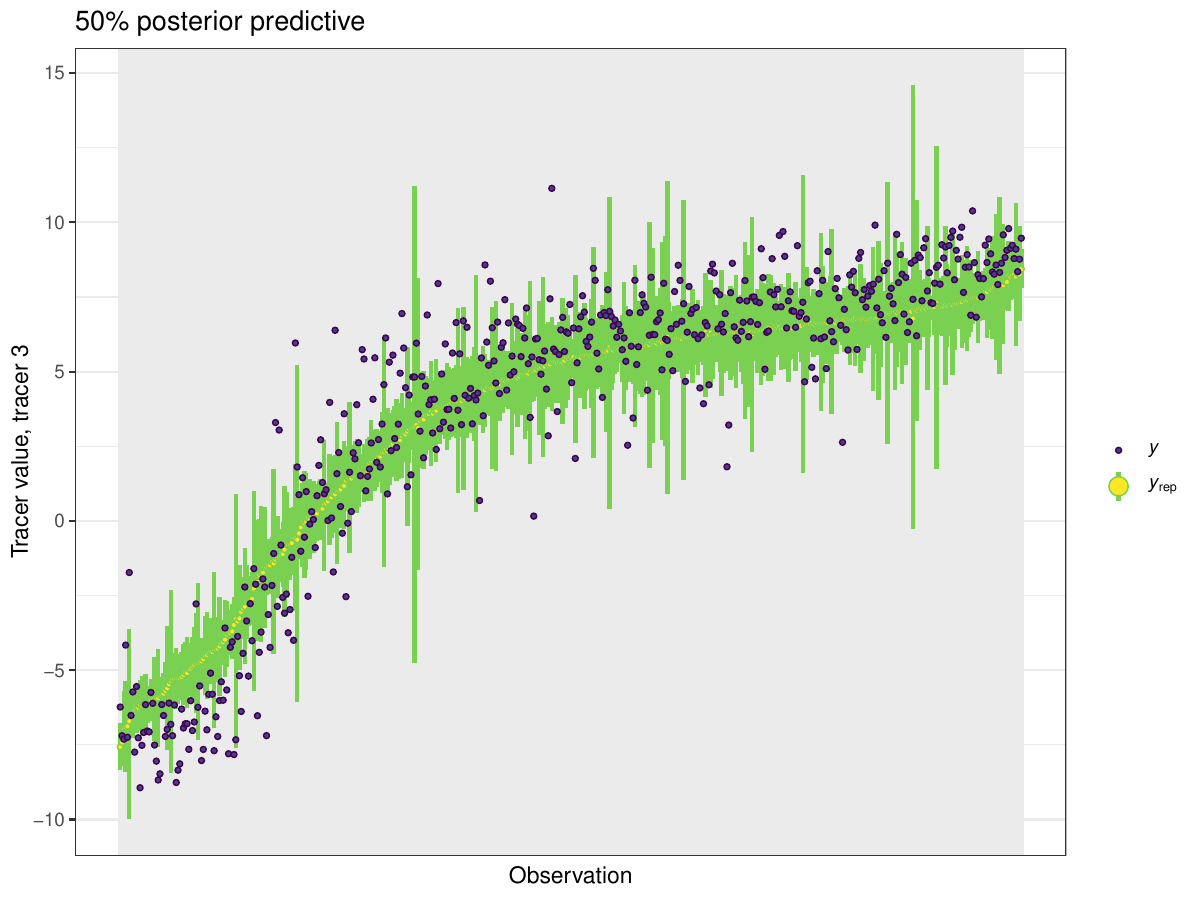}
    \caption{Tracer 3}
    \label{fig:t3_high}
  \end{subfigure}
   \hfill
 \begin{subfigure}[b]{0.45\textwidth}
    \includegraphics[width=\textwidth]{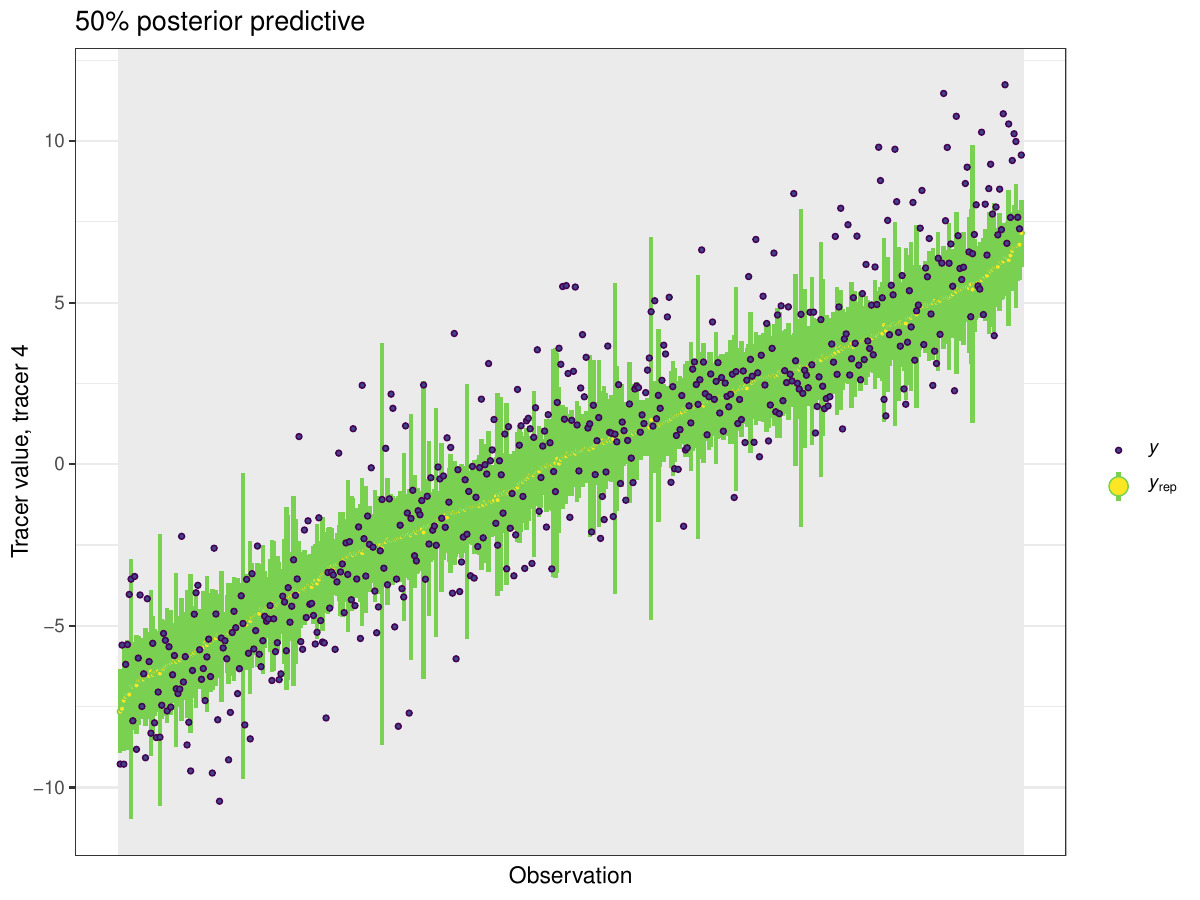}
    \caption{Tracer 4}
    \label{fig:t4_high}
  \end{subfigure}
   \hfill
  \caption{\label{fig:high_post_pred_app} Plot showing posterior uncertainty intervals at the 50\% level for the `High' simulated model run for tracers 1-4. The proportion of posterior values inside these values was 60\%.}
\end{figure}

  \begin{figure}[h!]
\centering
\includegraphics[width=0.75\textwidth]{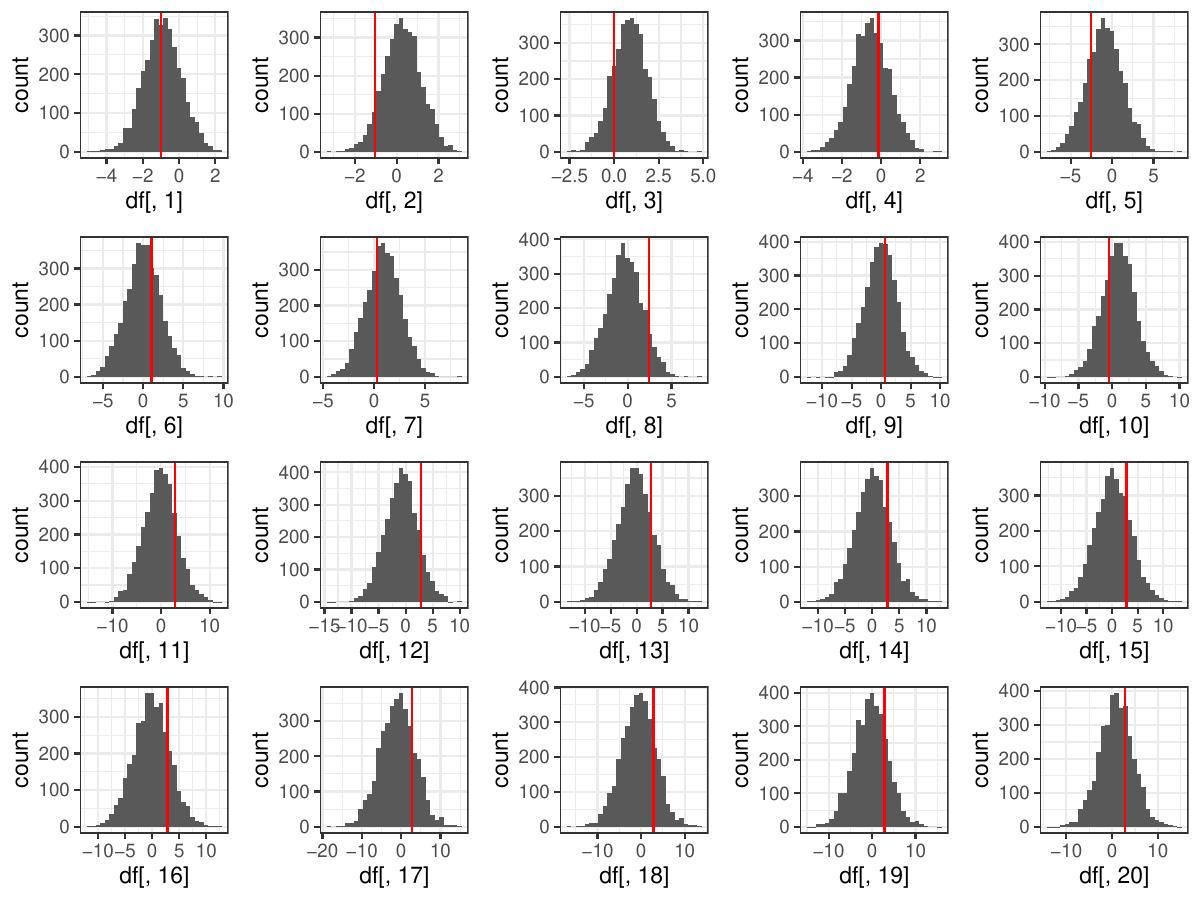}
\caption{\label{fig:beta_med} Histograms showing posterior samples for beta values generated via \texttt{cosimmr} for the `Medium' example, and red line showing `true' value of $\beta$ used to generate the mixture data.}
\end{figure}

\begin{figure}[h!]
\centering
\begin{subfigure}[b]{0.45\textwidth}
    \includegraphics[width=\textwidth]{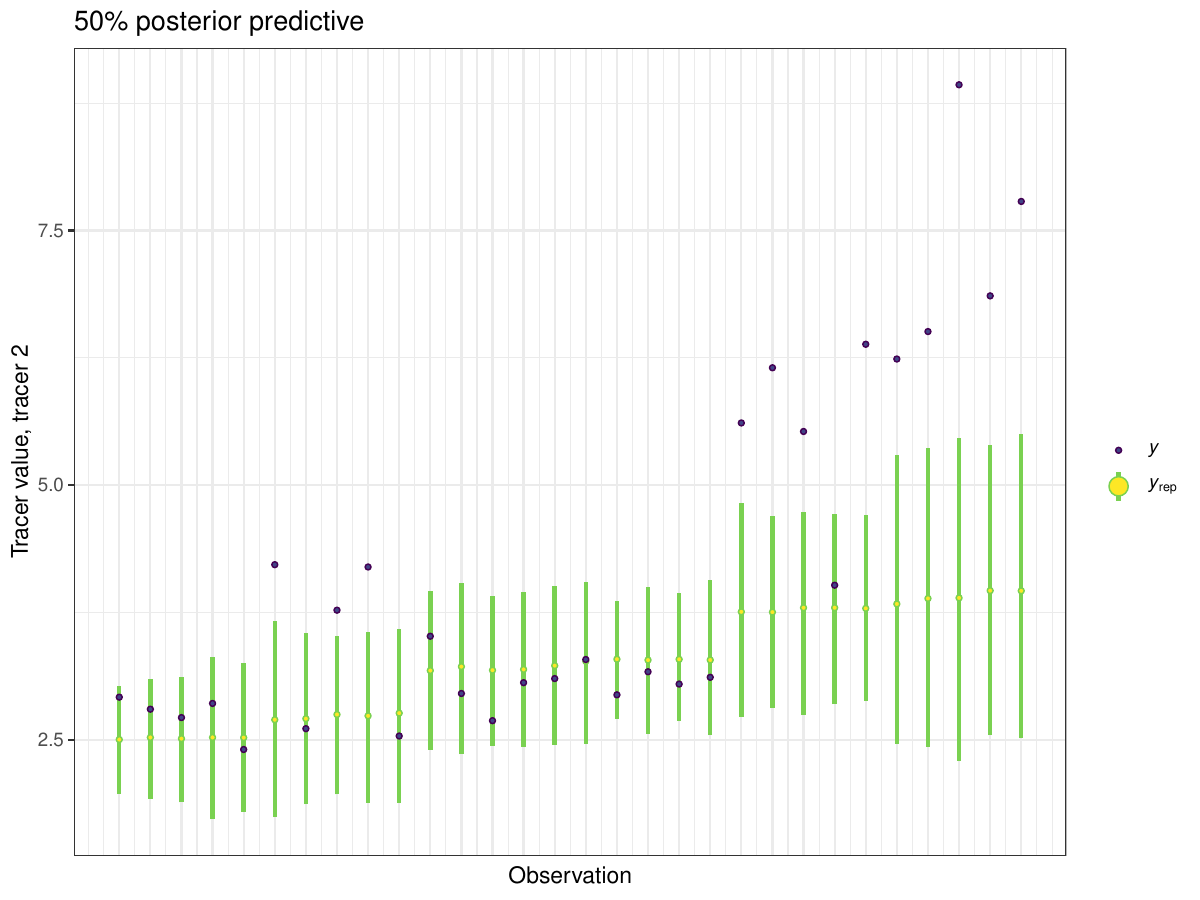}
    \caption{Tracer 2}
    \label{fig:iso2}
  \end{subfigure}
 \begin{subfigure}[b]{0.45\textwidth}
    \includegraphics[width=\textwidth]{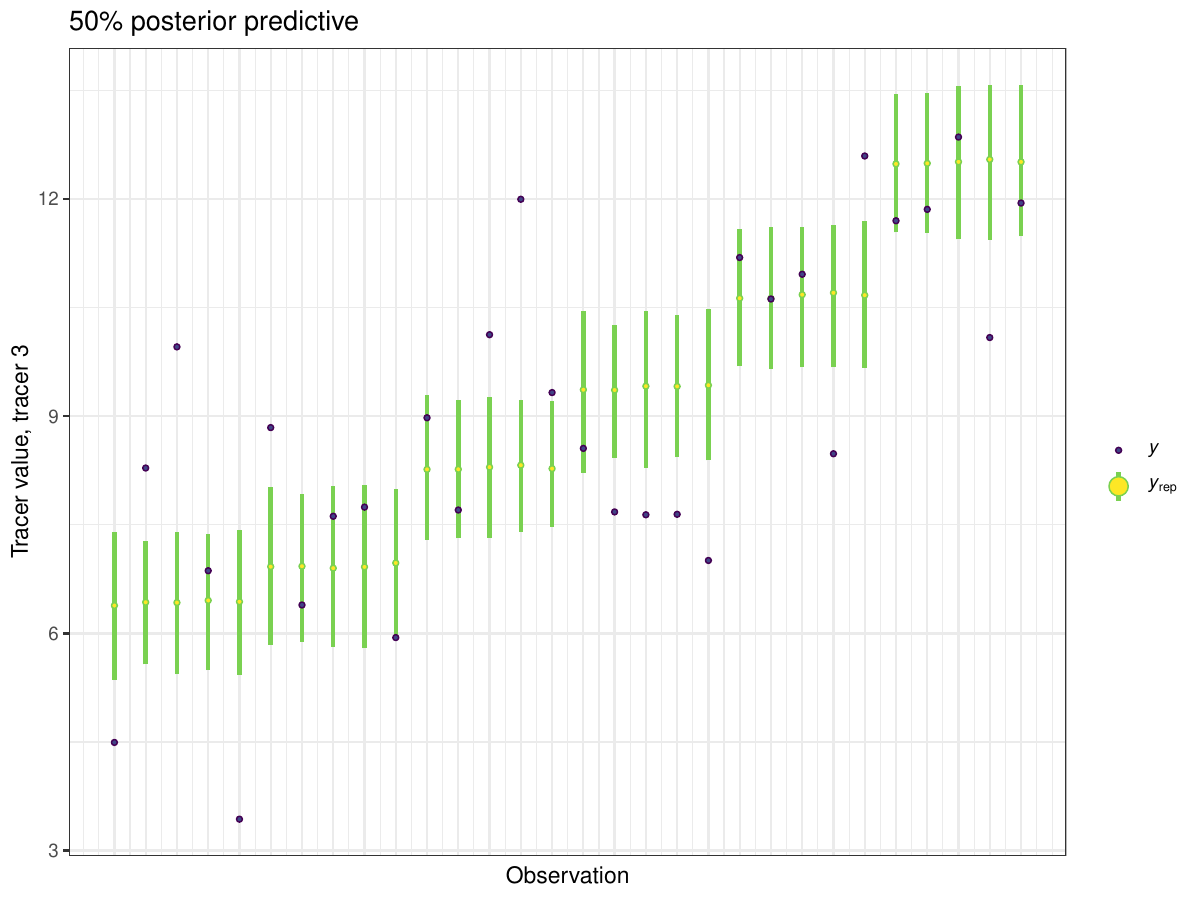}
    \caption{Tracer 3}
    \label{fig:iso3}
  \end{subfigure}
 \hfill
 \begin{subfigure}[b]{0.45\textwidth}
    \includegraphics[width=\textwidth]{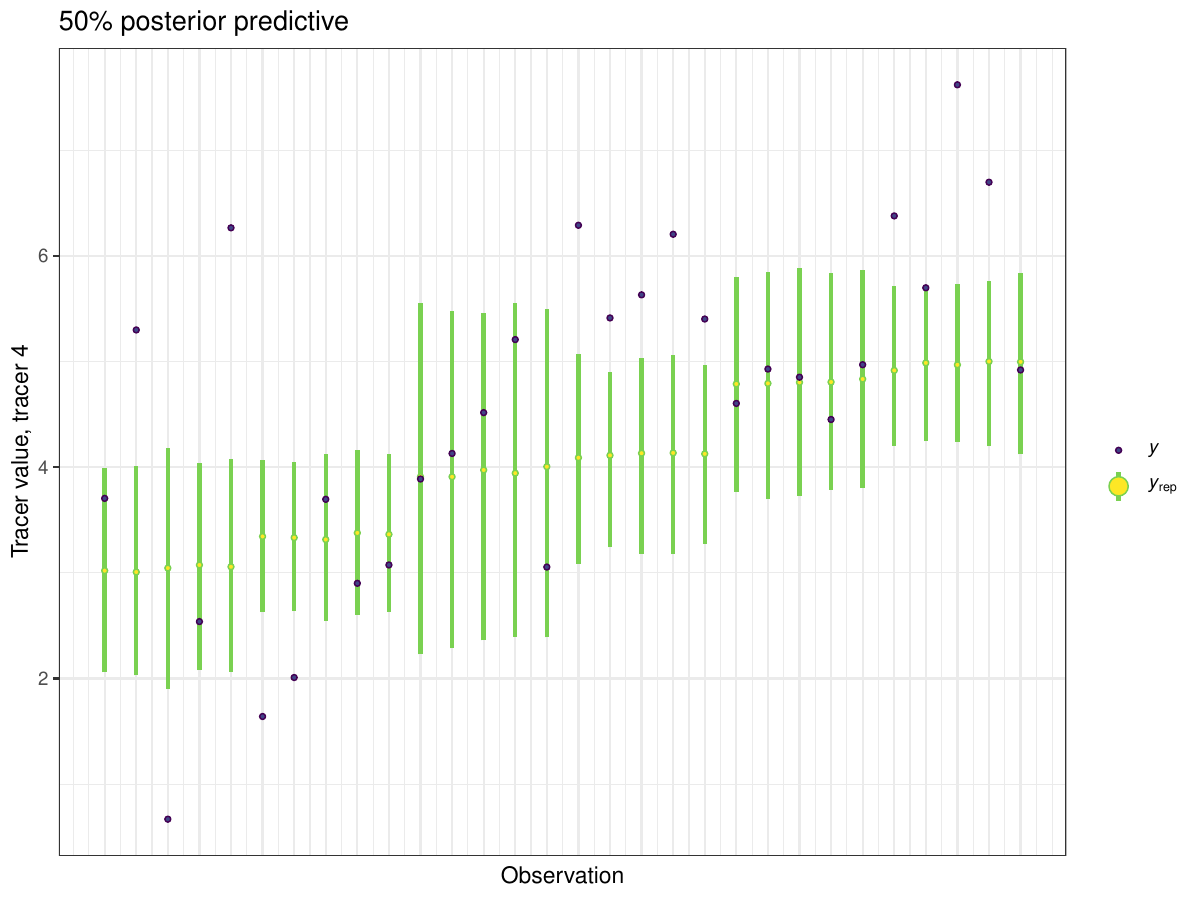}
    \caption{Tracer 4}
    \label{fig:iso4}
  \end{subfigure}
 \begin{subfigure}[b]{0.45\textwidth}
    \includegraphics[width=\textwidth]{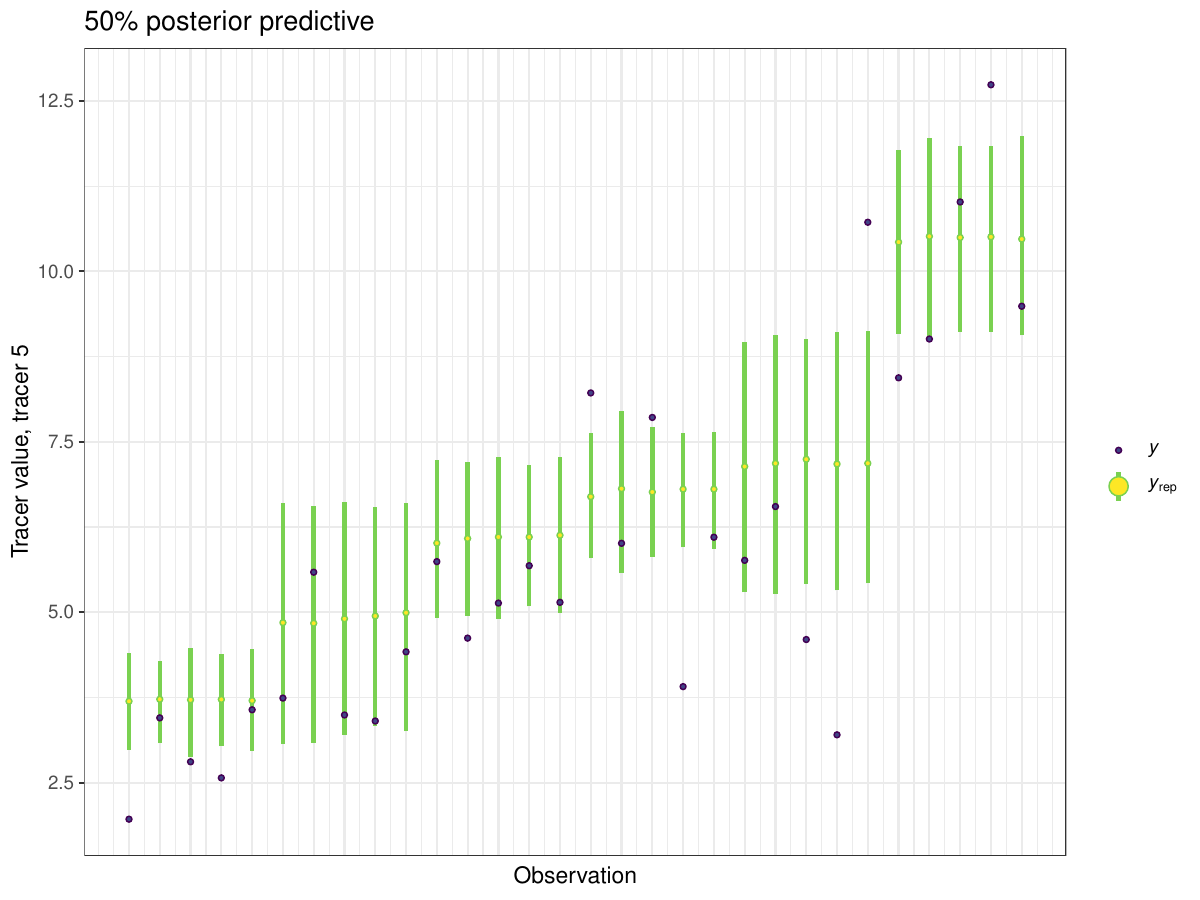}
    \caption{Tracer 5}
    \label{fig:iso5}
  \end{subfigure}
   \begin{subfigure}[b]{0.45\textwidth}
    \includegraphics[width=\textwidth]{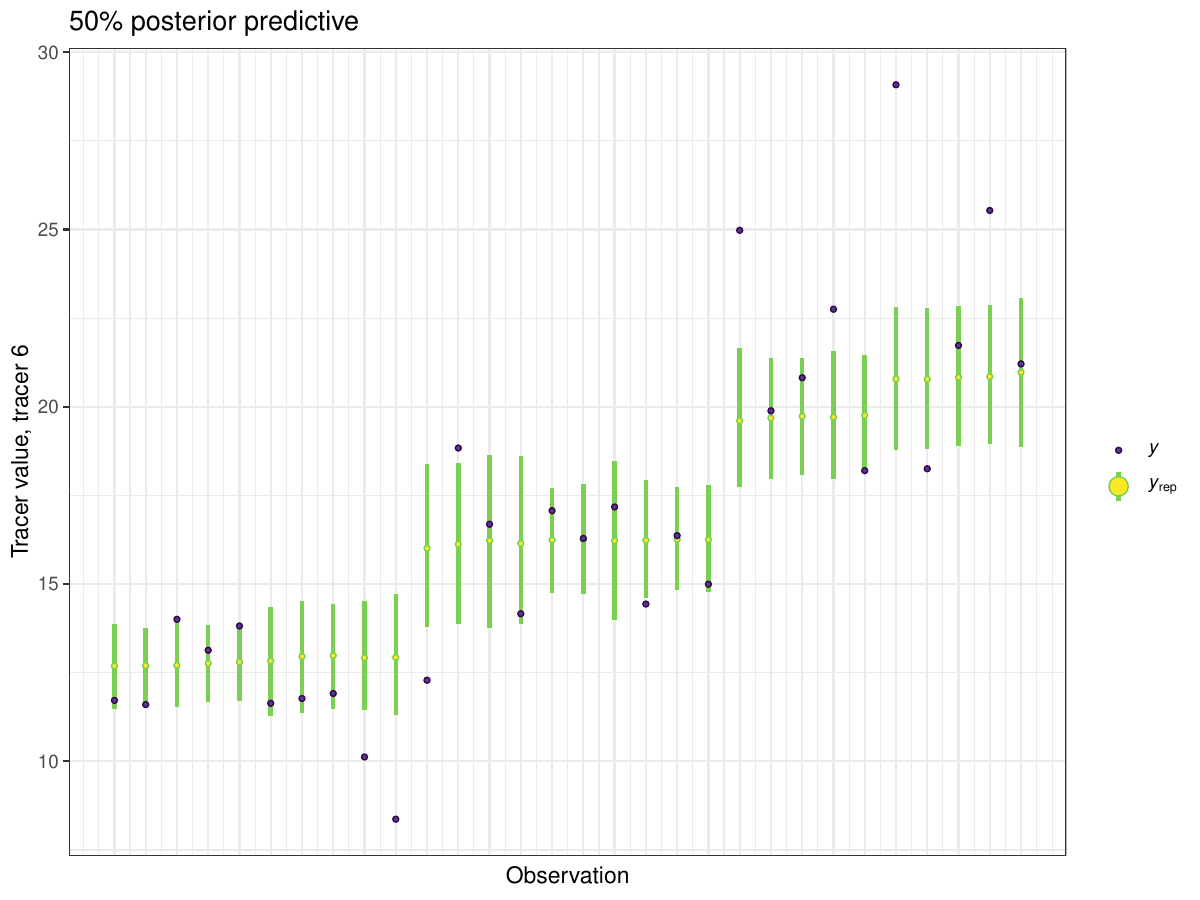}
    \caption{Tracer 6}
    \label{fig:iso6}
  \end{subfigure}
 \begin{subfigure}[b]{0.45\textwidth}
    \includegraphics[width=\textwidth]{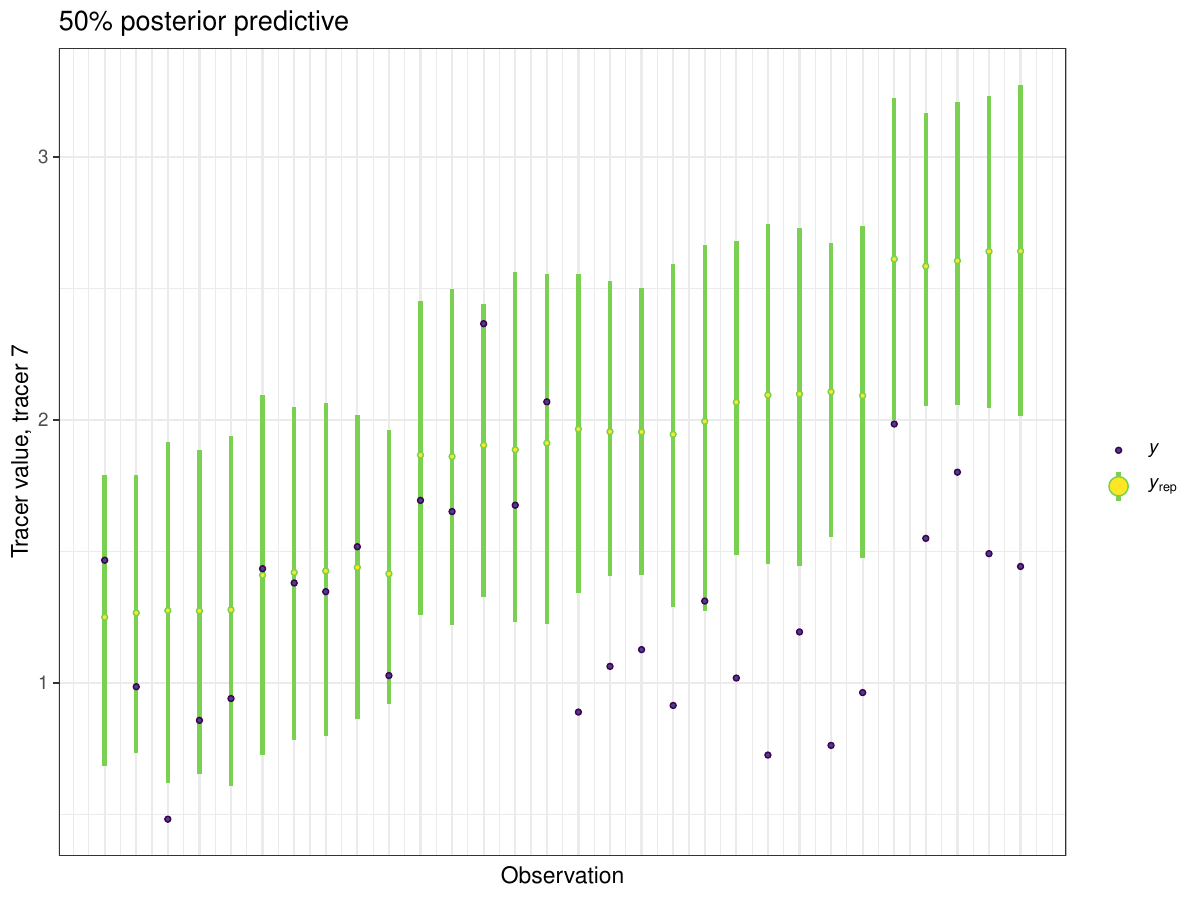}
    \caption{Tracer 7}
    \label{fig:iso7}
  \end{subfigure}
   \begin{subfigure}[b]{0.45\textwidth}
    \includegraphics[width=\textwidth]{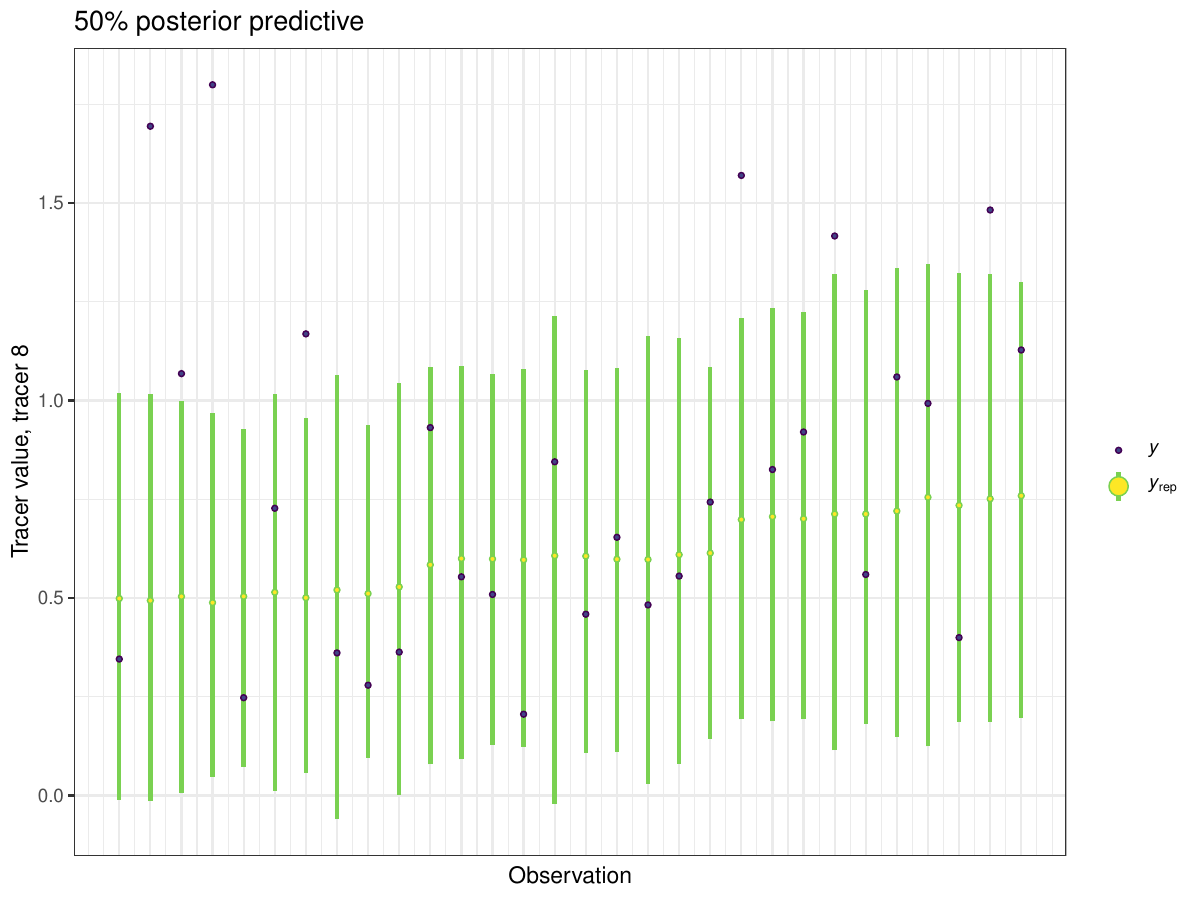}
    \caption{Tracer 8}
    \label{fig:iso8}
  \end{subfigure}
\caption{\label{fig:iso_appendix_post_pred} Plot showing posterior uncertainty intervals at the 50\% level for the Isopod data for tracers 2-8. The proportion of posterior values inside these values was 59\%.}
\end{figure}

\end{appendix} 
\end{document}